\documentclass[
 amsmath,amssymb,
pra,
]{revtex4-1}
 \usepackage{multirow}
\usepackage{graphicx}
\usepackage{dcolumn}
\usepackage{bm}
\usepackage[mathlines]{lineno}
\usepackage{color}

\begin{document}

\preprint{APS/123-QED}

\title{Accelerating the computation of quantum brachistochrone}

\author{Ding Wang$^1$}
\author{Haowei Shi$^1$}

\author{Yueheng Lan$^{1,2,}$}%

 \email{lanyh@bupt.edu.cn}
\affiliation{%
 $^1$School of Science, Beijing University of Posts and Telecommunications, Beijing 100876, China.\\
 $^2$State Key Lab of Information Photonics and Optical Communications,Beijing University of Posts and Telecommunications, Beijing 100876, China
}%

%
%

\date{\today}

\begin{abstract}
 Efficient control of qubits plays a key role in quantum information processing. In the current work, an alternative set of differential equations are derived for an optimal quantum control of single or multiple qubits with or without interaction.
 The new formulation enables a great reduction of the computation load by eliminating redundant complexity involved in previous formulations. A relaxation technique is designed for numerically detecting optimal paths involving entanglement. Interesting continuous symmetries are identified in the Lagrangian, which indicates the existence of physically equivalent classes of paths and may be utilized to remove neutral directions in the Jacobian of the evolution.  In the ‘ground state’ solution among the set of optimal paths, the time-reversal symmetry of the system shows up, which is expected to be universal for the symmetry-related initial and final state.
\end{abstract}

\maketitle

\section{\label{sec:level1}Introduction}
Ever since people realized that quantum computers are much more powerful than their classical counterparts\cite{Feynman1982,SHOR1994Algorithms}, researchers have been working hard on the problem for years, which has been making considerable progress and leads to a dramatic increase in the complexity of controllable quantum systems, from prototyped small quantum devices to machines with thousands of qubits (a typical example is the D-wave quantum annealing system \cite{Boix}). Undoubtedly, in order to reduce the impact of dissipation or decoherence that is always present in such complex systems, people continue to work on time-optimal quantum control, which is trying to minimize the necessary traversing time required to reach a quantum target (namely, a quantum state or a unitary operation) and proves to be important for building efficient gates in quantum computing architectures.
Therefore, a lot of research about the quantum speed limit (QSL) has been reported, which gives a lower bound of the time needed by a quantum system to travel from the initial to the final state and is estimated with the average energy or its variance~\cite{Campo2013,Deffner2013,Taddei2013,Mirkin2016}.
In addition, time optimal control provides a physical framework\cite{Schulte2005} for defining complexity of quantum algorithms, probably superior to traditional concepts~\cite{QuanInf}. For example Nielsen et al. proposed a new quantum computing criterion by using the Riemannian geometry, namely transforming the problem of finding an optimal quantum trajectory into a geometric one in the space of Hamiltonians\cite{Nielsen2006sci}, and revealed that the complexity of a quantum gate is related to the problem of optimal control~\cite{Nielsen2006pre}.

In a series of work about QSL~\cite{Campo2013,Deffner2013,Taddei2013,Mirkin2016}, the estimation of minimum evolution time is quite general and applicable to many different situations, in which, however the optimal evolution path is not given explicitly. In the literature, the GRAPE algorithm and the Krotov algorithm are very popular for locating optimal evolution path. In the GRAPE algorithm~\cite{Khaneja2005,Maximov2008}, the initial and the target state evolve forward and backward respectively for the same amount of time to obtain two state sequences and the control variables are updated by iteratively minimizing the difference between the two sequences. The GRAPE algorithm relieves the limitation in the number of control variables and shows high flexibility. However, the convergence is slowing down when the difference is close to some extreme values. Also, improper selection of step size often leads to a tortuous search route. In order to overcome this difficulty, the Krotov method~\cite{Maday2003,Maximov2008,Zhu1998,Shi1990} introduces a second term to penalize the undesired effects. A key feature of the algorithm is that all information is utilized at each time step which often consistently reduces the objective functional and accelerates the iteration. However, when  parameters are too small, its convergence rate deteriorates, and sometimes it becomes even worse that the algorithm turns unstable.
Another way to find some time-optimal solution and its precise characterization in case of different constraints on the Hamiltonian is through invoking interesting theories such as Pontryagin maximum principle and the geometry of the unitary group~\cite{KN2,Boozer2012,Hegerfeldt2013}.
A significant piece of work was reported by Carlini~\cite{Carlini2006}, who derived a quantum brachistochrone equation (QBE) with variational method which is widely used and plays a crucial role~\cite{Meiss1992,Bunimovich1995,Lan2004,Ghoussoub2007,Dong2014,Wang2018} in classical systems. The scheme produces a Schrodinger equation together with the optimal control strategy under an energy constraint. Except for some special cases~\cite{Carlini2007,Carlini2008,Carlini2014}, analytic solutions are hard to obtain and hence numerical solution is the only resort. A scheme based on a geometric point of view is designed recently~\cite{Xiaoting2015,Xiaoting2017}, which depicts the quantum brachistochrone paths as geodesics on the constraining manifold and solves the problem by solving a family of geodesic equations.
However, there is an infinite number of geodesic families present even for locally time-optimal solutions, which makes the scheme less convenient to use. Another drawback is the complexity of the commutators introduced by different terms in the Hamiltonian to the Euler-Lagrange equation (see Eq.(2) in Ref.\cite{Xiaoting2015}). As a result, the computation load for multiple qubits could get very high and hardly be applied in practice. The problem of how to reach an effective dimension reduction and achieve fast maneuver of quantum bits is worth further exploration.

In this paper, we uncover the source of computational complexity from the perspective of a variational principle, and in the process propose an alternative set of equations with much improved efficiency, especially for multiple bit with limited control parameters. We first slightly extend the formulation in Ref.~\cite{Carlini2006} and derive all the necessary equations for later convenience. By introducing a set of new variables, the redundancy originated from the commutation relation between operators gets eliminated, which enables a very efficient computation of optimal paths in the presence of multiple qubits.
A relaxation scheme is the designed and implemented to solve the problem involving quantum entanglement when interaction between different bits comes into paly. With several continuous symmetries identified in the Lagrangian formulation, it is possible to remove the annoying neutral directions in the numerical computation. In all the examples we tried, the simplest optimal paths seem to always bear the time-reversal symmetry if the initial and target state are related by a symmetry of the Hamiltonian.

The paper is organized as follows. In Sec.~\ref{sec:1}, a Lagrangian formation of the QBE is reproduced with an introduction of uncontrollable terms (the coefficients of which are fixed values). All the necessary variational equations are derived and written in convenient vector forms. The solution of these equations is checked analytically and verified numerically in Sec.~\ref{sec:rotation}, in the absence of interaction. To deal with multiple qubits with interaction, a new set of equations are written down in Sec.~\ref{sec:numerical} by change of variables from the original set, which achieves a great reduction of complexity. The evolution of the entanglement on an optimal path is always unimodal, which is demonstrated in all the examples in Sec.~\ref{sec:example}.  The results are summarized in the final section.
\section{the variational principle}
\label{sec:1}
Without loss of generality, the Hamiltonian of a quantum system that is of interest could be written as:
\begin{equation}\label{eq:H}
   H=\sum_{j=1}^m \xi_j(t) A_j\,,
\end{equation}
where $\xi_j(t)$'s are real numbers and $A_j$'s are Hermitian operators that satisfy $Tr A_j=0$ and $Tr(A_jA_k)=\delta_{jk}~(j,k \leq m)$.
The traceless Hamiltonian could be divided into two parts: the controllable part $\bm{\mathcal A}=\sum\limits_{j=1}^n \xi_j(t) A_j $ with each $\xi_j (j=1,...,n)$ subject to external control and the uncontrollable part $\bm{\mathcal B}=\sum\limits_{j=n+1}^m \xi_j(t) A_j$ with $\xi_j$'s being constant or changing in a pre-determined way, often used to model interaction between qubits.
For later convenience we use $\bm u$ and $\bm v$ to label coefficients of the controllable part $\bm u=(\xi_1\,,\xi_2\,,\cdots\,,\xi_n\,,0\,,\cdots\,,0)^t$ and the uncontrollable part $\bm v=(0\,,0\,,\cdots\,,\xi_{n+1}\,,\xi_{n+2}\,,\cdots\,,\xi_m)^t$, and hence $\bm \xi=\bm u + \bm v$
. In this work, we build a formulation applicable to the general case which includes both components.

\subsection{\label{sec:level2}A variational scheme for the QBE}
The action integral is written as~\cite{Carlini2006}:
\begin{widetext}
  \begin{equation}
     S(\psi\,,H\,,\phi\,,\lambda,,\lambda')= \displaystyle{\int} dt \dfrac{\sqrt{ \langle \dot{\psi}|(1-P)|\dot{\psi}\rangle}}{\Delta E}+\big(i\langle \dot{\phi}|\psi\rangle+\langle\phi|H|\psi\rangle+c.c.\big)+
     \lambda(\sum\limits_{j=1}^m\frac{\xi_j^2}{2}-\omega^2) +\sum\limits_{j=n+1}^m\lambda^{'}_j (\xi_j-Q_j)\
\,,\label{eq:S}
  \end{equation}
\end{widetext}
where $P(t)=|\psi(t)\rangle\langle\psi(t)|$ is the projection to the state $|\psi(t)\rangle$ which is a normalized wavefunction defined in the configuration space, $\phi$ is an auxiliary wavefunction and $c.c.$  denotes complex conjugates. $(\Delta E)^2 \equiv \langle\psi|H^2|\psi\rangle-\langle\psi|H|\psi\rangle^2$ is the energy variance. Parameters $\lambda$ and $\lambda_j^{'}$ are Lagrange multipliers and all $\{\lambda'_j\}_{j=1,2,...,n}$ are set to zero, $\omega$ is a given constant, the $Q_j$'s denote the coupling constants in the Hamiltonian and the Planck’s constant $\hbar$ is chosen to be 1 for simplicity.

Now we take the variation of the action Eq.~(\ref{eq:S}) to obtain the equation of motion and all the constraints imposed to the variables.
\paragraph{}
The variation with respect to $\phi$ gives
\begin{equation}\label{Eq1}
  i|\dot{\psi}\rangle=H|\psi\rangle\,,
\end{equation}
which is the usual Schrödinger equation. In the projective space $\mathbb{C}P^{n-1}$, the Fubini-Study line element is written as $ds^2=\langle d\psi|(1-P)|d\psi\rangle$, which measures the displacement in the angular direction. Here, it is easy to check that $ ds^2=(\Delta E)^2 dt^2$. Hence, the first part of Eq.~(\ref{eq:S}) refers to the total time of the process and the other parts embody differential or other constraints.
\paragraph{}
Here, we take the variation with respect to $\lambda$ and $\lambda'_j$ separately, resulting in
$\frac{1}{2}\sum\limits_{j=1}^m\xi_j^2=\omega^2$, which defines the size of the Hamiltonian as a finite energy constraint, and $\xi_j=Q_j$~($n<j \leq m$), which depicts the influence of generic interactions.

\paragraph{}
The variation with respect to $\psi$ gives
\begin{equation}\label{Eq4}
  i\dfrac{d}{dt}\bigg[\dfrac{H-\langle H\rangle}{2(\Delta E)^2}\bigg]|\psi\rangle-i|\dot{\phi}\rangle+H|\phi\rangle=0 \,,
\end{equation}
where $\langle \cdot \rangle$ refers to the average with respect to $|\psi\rangle$. Eqs.~$(\ref{Eq1})$ and $(\ref{Eq4})$ are seen in the literature\cite{Carlini2006} but the equations below appear new.
\paragraph{}
The variation with respect to $H$ could now be taken as multivariate change with respect to $\{\xi_j\}_{j=1,2,...m}$, resulting in
\begin{equation}\label{Eq6}
  D_j=\frac{1}{2\Delta E^2}\sum\limits_k F_{jk} \xi_k -\lambda \xi_j -\lambda^{'}_j ~~~~(\text{with}~ \lambda^{'}_j=0 ~\text{for}~ j\le n),
\end{equation}
where $F_{jk}=\langle A_jA_k+A_kA_j\rangle-2 \langle A_j\rangle\langle A_k\rangle$, $ D_j=\langle \phi|A_j|\psi\rangle +c.c.$ and $\sum\limits_{k}$ stands for $\sum_{k=1}^{m}$ which is also the case in all the following equations. The detailed derivation of Eq~(\ref{Eq6}) has been relegated to Appendix~\ref{sec:A}.
Next, we derive the equations of motion for $\lambda$ and $\{\xi_j\}_{j=1,...,n} $.

 A multiplication of both sides of Eq.~($\ref{Eq6}$) with $\{\xi_j\}_{j=1,...,n} $, the controllable components, and then a summation over j results in
 \begin{equation}\label{Eq7}
   \lambda=\dfrac{1-\bm D\cdot \bm\xi-\sum\limits_{j=n+1}^m \bigg(\sum\limits_k \dfrac{\xi_jF_{jk}\xi_k}{2\Delta E^2} - D_j\xi_j\bigg)}{\omega'^2} \,,
 \end{equation}
  where $\bm D\cdot \bm\xi=\sum\limits_{j=1}^{m} D_j \xi_j$ and $\omega'^2 = \sum\limits_{j=1}^{n}\xi_j^2$ is constant (the constraint on the controllable part of the Hamiltonian). Noticing that $\{\dot \xi_j\}_{j=n+1,...m}\equiv 0 $, we could obtain:
 \begin{equation}\label{Eq8}
   \dot{\lambda}=\dfrac{\sum\limits_{j=n+1}^{m}\bigg\{-\dfrac{\xi_j}{2\Delta E^2}\bigg[\sum\limits_k (\dot{F_{jk}}\xi_k+F_{jk}\dot{\xi_k})
-\dfrac{\tilde{\bm F}\cdot \dot{\bm\xi}}{\Delta E^2}\tilde{F_j}\bigg]+\dot{D_j}\xi_j\bigg\}}{\omega'^2}\,,
 \end{equation}
 where $\tilde{\bm F}=\bm\xi \cdot F$ and $\tilde{F_j}=\sum_k F_{jk}\xi_k$. Here we used the fact that $ \dfrac{d}{dt}\bm D\cdot\bm\xi=0$ (details in Appendix~\ref{sec:C}), $\bm\xi\cdot \bm C= \bm C\cdot\bm\xi=0 $ (details in Appendix~\ref{sec:D}) and
 \begin{equation}\label{Eq9}
   \dfrac{d}{dt}D_j=\dfrac{1}{2\Delta E^2}(\sum\limits_k F_{jk} \dot{\xi}_k-\dfrac{\tilde{\bm F}\cdot\dot{\bm\xi}}{\Delta E^2}\tilde{F_j})+\sum\limits_{k}C_{jk}D_k\,,
 \end{equation}
 where $C_{jk} $ satisfies $[H,A_j]=-i\sum_k C_{jk} A_k $, with $C_{jk}$ being a real number. For a detailed derivation of Eq.~($\ref{Eq9}$), please check  Appendix B. Feeding Eq.~(\ref{Eq9}) into Eq.~($\ref{Eq8}$) results in
 \begin{align}\label{Eq10}
   \dot{\lambda}= & \dfrac{\sum\limits_{j=n+1}^{m}\sum\limits_{k=1}^{m} \xi_jC_{jk}(D_k-\frac{\tilde{F}_k}{2\Delta E^2})}{\omega'^2}\nonumber \\
    =& -\dfrac{ \bm{v\cdot C \cdot \lambda}' }{\omega'^2} \,,
 \end{align}
with $\dot{\bm F}=\bm C\cdot \bm F +\bm F\cdot \bm C^{T}$ (Appendix \ref{sec:E}). Taking the time derivative of both sides of Eq.~(\ref{Eq6}), and together with Eq.~(\ref{Eq9}), we get
\begin{equation}\label{Eq11}
  \dot{\bm u}=- \dfrac{\bm C\cdot \bm D}{\lambda}+\dfrac{\bm C\cdot F\cdot \bm\xi
  }{2\lambda\Delta E^2}+\dfrac{(\bm{v\cdot C \cdot \lambda}')}{\lambda\omega'^2}\bm u\,.
\end{equation}
With Eq.~(\ref{Eq9}), Eq.~(\ref{Eq10}), Eq.~(\ref{Eq11}) and the Schrödinger equation Eq.~(\ref{Eq1}), the optimal path can be calculated now.

Here let's discuss a special case with $H\in\mathcal{A} $, namely $\bm \xi=\bm u$, which means that all parts are controllable and thus $\bm v =0$.
It is obvious that this case is simple since
\begin{align}
D_j&=\frac{1}{2\Delta E^2}\sum\limits_k F_{jk} u_k -\lambda u_j,  \label{Eq12}\\
\dfrac{d}{dt}D_j&=\dfrac{1}{2\Delta E^2}(\sum\limits_k F_{jk} \dot{u}_k-\dfrac{\tilde{\bm F}\cdot\dot{\bm u}}{\Delta E^2}\tilde{F_j})+\sum\limits_{k}C_{jk}D_k .\label{Eq13}
\end{align}
Hence Eq.(\ref{Eq10}) implies that $\dot\lambda=0 $. Eq.~(\ref{Eq11}) could be reduced to
\begin{equation}\label{Eq14}
\dot{\bm u}=- \dfrac{\bm C\cdot \bm D}{\lambda}+\dfrac{\bm C\cdot \bm F\cdot \bm u
}{2\lambda\Delta E^2}.
\end{equation}
Considering $\bm u\cdot \bm C=0$ (Appendix~\ref{sec:D}) and feeding Eq.~(\ref{Eq12}) into Eq.~(\ref{Eq14}) leads to a trivial solution  $\dot {\bm u}=0 $, which indicates that $\bm u$ is a constant vector and the evolution corresponds to a simple rotation with a fixed axis on the Bloch sphere.
\subsection{Gauge symmetries in the Lagrangian}
\label{sec:symm}
In this section, we discuss the gauge symmetries in the Lagrangian Eq.~(\ref{eq:S}), which create neutral directions in the evolution and may bring complication in numerical calculation. First, consider a transform $\hat{U}_1: \phi \to \phi+ia\psi$, $a\in \mathcal{R}$, with which the Lagrangian becomes
\begin{equation}\label{Eq15}
S'=S+\int dt ~(a\langle \dot{\psi}|\psi\rangle-ia\langle\psi|H|\psi\rangle+c.c.)
\,,
\end{equation}
The expression $\langle\dot{\psi}|\psi\rangle+c.c$ is a total differentiation, which could be integrated out. With the Hermiticity of $H$, the extra integral in Eq.~(\ref{Eq15}) vanishes and hence $\hat{U}_1$ is a symmetry transform. Next, we consider another transform $\hat{U} _2:\phi\rightarrow\dfrac{\lambda+b}{\lambda}\phi,~\lambda\rightarrow\lambda+b,~\bm\lambda'\rightarrow\dfrac{\lambda+b}{\lambda}\bm\lambda',~ b\in \mathcal{R}$ which sends Eq.~($\ref{eq:S}$) to
\begin{align}\label{Eq16}
S(\psi,H,\phi,\lambda,\lambda')&=\int dt \dfrac{\sqrt{\langle \dot{\psi}|(1-P)|\dot{\psi}\rangle}}{\Delta E}\nonumber\\
  &+\dfrac{\lambda+b}{\lambda}\left[ (i\langle \dot{\phi}|\psi\rangle+\langle\phi|H|\psi\rangle+c.c.)
+\lambda(\sum_{j=1}^{m}\frac{\xi_j^2}{2}-\omega^2)+\sum_{j=n+1}^{m} \lambda'_j (\xi_j-Q_j) \right]
\,,
\end{align}
which only changes the values of the invisible Lagrange multipliers $\phi,~\lambda,~\bm\lambda'$ while the equation for the observables $\psi,~\bm u$ remain intact. Actually, it is most obvious in the simplified formulation Eq.~(\ref{eq:evolution2}) below, where $\bm\Omega,~\bm{u},~\dot{\bm\Omega},~\dot{\bm u}$ keep invariant under the transform $\hat{U}_2$. Hence $\hat{U}_2$ is a symmetry transform.

\section{The evolution of Hamiltonian as a rotation }
\label{sec:rotation}
The equation $\bm\xi\cdot\bm C =0 $ tell us that $\bm v\cdot\bm C =-\bm u \cdot \bm C $. Substituting the Eq.~(\ref{Eq6}) into Eq.~(\ref{Eq11}) results in
\begin{equation}
\dot{\bm u}=(I-P_{u})\cdot\dfrac{\bm C\cdot\bm\lambda'}{\lambda}
\,,\label{eq:dua}
\end{equation}
where $I$ is the identity matrix and $P_{u}$ is the projection operator $\dfrac{|\bm{u}\rangle\langle \bm u|}{|\bm u|^2}$ ($\omega'^2 = \sum\limits_{j=1}^{n}\xi_j^2=\sum\limits_{j=1}^{n}u_j^2=|\bm u|^2$). Here we use the Dirac notation to emphasize the projection, whereas $\bm u$ is actually real. With the fact that $\bm{\dot{v}}=0$, it is easy to check that $\bm\xi\cdot \bm{\dot{\xi}}=\bm u\cdot \dot{\bm u}=0$.
And the uncontrollable part of the time derivative of Eq.~(\ref{Eq6}) similarly gives the evolution of $\bm\lambda'$
\begin{equation}
  \dot{\bm\lambda}'=
  \bm C\cdot\bm\lambda'
  +\dfrac{\bm v\cdot \bm C\cdot\bm\lambda'}{\omega'^2}\bm v=(I+\dfrac{|\bm v|^2}{|\bm u|^2}P_{\bm v})\cdot \bm C \cdot\bm\lambda'
.\label{eq:dlambda'}
\end{equation}
where $P_{v}=\dfrac{|\bm v\rangle\langle \bm v|}{|\bm v|^2}$ and $\dfrac{|\bm v|^2}{|\bm u|^2}$ is a real constant.
After,
defining $\bm{\Omega}=-\dfrac{2}{\sqrt{Tr I}}\dfrac{\bm{\lambda'}}{\lambda}$, the angular velocity of the rotation of $\bm{u}$ can be described with $\bm\Omega$. Furthermore,  the QBE problem can be characterized with a simpler set of equations
\begin{equation}
\begin{aligned}
\dot{\bm u}&=-\dfrac{\sqrt{Tr I}}{2}(I-P_{u})\cdot \bm C\cdot\bm\Omega \\
\dot{\bm \Omega}&=(I+\dfrac{|\bm v|^2}{|\bm u|^2}P_{\bm v})\cdot \bm C\cdot\bm\Omega-\dfrac{\sqrt{Tr I}}{2\omega'^2}(\bm v\cdot \bm C\cdot\bm\Omega)\bm\Omega
\end{aligned}
\label{eq:evolution2}
\end{equation}
along with the Schrödinger equation $\dot{\psi}=-iH\psi$.

According to Appendix \ref{sec:D}, $\bm C$ can be conveniently determined, which in specific circumstances may provide possible routes to analytic solutions. As a specifical case, in the case without entanglement, the phase space can be divided into separate Bloch spheres of qubits.
Then, according to Eq.~(\ref{eq:evolution2}), the equation of a single qubit in 3-dimensional space is appreciably simplified to $\dot{\bm u}=\bm \Omega\times \bm\xi-\hat{\bm u}\cdot(\bm\Omega\times \bm\xi)~\hat{\bm u}=\bm\Omega\times \bm u$. Meanwhile, the uncontrollable part is generally limited to no more than one dimension per qubit, or there will be no solution for most boundary conditions. As a result, $\bm\Omega$  is in the direction of $\bm\lambda'$ which is along $\bm v$ in the case of three dimensions. Here the cross product holds for each $SO(3)$ Bloch sphere, which indicates that $\bm u$ is rotating with the angular velocity $\bm\Omega$.

As $\bm\Omega, \bm v$ is in the same direction for a single qubit. After multiplying $\bm\Omega$ with Eq.~(\ref{eq:evolution2}) we see that $|\bm\Omega|$, the rotation speed of $\bm u$, is a constant vector. In fact, in this case $\bm\lambda'=0$ and thus $\bm\Omega=0$, $\dot{\bm u}=0$ as given in the previous section.
It is easy to see that the spin vector $\langle\bm\sigma\rangle$ ($\sigma_{j=\{x,y,z\}}$ are the Pauli matrices) is rotating around $\bm u$, and the evolution of wave function is analytically computed with rotation matrices.

\subsection{Analytical solution of a single qubit}\label{sec:1bit_ana}
The evolution path of $\psi$ can be analytically addressed, as a rotation with the angular velocity $\bm\omega_\mathrm{eff}$ around an axis $\bm u$ rotating itself at $\bm\Omega$.
The effective angular velocity $\bm\omega_\mathrm{eff}$ of spin vector is determined with
\begin{equation}
\bm\omega_\mathrm{eff}=2\bm B=\dfrac{2}{\sqrt{TrI}}\bm u=\sqrt{2}\bm u
\,.
\end{equation}
where $B$ is an effective magnetic field and the Hamiltonian $H=\bm\sigma\cdot\bm B=\sum\limits_j A_ju_j$. This is derived from the fact that the SU(2) generator of a rotation with $\Delta\phi$ is $e^{i\bm\sigma\frac{\Delta\phi}{2}}$. In the frame of reference rotating with $\bm\Omega$, the motion of $\psi$ is viewed as a fixed-axis rotation with the angular velocity $\bm\Omega'\!=\bm\omega_\mathrm{eff}\!-\bm\Omega$.

For instance, consider the numerically simulated case~\cite{Carlini2006} where $\psi(0)=|+x\rangle$, $\psi_{target}=|-x\rangle$, $\bm B(0)$ is $\omega$ in the $-y$ direction and $\bm\Omega$ lies in the $z$ direction with $Tr(H\sigma_z)=0$. Explicitly, $\langle\bm\sigma\rangle(0)=(1,0,0),
\bm B(0)=(0,-\omega,0),
\bm\Omega\equiv(0,0,\Omega)$.
Now the evolution turns out to be simply a rotation around $\bm\Omega'$ followed by a reversal rotation around $\bm\Omega$ back to the static frame. The rotation matrix $R(\phi,\theta,\Psi)=R_{\bm\Omega}^{-1}(\phi)R_{\bm\Omega'}(\theta)=R_z^{-1}(\phi)R_x^{-1}(\Psi)R_y(\theta)R_x(\Psi)$ (see Appendix~$\ref{sec:G}$ ). Here $(\phi,\theta,\Psi)$ is valued $(\Omega t,\Omega' t,\mathrm{tan^{-1}}\dfrac{\Omega}{\omega_\mathrm{eff}})$, where $\Omega'=\sqrt{\Omega^2+\omega_\mathrm{eff}^2}$. Then it is easy to check that
\begin{equation}
\langle\bm\sigma\rangle(t)=R(\phi,\theta,\Psi)\cdot \langle\bm\sigma\rangle(0)=
\begin{pmatrix}
&~\mathrm{cos}\!~\Omega t\!~\mathrm{cos}\!~\Omega't+\dfrac{\Omega}{\Omega'}~\mathrm{sin}\!~\Omega t\!~\mathrm{sin}\!~\Omega't\\
&-\mathrm{sin}\!~\Omega t\!~\mathrm{cos}\!~\Omega't+\dfrac{\Omega}{\Omega'}~\mathrm{cos}\!~\Omega t\!~\mathrm{sin}\!~\Omega't\\
&\dfrac{\omega_\mathrm{eff}}{\Omega'}\mathrm{sin}\!~\Omega't
\end{pmatrix}
\,,
\end{equation}
which gives a family of analytic solutions if we choose $(\Omega T, \Omega' T, \dfrac{\Omega}{\omega_{\mathbf{eff}}})=(k\pi, l\pi, \dfrac{k}{\sqrt{l^2-k^2}})$, where $l, k$ are non-negative integers and $l+k$ is odd. The case $k=0$ corresponds to the geodesic solution with no anisotropic contraints and cases of $k\not=0$ correspond to the oscillating solutions with $l-1$ nodes, and setting L= $l-1$ for later convenience. The oscillating curve in FIG.~\ref{Fig2} plots the solution with $k=1,L=1$.

\subsection{Analytical solution for multiple qubits}
First we give an example of two qubits with $H=B_x^{(1)}\sigma_x^{(1)}+B_y^{(1)}\sigma_y^{(1)}+B_x^{(2)}\sigma_x^{(2)}+B_y^{(2)}\sigma_y^{(2)}$.
 Similarly, our purpose is to flip both qubits (where $\psi(0)^{(1)}=|+x\rangle$ and $\psi_{target}^{(1)}=|-x\rangle$ for qubit one, $\psi(0)^{(2)}=|+y\rangle$ and $\psi_{target}^{(2)}=|-y\rangle$ for qubit two). Obviously, these two qubits are both reversing their states, and because $x-$ and $y-$ directions are totally equivalent there qubits should share the same energy at the ground state(namely $\omega_{\mathbf{eff}}^{(1)}=\omega_{\mathbf{eff}}^{(2)}$). Fig.~\ref{Fig3} shows a local optimal solution of the constrained two-qubits system with $L=1$. $\bm\Omega^{(1)}=(0,0,-\dfrac{2}{\sqrt 3}\omega),~\bm\Omega^{(2)}=(0,-\dfrac{2}{\sqrt 3}\omega,0)$. It is appearent that the evolution of $\bm B$ turns out to be a rotation as expected.

Actually, the local solutions can be derived analytically as long as the entangled terms are missing in the Hamiltonian. According to the rotation picture, the analytical approach can be applied to non-entangled many-qubit systems with an arbitrarily large number of qubits. Here in Fig.~\ref{Fig3} the solution corresponds to $(k^{(1)},L^{(1)})=(k^{(2)},L^{(2)})=(1,1)$ as the oscillating solution in Fig.~\ref{Fig2}, so they naturally share the same geometric property.
Later on, these solutions would be good initial guesses in the case of weak entanglement (see \ref{sec:relaxation}).

\section{An Accelerated Approach to the QBE}
\label{sec:numerical}
In practice, there are usually a large number of uncontrollable terms in the Hamiltonian. Sometimes for the convenience of computation and design, almost all the interaction terms are made to vanish. At this moment the dimension of control parameters is appreciably low, however, the computation complexity of the conventional numerical methods remains as high. To address and eliminate the redundant complexity, here we propose an alternative approach, reintroducing the Lagrange multiplier $\phi$ previously replaced by $\bm D$.

In general situations we see that the tremendous complexity is mainly brought by the factor $\bm C\cdot \bm D$, since the commutation relation~(\ref{b6}) will incur a large number of or infinitely many more $A_k$'s and thus $D_k$'s in the QBE equations when there are multiple interacting qubits. It is frustrating that most of the previous researches~\cite{Carlini2006,Xiaoting2015} are confronted with similar problems but no special attention has been paid to this problem. Here, we introduce an effective deduction scheme based on the fact that $\bm C\cdot\bm D$ is immediately derivable from $\phi$. Technically, consider
\begin{equation}\label{phi2}
  \phi'=\phi-\dfrac{1}{2\Delta E^2}(H-\langle H\rangle)\psi,
\end{equation}
 then Eq.~(\ref{Eq6}) turns to
\begin{equation}\label{eq:D2}
  D'_j=\langle\phi'|A_j|\psi\rangle+c.c.=
    -\lambda \xi_j-\lambda'_j, ~~~~(\text{with}~\lambda'_j=0~\text{for}~j\le n).
\end{equation}
Thus ${\bm C\cdot D'_j}={\bm C\cdot\langle\phi'|A_j|\psi\rangle}+c.c.$.
Furthermore, it is much easier to evolve $\phi'$ instead of $\phi$ since Eq.~(\ref{Eq4}) yields
\begin{equation}\label{eq:dphi}
  i\dot{\phi}-H\phi=\frac{i}{2\Delta E^2}[(\dot{H}-\langle\dot{H}\rangle)\psi-\dfrac{\tilde{\bm F}\cdot\dot{\bm\xi}}{\Delta E^2}(H-\langle H\rangle)\psi]\\ \,,
\end{equation}
which gives
\begin{equation}\label{eq:dphi'}
  \begin{aligned}
    i\dot{\phi}&'=H\phi'\,.\\
  \end{aligned}
\end{equation}
In analogy to previous derivations in Sec.~\ref{sec:rotation}, it is easy to check that Eq.~(\ref{eq:evolution2}) now holds in an alternative form
\begin{equation}\label{eq:evo2}
  \begin{aligned}
    \dot{\bm u}&=(I-P_{u})\cdot\dfrac{\bm C\cdot\bm\lambda'}{\lambda},\\
    i\dot{\phi}&'=H\phi',\\
    \dot{\lambda}&=-\dfrac{ \bm{v\cdot C \cdot \lambda}' }{2\omega^2}, \\
    \dot{\psi}&=-iH\psi \,,\\
  \end{aligned}
\end{equation}
where from Eq.~(\ref{eq:D2}), $\bm C\cdot\bm\lambda'= -\bm C\cdot\bm D'$ with $\sum\limits_{j}C_{ij}D'_j={\langle\phi'|[H,A_i]|\psi\rangle}+c.c.$, which could be evaluated directly with the computed states $\phi^{\prime}$ and $\psi$ and the known $H$ and $A_i$'s.
In conclusion, the whole QBE problem is fully described by Eq.~(\ref{eq:evo2}). Terms like $\bm C\cdot\bm F\cdot\bm\xi/2\Delta E^2$ in $\bm C\cdot\bm D$ are eliminated now because of the reintroduction of the state variable $\phi'$, and hence the previous large number of equations needed to evaluate $\bm C\cdot\bm D$ are simply replaced by the evolution of $\phi'$. Since the number of degrees of freedom (DOF) of vector $\phi’$ is much smaller than that of matrix $\bm C$, the computation will be significantly accelerated.

Based on Eq.~(\ref{Eq6}),~(\ref{eq:D2}) and the fact that $Tr H=0$ (i.e.~$\xi_0=0$), we immediately obtain
\begin{equation}\label{Eq17}
\begin{aligned}
2Re(\langle\phi|\psi\rangle) &=D_0=\sum_j\dfrac{F_{0j}}{2\Delta E^2}\xi_j-\lambda\xi_0\equiv 0\,,\\
 Re(\langle\phi'|\psi\rangle) &=D'_0=-\lambda\xi_0\equiv0\,,
\end{aligned}
\end{equation}
where $F_{0j}=0$ and $A_0=\sigma_0$, which is the unit matrix.
Hence the component of $\phi$ or $\phi'$, being parallel to $\psi$, vanishes. Notice that $\bm u$, in the absence of uncontrollable components, is parallel to $Re(\langle\phi'|\bm\sigma|\psi\rangle)$ according to Eq.~(\ref{eq:D2}). So $\bm u\cdot\langle\bm\sigma\rangle \propto Re(\langle\phi'|\bm\sigma|\psi\rangle)\cdot\langle\bm\sigma\rangle=0$ (proved in Appendix~\ref{sec:F}) for every qubit.

Now a two-point boundary value problem with Eq.(\ref{eq:evo2}) is left for us to solve. To begin with, we count the number of independent variables. The number of different variables in Eq.~(\ref{eq:evo2}) is $N=N_u+N_{\phi'}+N_\lambda+N_\psi+N_{\lambda'}$, while the number of the constraints is $N'=N_u+N_{\lambda'} + 2N_{\psi}+1$ given by Eq.(\ref{eq:D2}) (giving $N_u+N_{\lambda'}$), the energy constraint $\sum\limits_{j=1}^n\xi_j^2=\omega'^2$, the prescrible initial and final states $\psi(0)=\psi_0$ and $\psi(T)=\psi_{target}$ (giving $2N_{\psi}$), then the number of independent variables is $N-N'=0$ (where $N_\lambda=1$ and $N_{\phi'}=N_{\psi}$).

In the current paper the \emph{shooting method} (\ref{sec:shoot}) and the idea of \emph{relaxation} play a major role in finding the optimal path when Hamiltonian $H$ contains interaction terms. We give some details below.

\subsection{Shooting Method}
\label{sec:shoot}
The shooting method aims to implement the multidimensional Newton-Raphson method to locate zeros of $N_{target}$ functions
with $N_{target}$ varibales, which are obtained by integrating $N$ differential equations in our examples.

For a general dynamical system, the ordinary differential equation (ODEs) is written as
\begin{equation}\label{v1}
\dot{\bm x}=\bm v(\bm x(t)),
\end{equation}
where $\bm x\in \mathbb R^d$, $t \in \mathbb R$, and the state of the system obtained by integrating Eq.(\ref{v1}) at time $t$ is $\bm x(t)= f^t(\bm x)$. The corresponding Jacobian matrix is $J(\bm x,t)=\frac{\partial\bm x(t)}{\partial\bm x(0)}$, which could be obtained by integrating
\begin{equation}\label{J}
    \frac{dJ}{dt}=AJ,~~~A_{ij}=\frac{\partial{\bm{v}_i}}{\partial{\bm{x}_j}}, ~~~\text{with }J(\bm x,0)=\bm 1.
\end{equation}
Thus, the initial displacement $\delta \bm x(0)$ becomes $\delta \bm x(T)$ at time $T$, namely
\begin{equation}
\delta \bm x(T)=J(\bm x,T) \cdot \delta \bm x(0)
\,.\end{equation}
Define a discrepancy vector $\bm X=\bm x(T)-\bm x_{target}$. To procure a vanishing discrepancy, the correction $\Delta x_0$ should satisfy
\begin{equation*}
J\cdot \Delta\bm x_0=-\bm X,
\end{equation*}
to reach a better approximation
\begin{equation}\label{dx}
\bm x_{0,new}=\bm x_{0,old}+\Delta \bm x_{0}
\,.\end{equation}
Typically, the Jacobian matrix $J$ is numerically evaluated by
  \begin{equation}
    J_{ij}=\dfrac{\bm X_i(\bm x_{0,1},\ldots,\bm x_{0,j}+\Delta \bm x_{0,j},\ldots,T)-\bm X_i(\bm x_{0,1},\ldots,\bm x_{0,j},\ldots,T)}{\Delta\bm x_{0,j}}\,.
  \end{equation}
In the current case, the adjustable part is the free parameter $\phi$ and the discrepancy is $\psi (T)-\psi_{target}$.

Let's slow down here and have a discussion on possible additional constraints and the treatment of the symmetries mentioned earlier. Firstly, based on Eq.(\ref{Eq17}), $Re(\langle\delta\phi'|\psi\rangle)=0$,
which with the help of the symmetry transformation $\hat{U}_1$ could always be satisfied by taking the difference $\phi^{\prime}_{0,new}-ia\psi$, where $ia\psi=|\psi\rangle\langle\psi|\phi'_{0,new}\rangle$. The unitary evolution of $\psi$ will produce a singularity for the Jacobian matrix $J$. To cope with this problem, our strategy is to totally ignore this eigen-direction since it corresponds to irrelevant symmetry directions.  The constraint $\sum_{j=1}^{n}\xi_j^2=\omega'^2$ could be garanteed by the symmetry $\hat{U}_2$ since $\lambda_0$ and thus $b$ could be obtained by substituting $\phi^{\prime}_{0,new}$ into Eq.~(\ref{eq:D2}) with $j \le n$ when starting a new iteration.

It is likely that the discrepancy may significantly fluctuate in the first few iterations. So it is important to properly choose the target hyperplane for the shooting on which the target state lies.
The final state $\psi(T)$  is adjusted to 'hit' the target state $\psi_{target}$ by adjusting the initial position Eq.(\ref{dx}) and the flight time $T$.
\subsection{Multiple Qubits: `Relaxation'}
\label{sec:relaxation}
When it comes to large-scale qubit systems, the shooting method per se can be too costly and need a good trial solution that is sufficiently close to the exact solution. In order to overcome those shortcomings we suggest a `relaxation' method to meet the boundary conditions step by step. First one stars from the case without interaction, solving the problem with separate qubits.

With interaction present, it is possible to solve the problem progressively. First, we invite part of the interactions back and use the non-interacting solution as the initial condition ({\em i.e.},taking the first-step coupling constant $J=0.01$ in our examples), so a new solution in case of the weak interaction is obtained, which can be used as the initial condition for the case with a little bit stronger interaction. Repeat this process until the full interaction is restored and an optimal path is then obtained. This is a relaxation process since the effective Hamiltonian in the computation is gradually relaxed to the desired one.
\begin{figure}[b]
  \includegraphics[width=7cm]{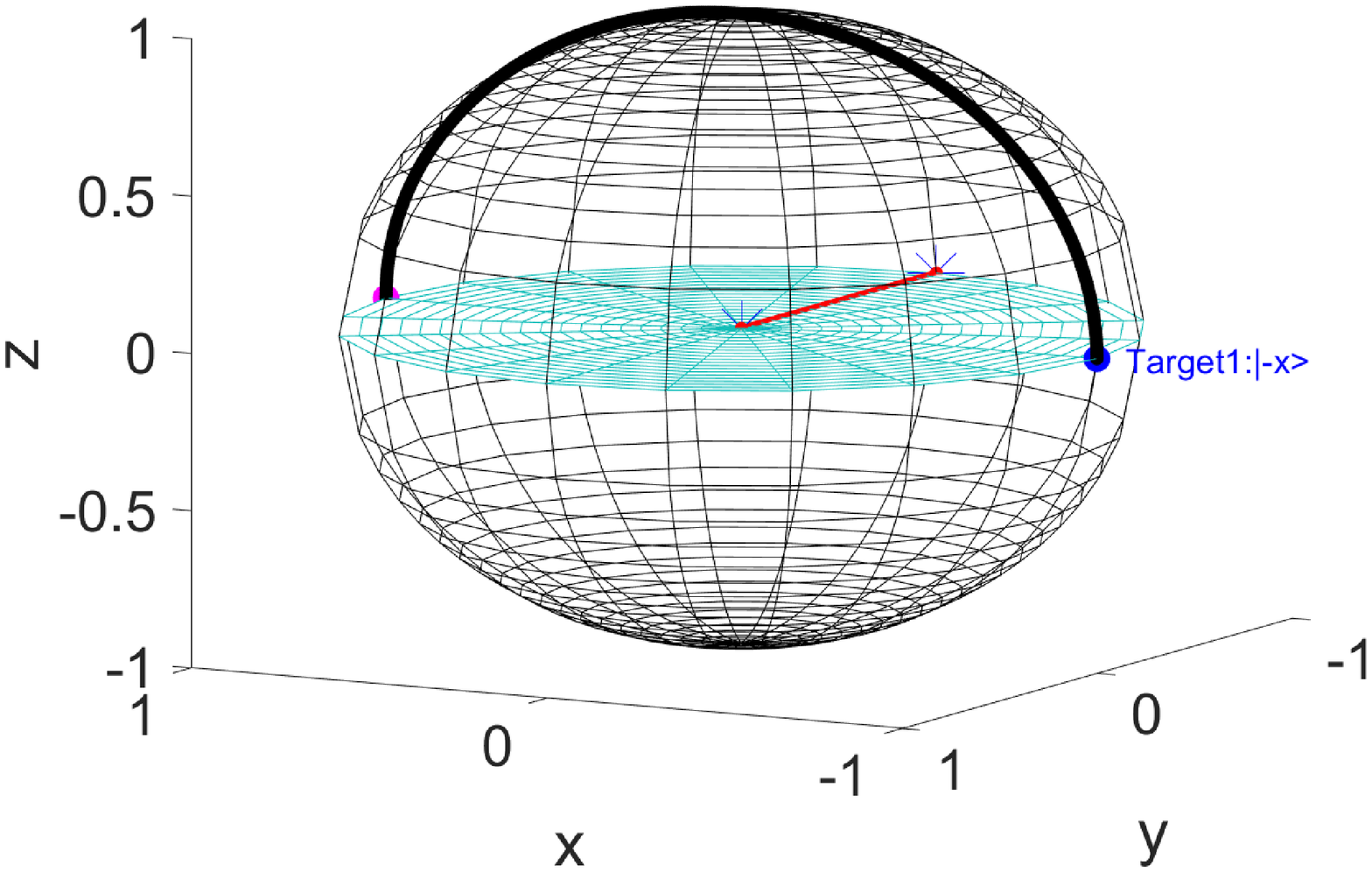}
    \caption{\label{Fig1} Evolution of Qubit 1 on a Bloch sphere. The black line represents the state $\psi(t)$ that evolves from the initial state (purple point on the left) to the final state (blue points on the right) on the Bloch sphere. The green grid inside the sphere marks the equatorial plane, and the red part is the area swept by the magnetic field vector. Similar representations are used in later diagrams.}
  \includegraphics[width=7cm]{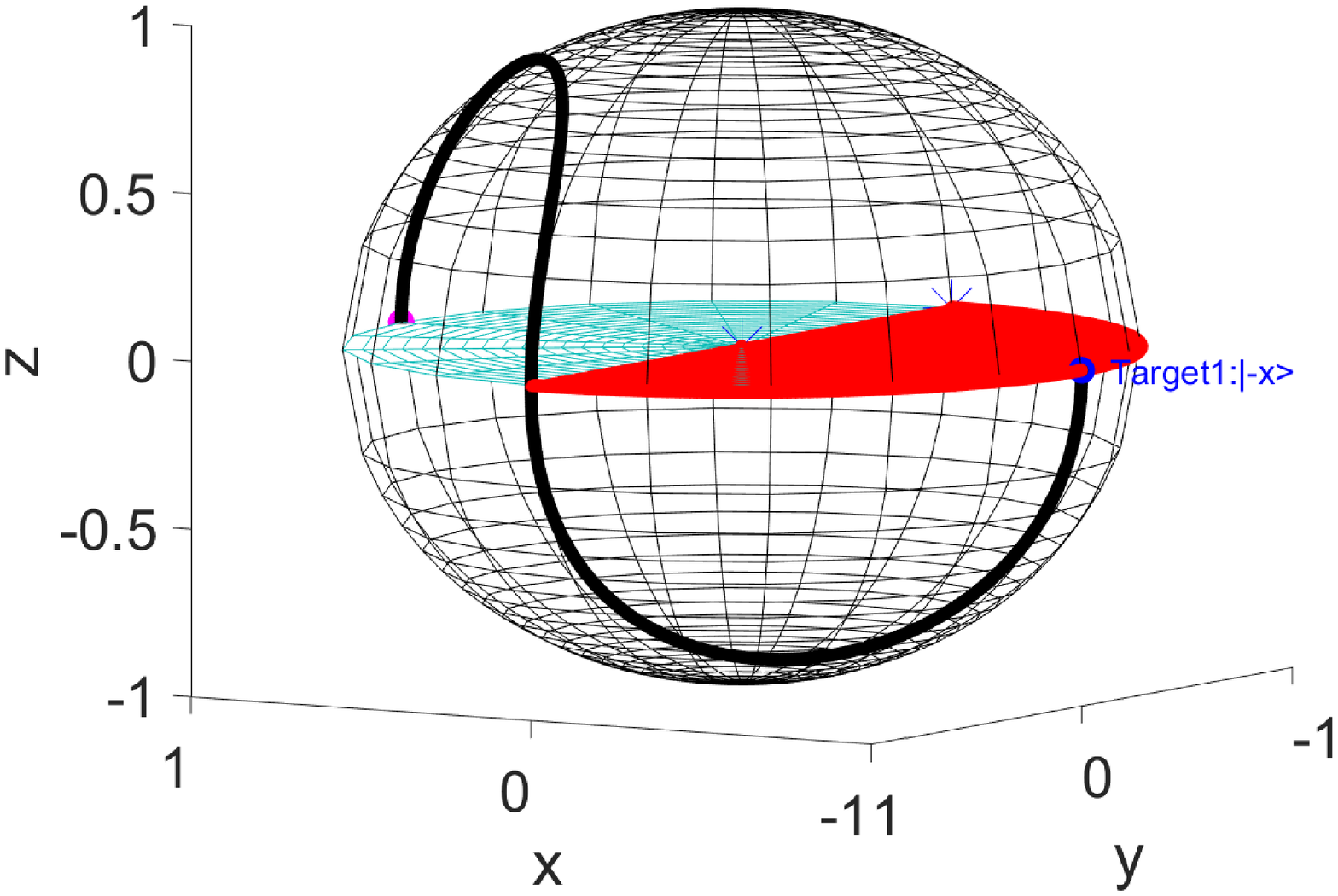}
  \includegraphics[width=7cm]{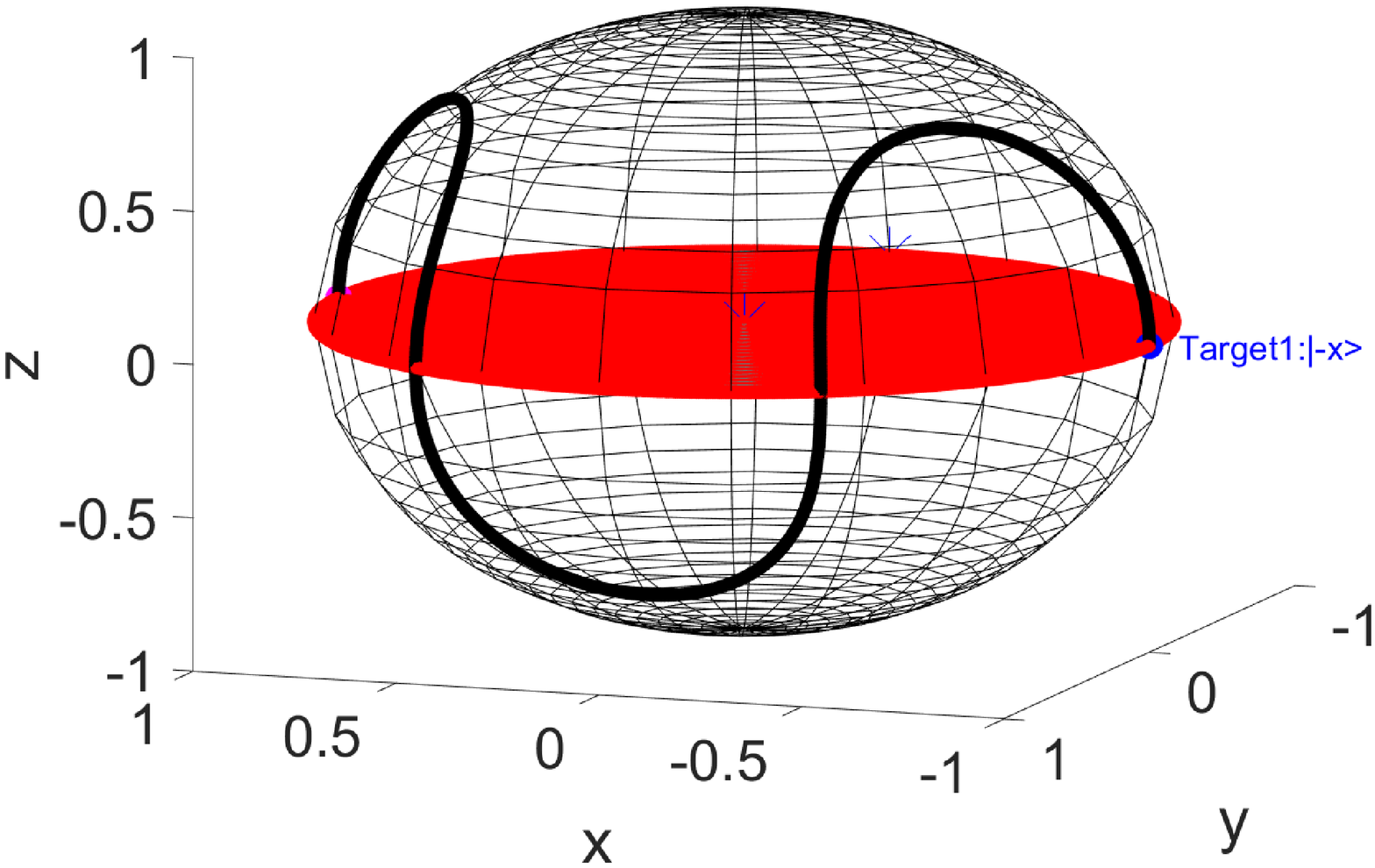}
    \caption{\label{Fig2} The evolution of $\psi(t)$ of a single qubit with $\omega_{eff}=2|\bm B|=2 $ and L=1 (left), 2 (right) (namely, the black path intersects the equatorial plane L times apart from the initial and final states).}
\end{figure}

\section{examples}
\label{sec:example}
\subsection{One-Qubit Bloch Sphere}
\label{sec:1bit}
So far most of the practical quantum control systems are based on spin magnetic moment. Here we first try to search for local optimal paths of single-qubit evolution on the Bloch sphere. The goal is to flip the spin state of a single Fermion from $|+x\rangle$ to $|-x\rangle$.

Assuming $\hbar=1$, the Hamiltonian in the isotropic condition is
\begin{equation*}
H=-\bm\mu_S\cdot\bm B=-\dfrac{\gamma}{2}\bm\sigma\cdot\bm B=-\dfrac{\gamma}{2}(B_x\sigma_x+B_y\sigma_y+B_z\sigma_z)
\,,\end{equation*}
where $\gamma=g\dfrac{q}{2m}$ is the gyromagnetic ratio. For electrons $\gamma\approx -2$ then simply $H=\bm B\cdot\bm\sigma$.
From previous discussions, we know that $\dot{\bm u}=0$, i.e. $\dot{\bm B}=0$. So the optimal path is naturally a geodesic curve (see Fig.~\ref{Fig1}, which shows one of the many geodesic solutions). The simulated paths are displaced as well in Fig.~\ref{Fig1}. The red line on the equatorial plane (green grid inside the sphere) indicates the evolution of the external magnetic field $\bm B$, which does not change throughout the evolution in the case.

In principle, the constraints $ Tr(H\sigma_j)_{j=x,y,z}=0$ allow oscillating solutions with $\dot {\bm B}\not=0$.
Let's take the case of $H=B_x\sigma_x+B_y\sigma_y, ~B_z\equiv 0$ as an example.
In numerical computation, we use the shooting method to identify the initial value $\phi'_0$ for the prescribed boundary condition $\psi(0)=|+x\rangle,~\psi(T)=|-x\rangle$.
Fig.~\ref{Fig2} shows the resulting optimal path which oscillates with one node on the equator, where $\bm\Omega=(0,0,-\dfrac{2}{\sqrt{3}})$. The evolution of $\bm B$ (the area swept by the red line in Fig.~\ref{Fig2}) corroborates the theory of rotation in Sec.~\ref{sec:rotation}. Of course, it is not hard to get a longer locally optimal path with more than one node, which we will not show here.

\subsection{Two-Qubit Bloch Sphere}\label{sec:2bitw}
  \begin{figure*}[]
    \includegraphics[width=16cm]{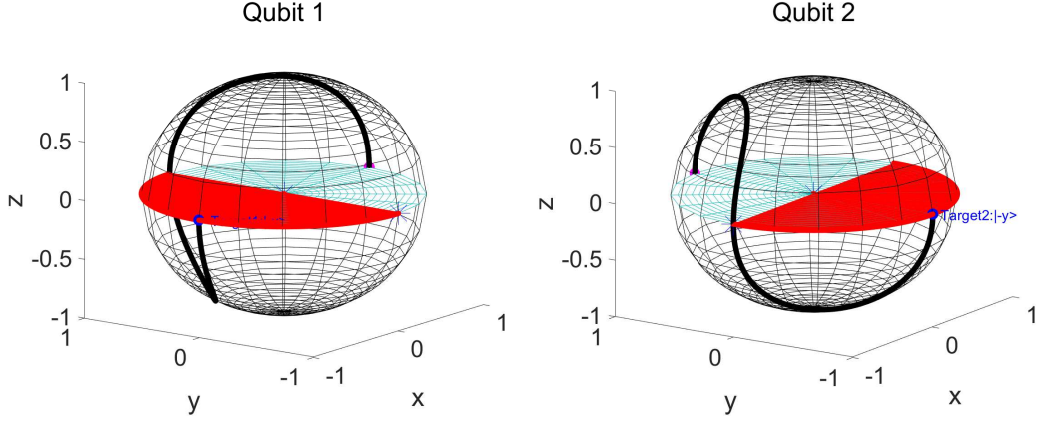}
      \caption{\label{Fig3} An optimal path with $L=1$, $\omega^{(1)}_{eff}=2$, $\omega^{(2)}_{eff}=2$ connecting the initial state $|+x\rangle^{(1)},|+y\rangle^{(2)}$ to the final state $|-x\rangle^{(1)},|-y\rangle^{(2)}$ of Qubits 1 and 2 plotted on two bloch spheres.}
  \end{figure*}
In the case of two qubits without interaction, two bloch spheres may be used to represent a non-entangled state and the geodesic solution is simple according to the discussion in Secion~\ref{sec:1bit} above. Here, we focus on the oscillating solutions under the anisotropic constraint featuring $ B_z=0$. We give an example with
\begin{equation}
  H=B_x^{(1)}\sigma_x^{(1)}+B_y^{(1)}\sigma_y^{(1)}+B_x^{(2)}\sigma_x^{(2)}+B_y^{(2)}\sigma_y^{(2)}.
\end{equation}
Fig.~\ref{Fig3} depicts the local optimal path obtained by our method, which oscillates with $L=1$. Also, we can obtain longer paths, not shown here.

\begin{figure*}[]
    \includegraphics[width=4.3cm]{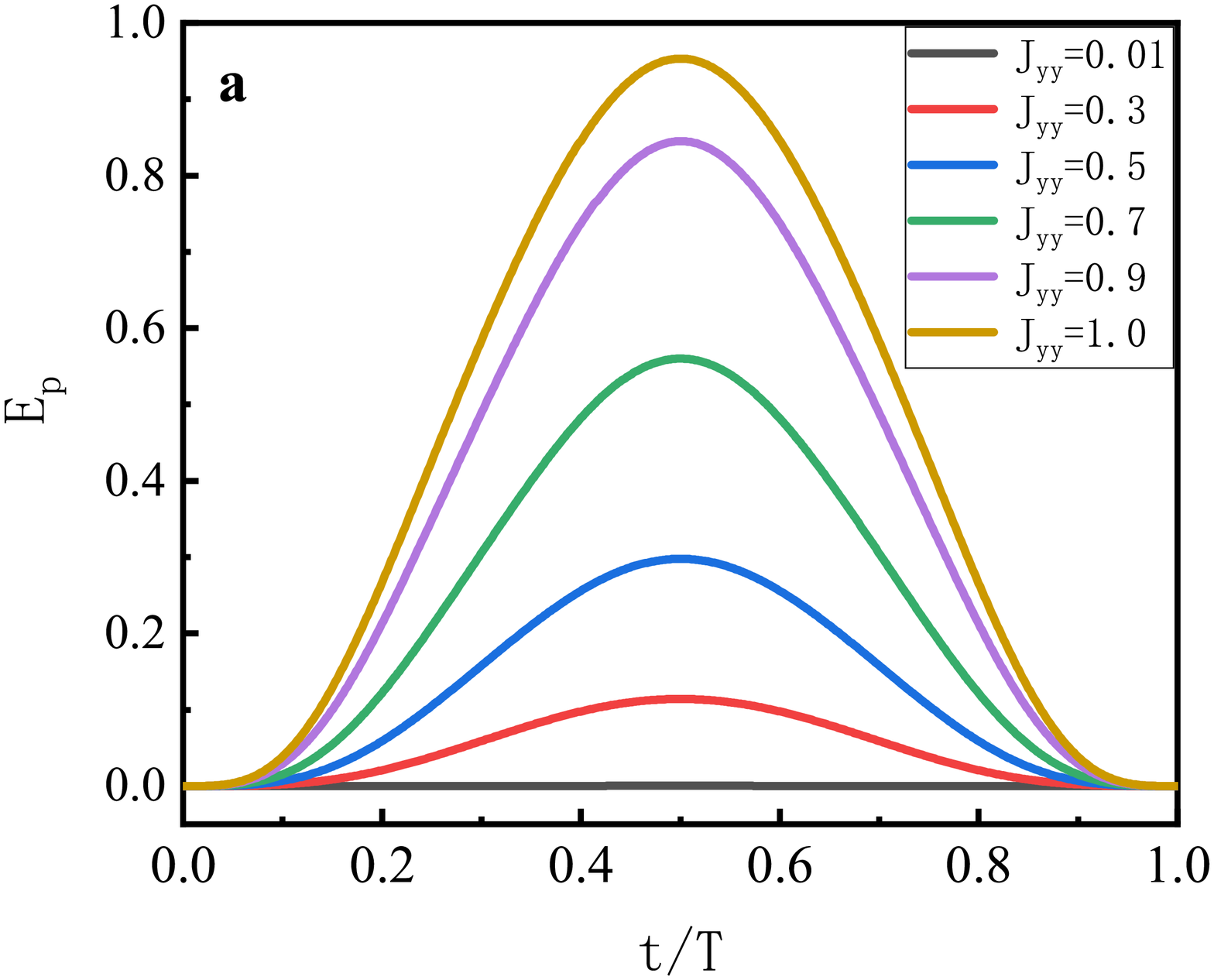}
    \includegraphics[width=4.3cm]{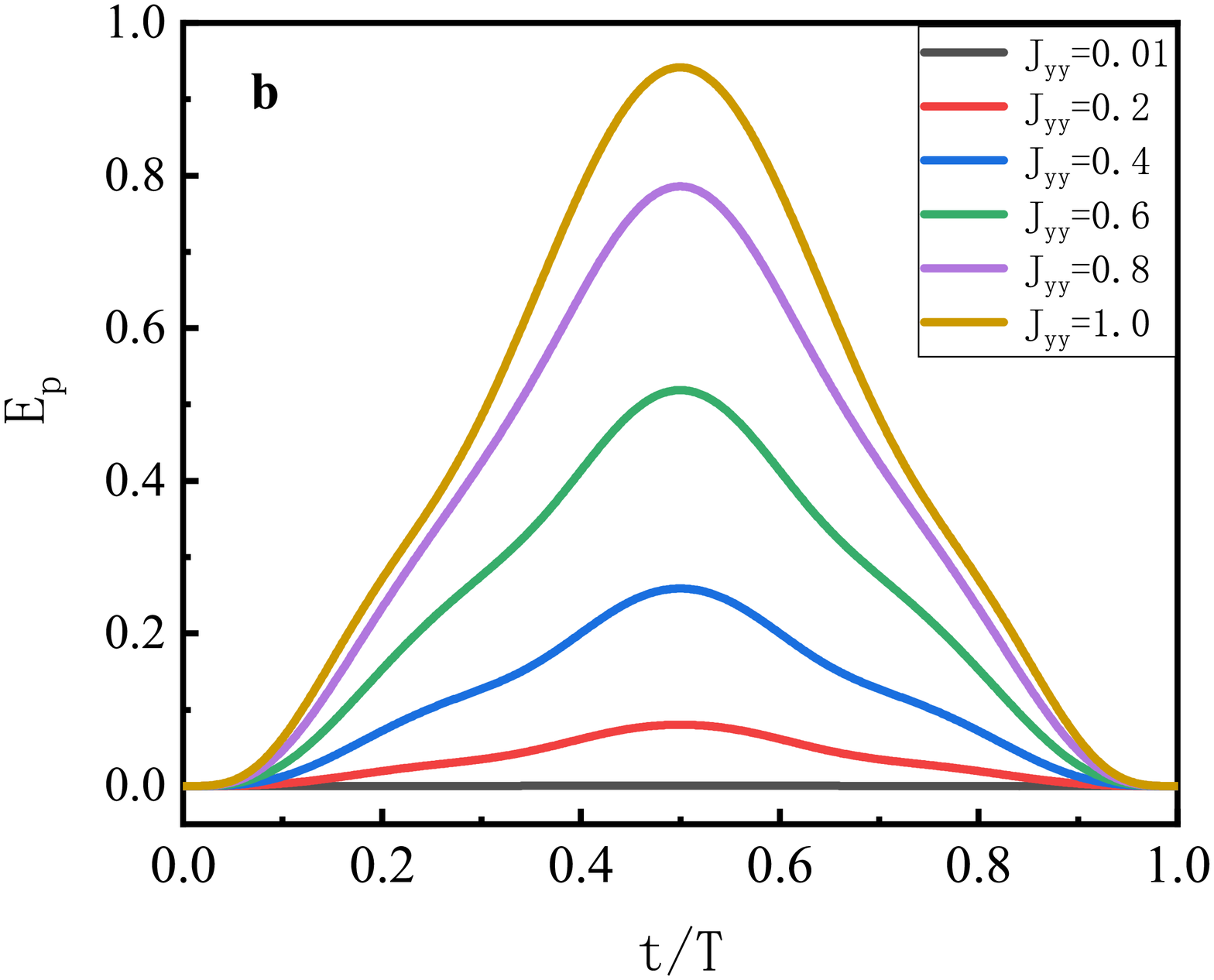}
    \includegraphics[width=4.3cm]{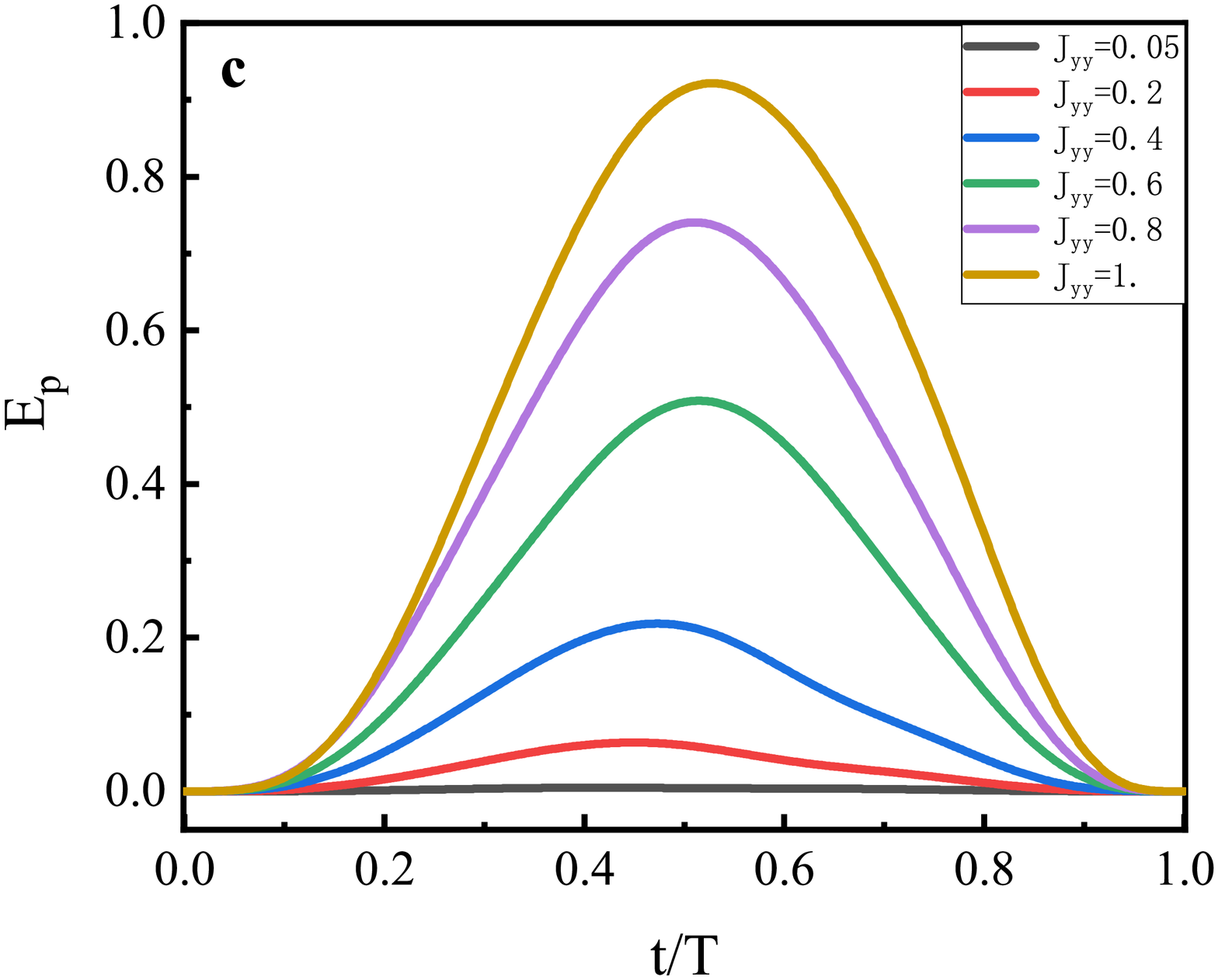}
    \includegraphics[width=4.3cm]{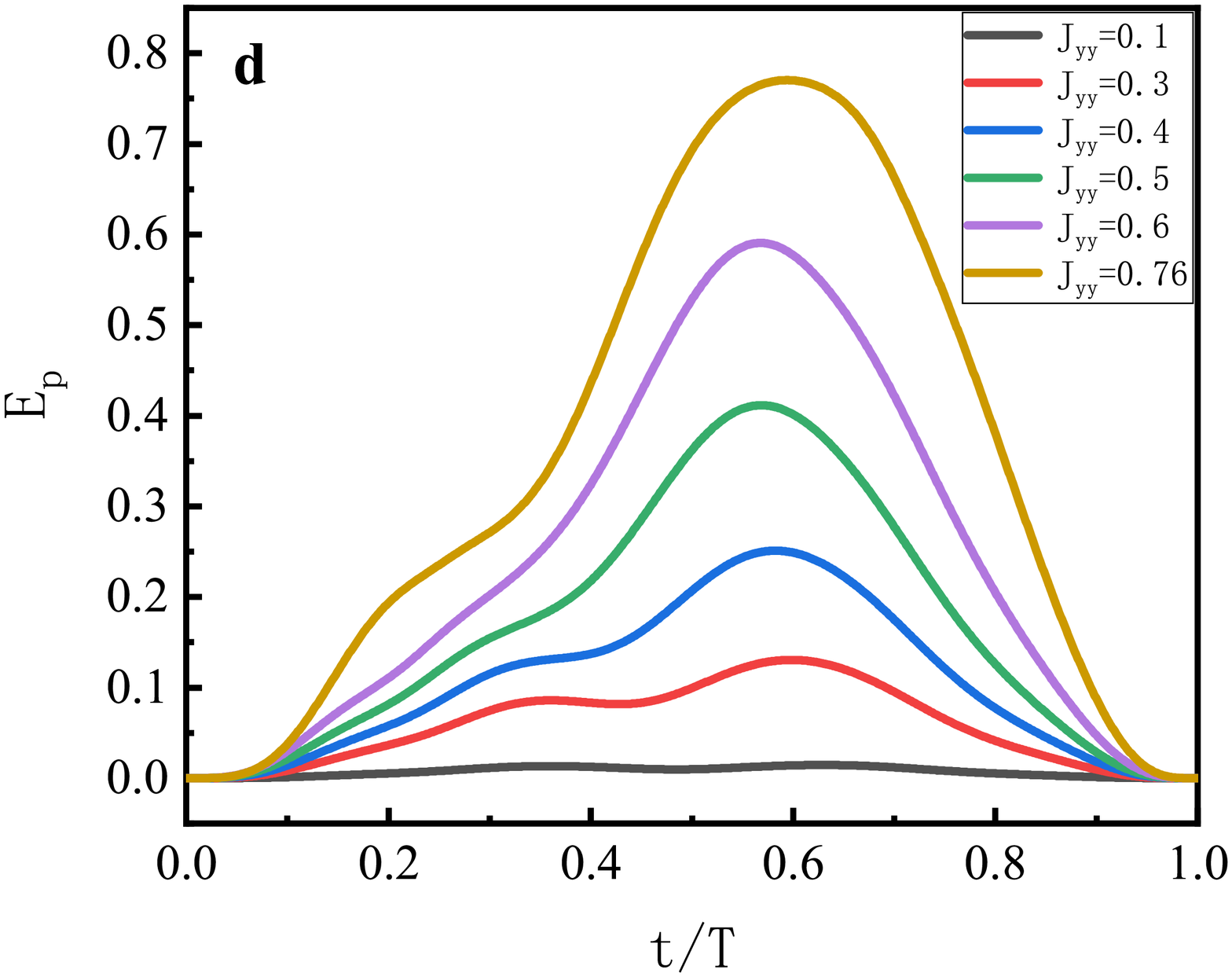}
     \caption{\label{Fig4}  The time evolution of the entanglement Ep of $\psi(t)$ for different $J_{yy}$ and L=1(a), 3(b), 2(c), 4(d). t/T represents the normalized time where T is the total time. The initial state are $|+x\rangle^{(1)},|+y\rangle^{(2)}$ and the final state are $|-x\rangle^{(1)},|-y\rangle^{(2)}$ for Qubits 1 and 2}
  \end{figure*}

  \begin{figure}[]
  \includegraphics[width=7cm]{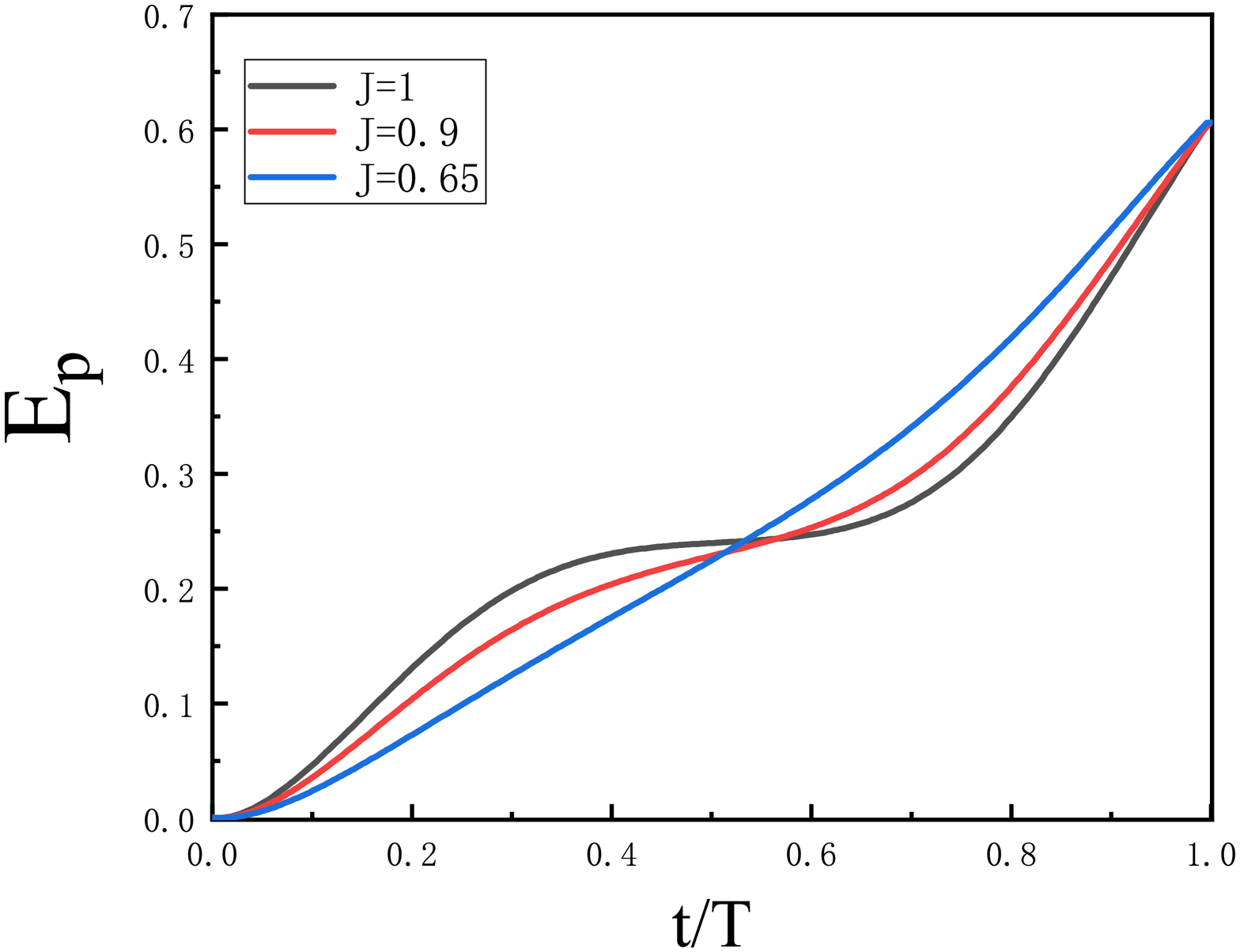}
  \caption{\label{Fig11} The evolution of the entanglement $E_p(t)$ of wavefunction $\psi(t)$ from the initial state $\psi(0)=(0.5, 0.5i, 0.5, 0.5i)^T $  to the final state(-0.216558-0.450669i, 0.528077+0.256587i, 0.070204+0.391403i, -0.294949+0.400223i) with J=0.65, 0.9, 1 and L=1.
  }
  \end{figure}
  \begin{figure*}[]
    \includegraphics[width=6cm]{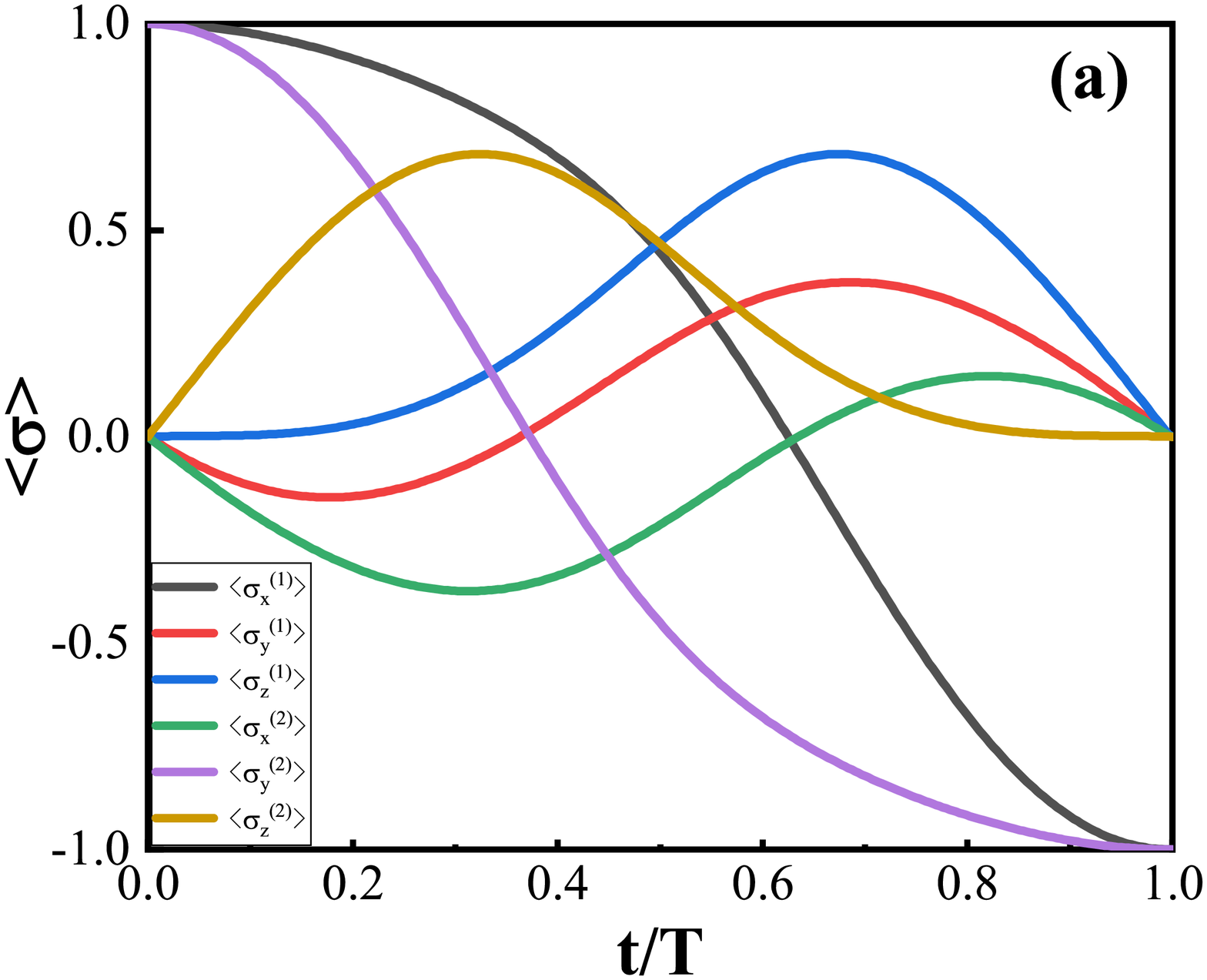}
    \includegraphics[width=6cm]{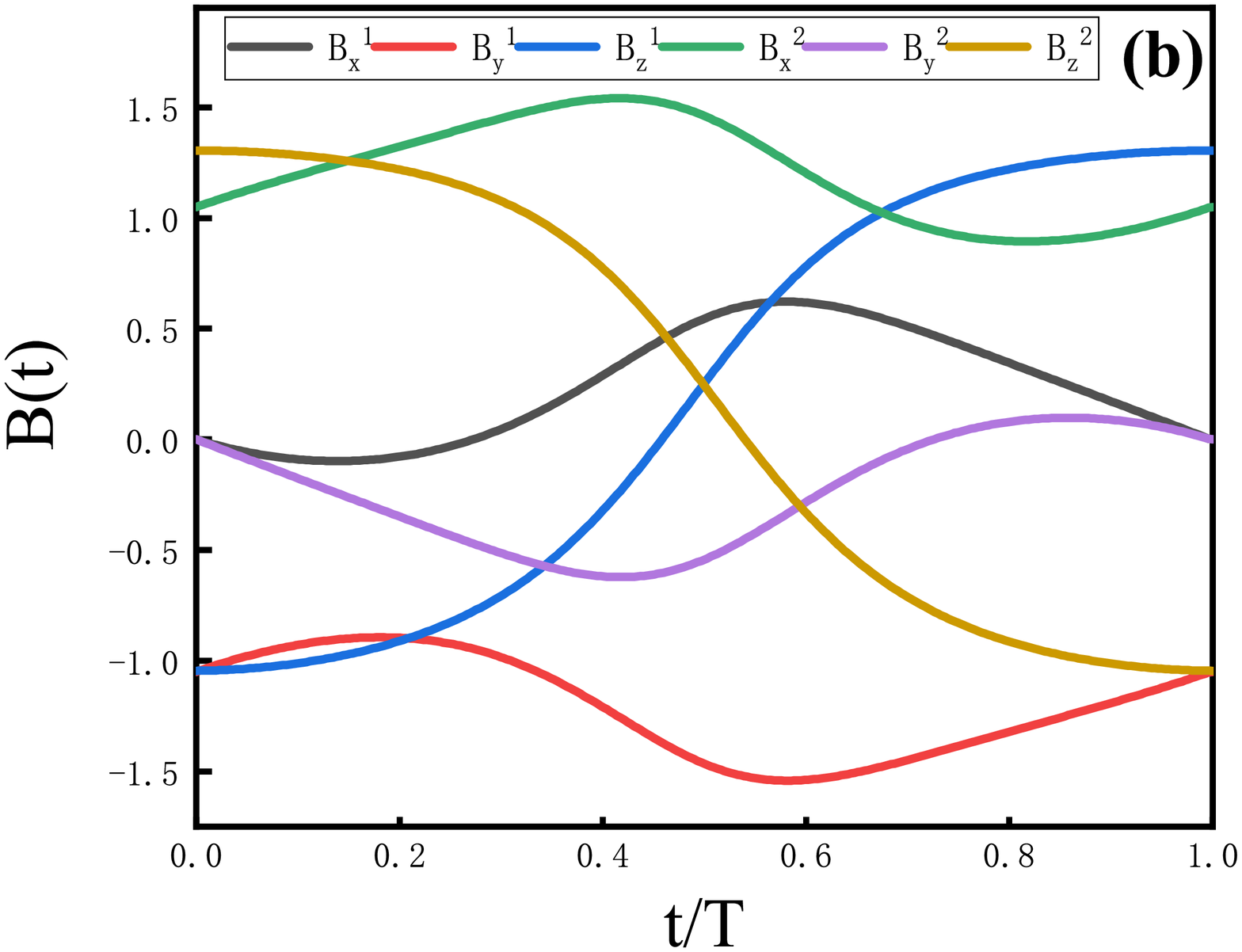}
    \includegraphics[width=6cm]{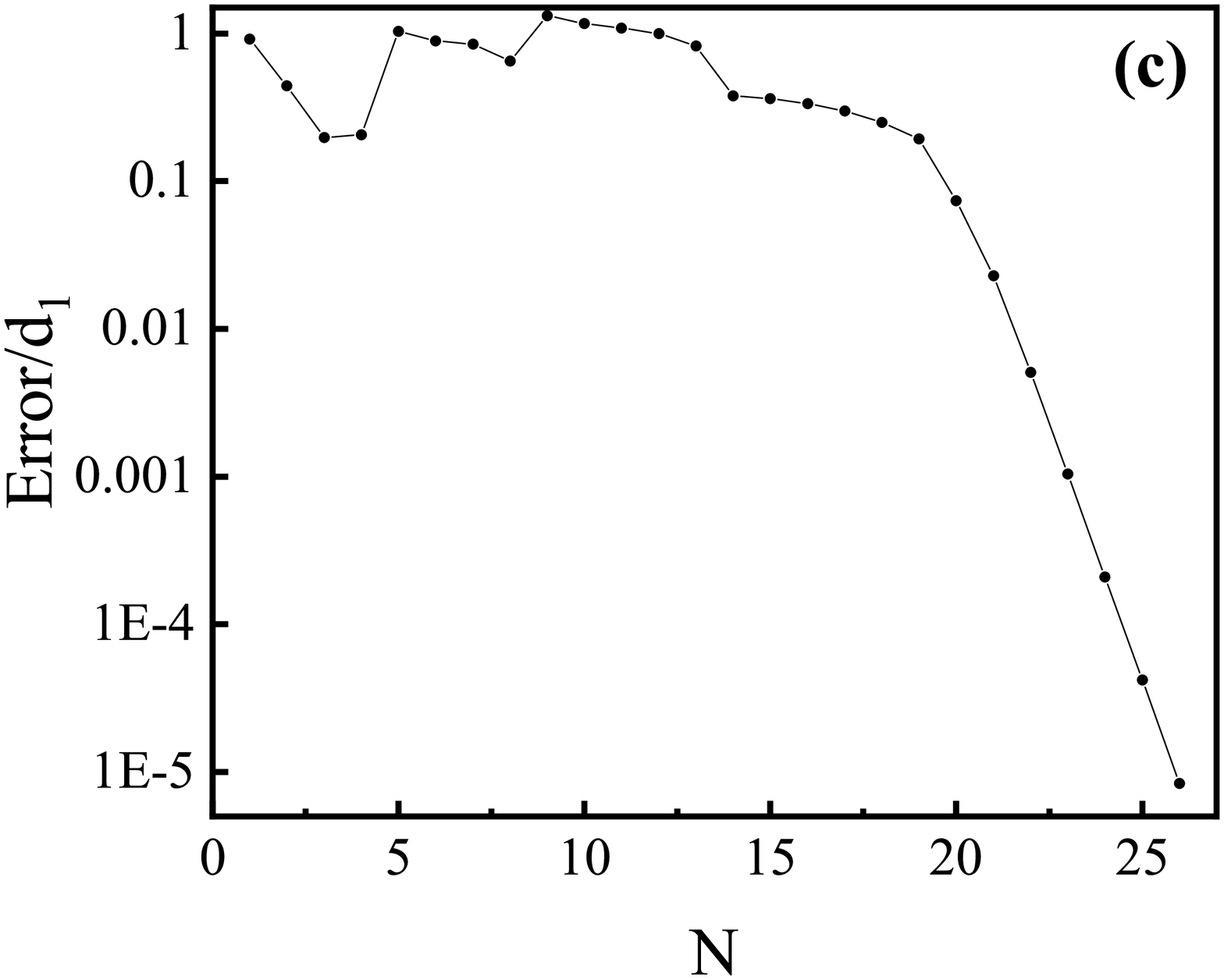}
    \includegraphics[width=6cm]{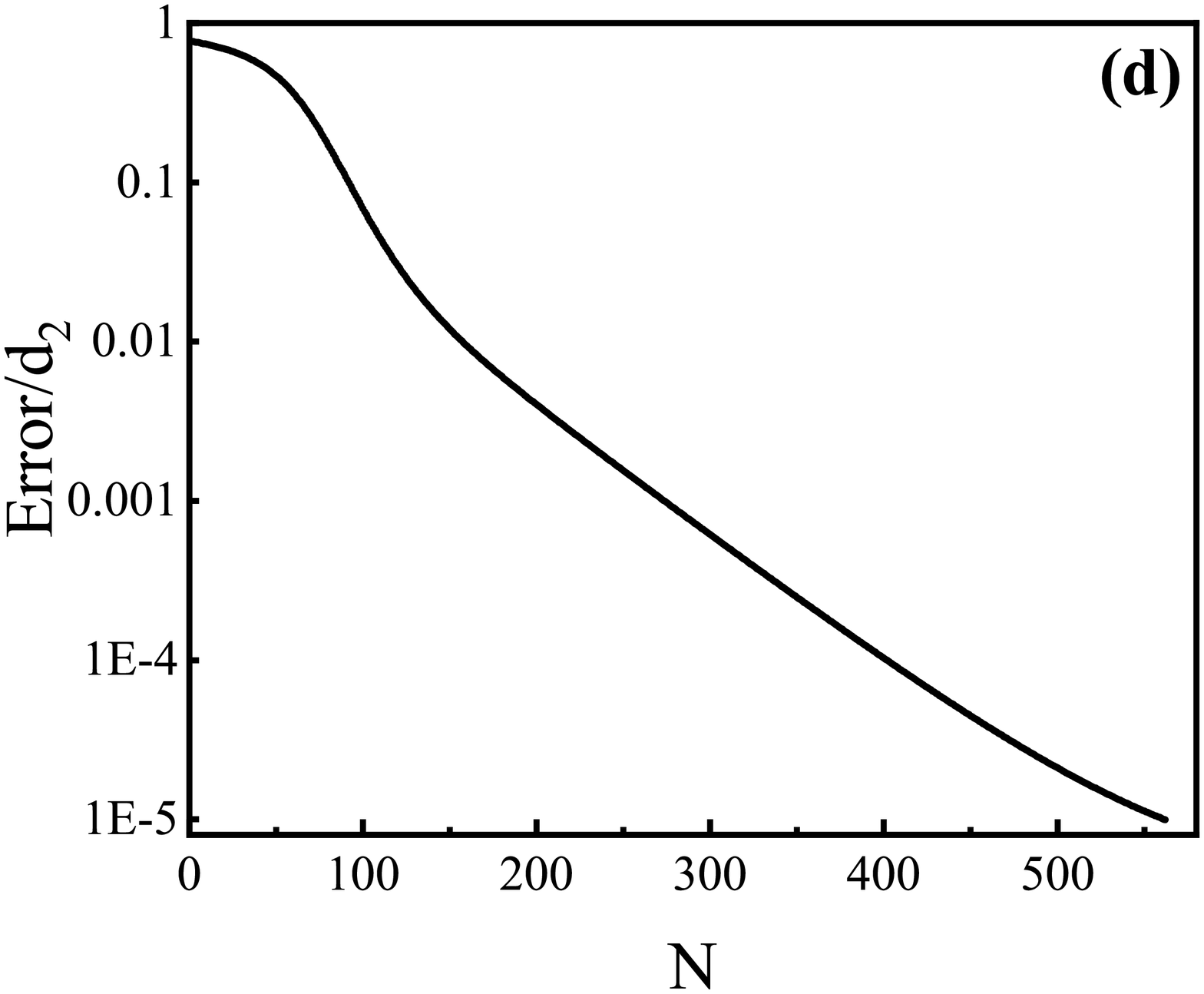}
     \caption{\label{Fig12}  The evolution and the convergence. The evolution of the observed measurement $\langle\pmb{\sigma}(t)\rangle$ (a) and the control magnetic field $\bm{B}$(t) (b), where $\langle\sigma(t)_m^{(l)}\rangle=\langle\psi(t)|\sigma(t)_m^{(l)}|\psi(t)\rangle~(m=x, y, z; l = 1, 2)$ represents the observed value of the spin in the m-direction on the $l~th$ bit. The convergence of our method (c) and the GRAPE method (d), where Error $d_1=|\psi(T)-\psi_{target}|$ and $d_2=|1-\langle\psi_{target}|\psi(T)\rangle|$, N is the number of iteration steps and J=1, L=1.}
  \end{figure*}
  \begin{figure}[]
  \includegraphics[width=6cm]{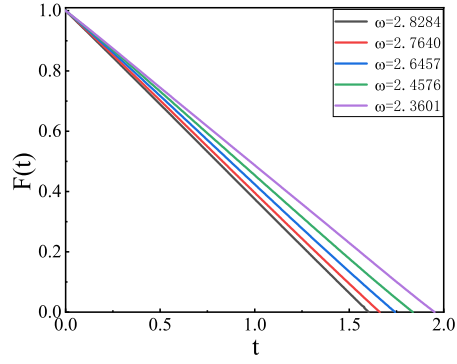}
  \caption{\label{Fig13} The Fubini-Study distance F(t) between $\psi(t)$ and $\psi_{target}$ with $\omega =2.8284,~2.7803,~2.6926,~2.5554,~2.3601$. The distance between neighboring states along a trajectory is defined by using the Fubini-Study line element $ds=\sqrt{\langle d\psi|(1-P)|d\psi\rangle}$, where $d\psi$ is the change of the state and $P=|\psi><\psi|$ is a projection operator
  }
  \end{figure}

  \begin{figure*}[]
    \includegraphics[width=6cm]{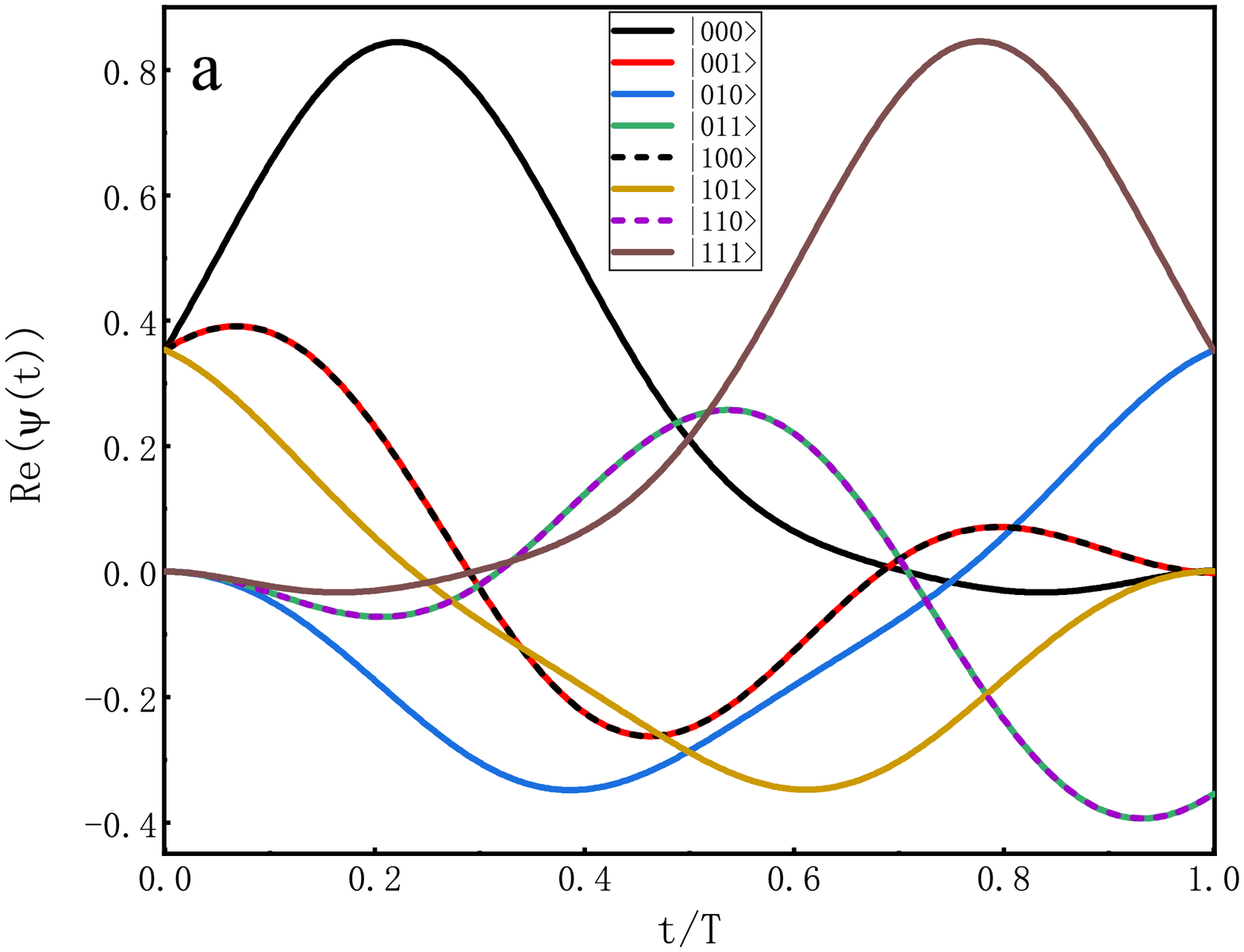}
    \includegraphics[width=6cm]{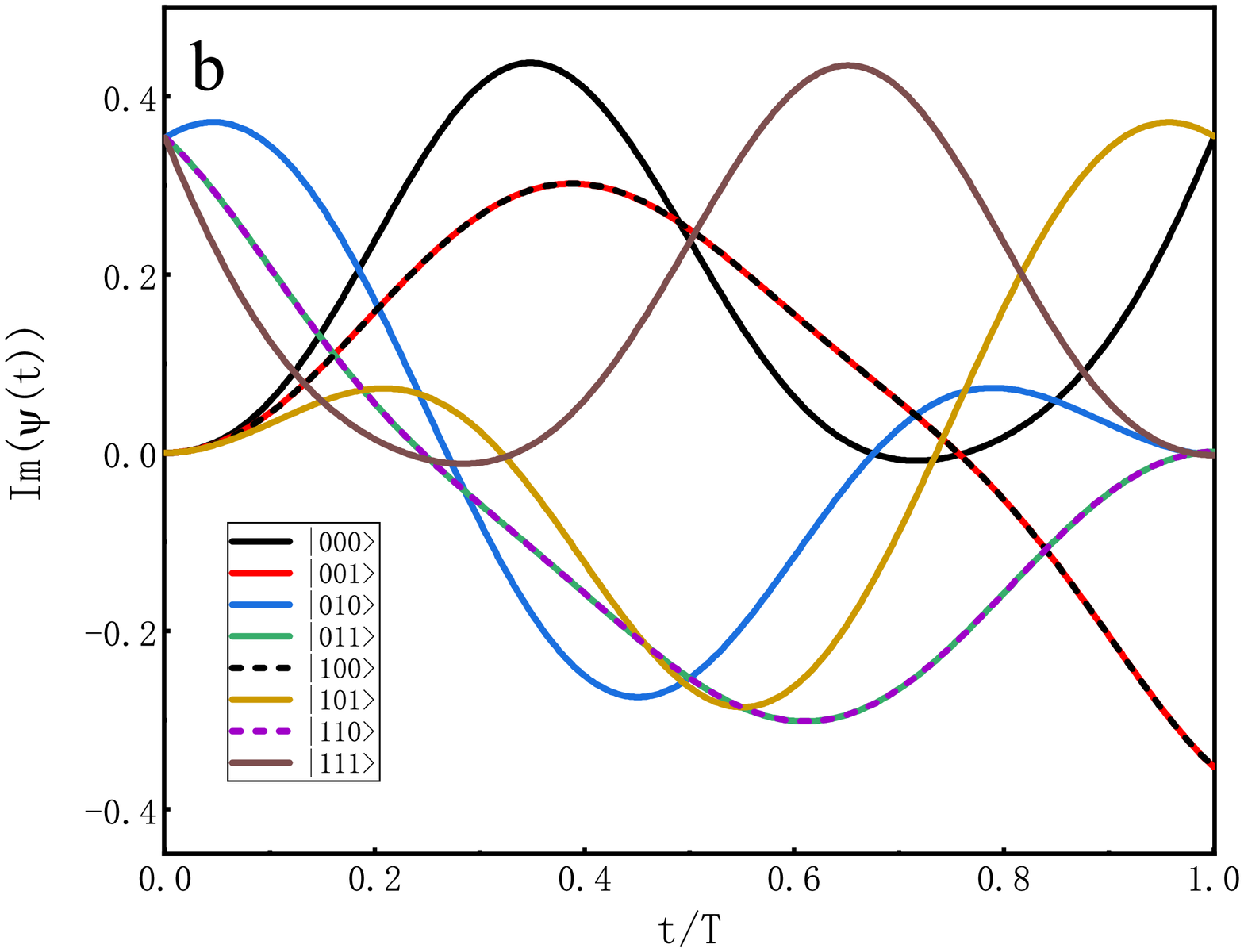}
     \caption{\label{Fig14}  The evolution of the wave function $\psi(t)$ with $J=0.2$ and $L=1$. $\text{Re}(\psi(t))$ and $\text{Im}(\psi(t))$ represent real part (a) and imaginary part (b) of wave function $\psi(t)$ respectively. The notation $|000\rangle \sim|111\rangle$ in the legend represents the eight components of $\psi(t)$.}
  \end{figure*}

\subsection{Two or Three Qubits with Interaction}
\label{sec:2bitw/o}

In the case of a single qubit or double qubits without interaction, it is easy to obtain the solutions analytically, which are very regular.
For two and three qubits with interaction, however, it is very difficult to obtain such an analytical solution, and thus we resort to the numerical method discussed in~\ref{sec:shoot}.In this section, we give three different cases are investigated together with  a general description of the consequences brought by possible entanglement in the wave functions.

Case 1, the initial and final spin states are the same as in section~\ref{sec:2bitw}, but there is an extra interaction term in the Hamiltonian
\begin{equation}\label{Hj}
 H=B^{(1)}_x\sigma^{(1)}_x + B ^{(1)}_y\sigma^{(1)}_y+B ^{(2)}_x\sigma ^{(2)}_x + B ^{(2)}_y\sigma ^{(2)}_y+J_{yy}\sigma^{(1)}_y\sigma^{(2)}_y
\end{equation}
where $J_{yy}=1 $ is a fixed interaction constant.
In the numerical computation, in order to utilize a reasonable initial condition, we gradually increase the value of $J_{yy} $ from 0.01 to 1 with a step-size 0.01.
After the interaction term $J_{yy} $ is introduced, the wave function $\psi(t)$ is no longer separable, but becomes entangled.
Following the work in the literature~\cite{Wootters1998,Charles1996}, we may define an entanglement index $E_p$ for the two subsystems A and B, $E_p(|\psi\rangle_{AB})\equiv -Tr_A(\rho_A log_2\rho_A)=-Tr_B(\rho_B log_2\rho_B)$, where $\rho_A=Tr_B(|\psi\rangle_{AB}\langle\psi|)$ is the reduced density matrix of the pure state $|\psi\rangle_{AB}$ over subsystem A, and $\rho_B$ has a similar definition.
Fig.~\ref{Fig4}(a-d) depict the entanglement of $\psi(t)$ with L=1, 3, 2, 4, which have a common feature: the entanglement of the wave function connecting the two separable states (the initial and target states) increases first, reaches a maximum and then decreases back to zero. The entanglement evolution profile for L=1, 3 (Fig.~\ref{Fig4} a and b) is symmetric and the maximum is exactly at the symmetry point, while for L =2, 4 (Fig.~\ref{Fig4} c and d), the profile is skewed. Given a suitable initial value of $\phi'_0 $, our numerical method converges exponentially when approaching an optimal path connecting the initial and target states (See Fig.~\ref{Fig12}(c)).
Next, we keep the initial state unchanged and set the target state as an entangled one with $E_p=0.6061$. The wavefunction  $\psi(t)$ connecting the two states is computed with J=1, 0.9, 0.65, the entanglement evolution profile of which is depicted in Fig.~\ref{Fig11}.

Case 2, we consider a two-qubit model, which is used in~\cite{Xiaoting2015}, with the following Hamiltonian
\begin{equation}\label{case3}
H=\sum\limits_{l,m}B_m^{(l)}\sigma_m^{(l)} +J\sum\limits_{m}\sigma_m^{(1)}\sigma_m^{(2)},
\end{equation}
where $\sigma_m^{(l)}~\text{with}~(m=x,y,z,l=1,2)$ are the Pauli matrices for the $l^{th}$ qubit. the initial and final spin states are the same as in section~\ref{sec:2bitw} and the coupling constant J=1. The results obtained by our method are portrayed in Fig.~\ref{Fig12}(a) and (b), where (a) depicts the components of the observable   and (b) shows that the control magnetic field B(t).

Next, we compare the efficiency of finding such a connecting wavefunction among different methods, such as the GRAPE and the “Geodesic-search”~\cite{Xiaoting2015}. On the same machine and programming language, our method finds such a connection wavefunction and the modulation magnetic field (shown in Fig.~\ref{Fig12}(b)) in less than one minute. The GRAPE needs about 3.3 minutes and the “Geodesic-search” needs about 18 minutes. Fig.~\ref{Fig12}(c) and (d) compare the convergence upon iteration between our method and the GRAPE for the same convergence criterion $10^{-5}$. Obviously, although the computation with our method oscillates at the initial stage, it converges much more quickly and displays good stability as well. However, the convergence of the GRAPE is quite slow (with more than 500 iterations), which explains its relatively long computation time. For more complex multi-qubit systems, we believe that these advantages will become more obvious.
In this case, the optimal time is computed as $T=\frac{\sqrt{N}\pi\hbar}{\omega}$ for a quantum state evolving between two end states of a diameter on the Bloch sphere, which is consistent with the lower bound computed by Margolus and Levitin~\cite{Margolus1998,Levitin2009}. The Fubini-Study distance to the target state seems to decrease linearly with time as shown in Fig.~\ref{Fig13}, and we note that with the decrease of the energy $\omega$, the traveling time T increases.

Case 3, we study a model with three qubits for which the Hamiltonian is
\begin{equation}\label{H3}
H=\sum\limits_{l,m}B_m^{(l)}\sigma_m^{(l)}+ J(\sigma_y^{(1)}\sigma_y^{(2)}+\sigma_y^{(2)}\sigma_y^{(3)})
\end{equation}
where $\sigma_m^{(l)}~\text{with}~(m=x,y,z,l=1,2,3)$ are the Pauli matrices for the $l^{th}$ qubit. We choose the initial state $|+x\rangle^{(1)}\otimes|+y\rangle^{(2)}\otimes|+x\rangle^{(3)}$  and the final state $|-x\rangle^{(1)}\otimes|-y\rangle^{(2)}\otimes|-x\rangle^{(3)}$ . The obtained optimal wavefunction $\psi(t)$ connecting them with the interaction strength J=0.2 and L=1 is depicted in Fig.~\ref{Fig14}.

 \begin{figure*}[]
   \includegraphics[width=4.3cm]{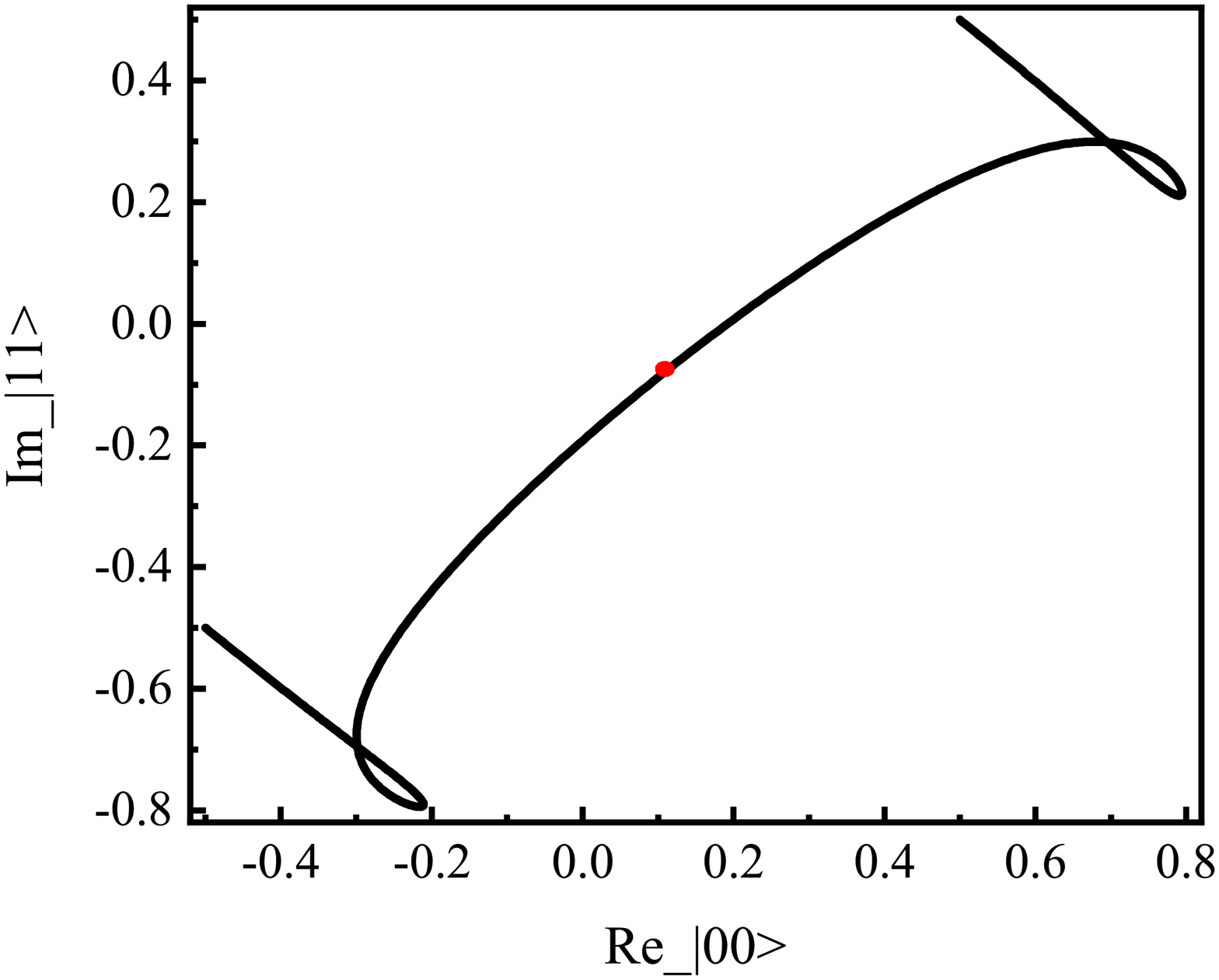}
    \includegraphics[height=3.5cm,width=4.3cm]{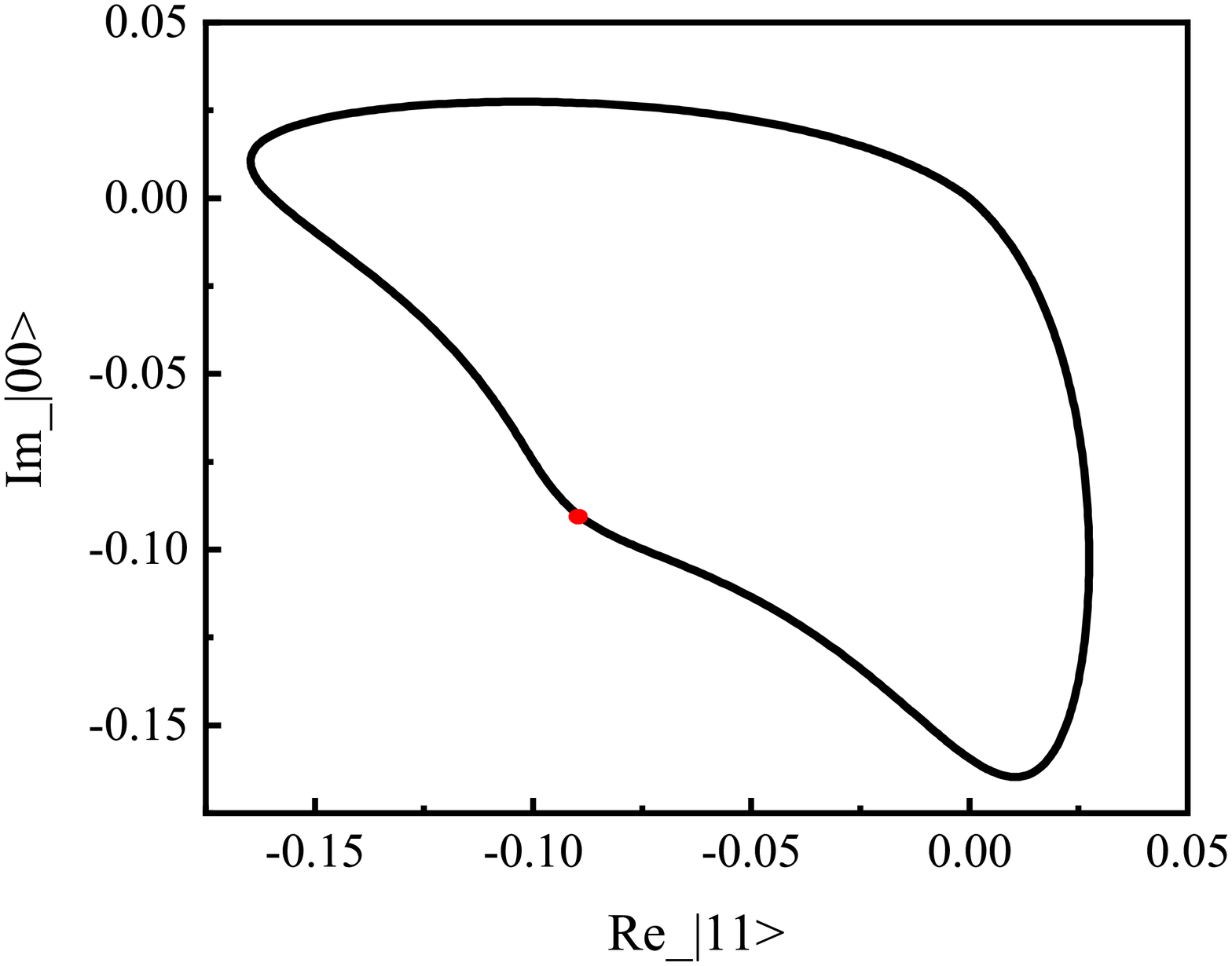}
     \includegraphics[width=4.3cm]{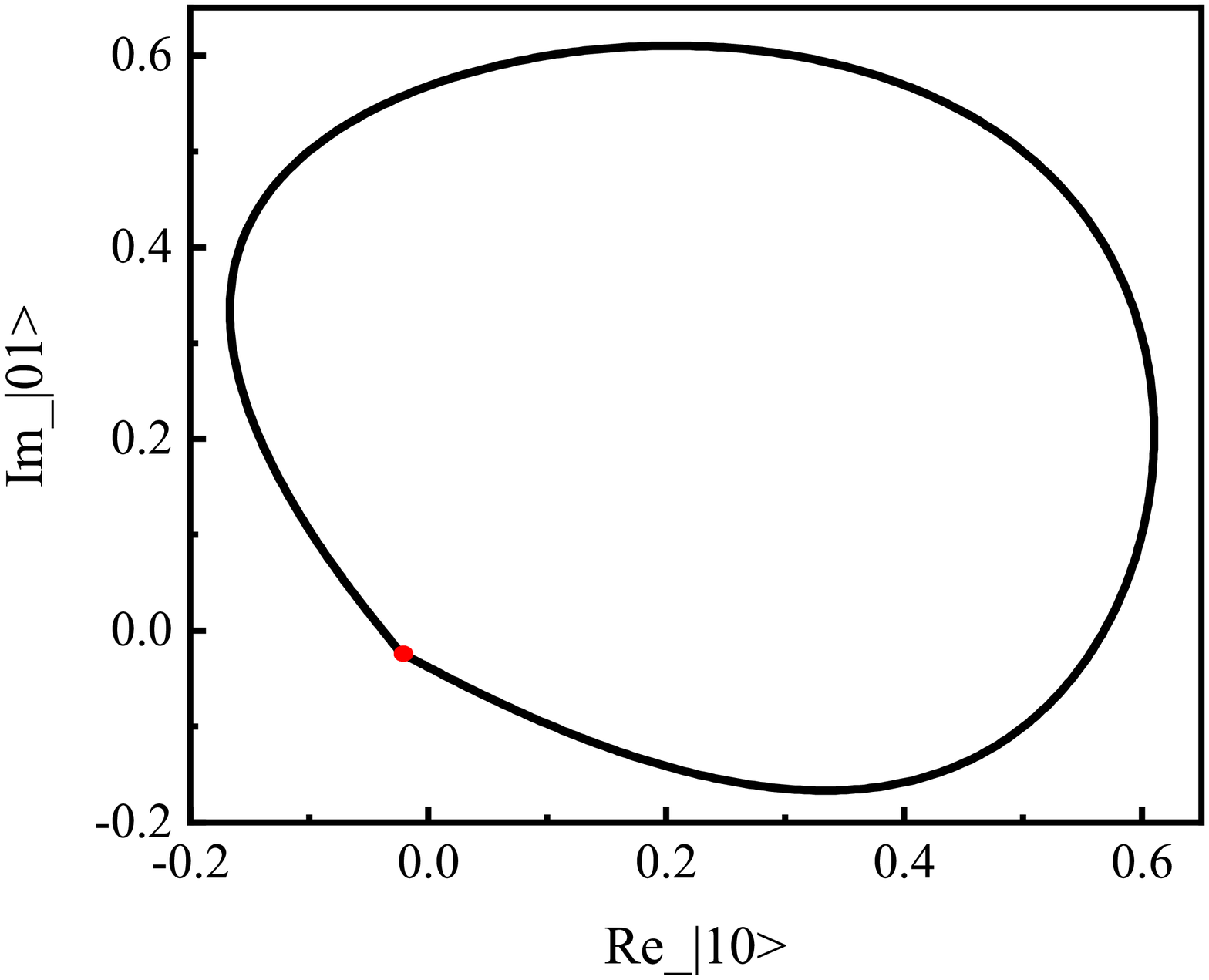}
    \includegraphics[width=4.3cm]{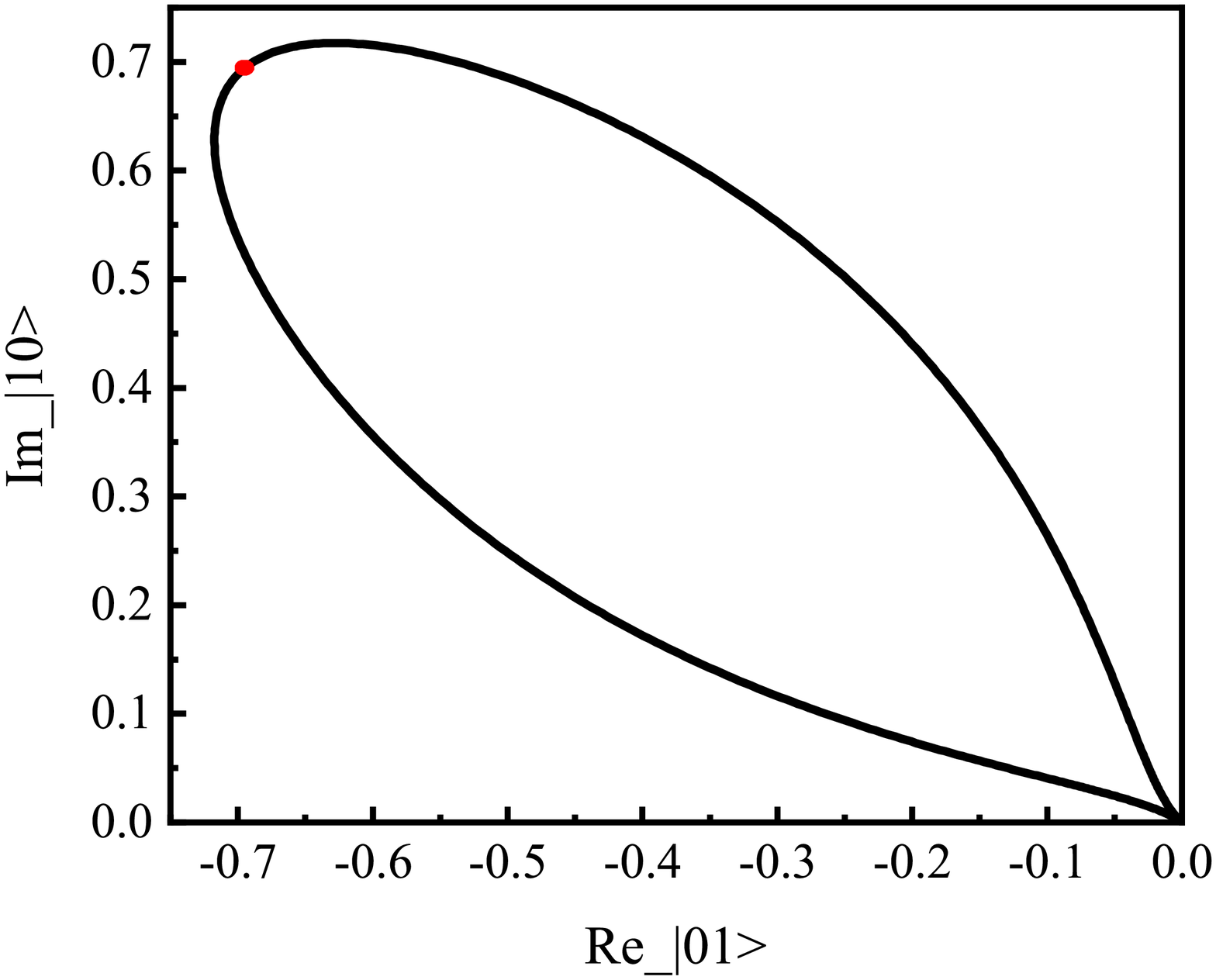}
      \caption{\label{Fig6} The wave function $\psi(t)$ for L=3 and $J_{yy}=1$. The red points are invariant under the symmetry operation F. $|00>,|01>,|10> $ and $|11>$ are the four components of $\psi(t) $, ‘Re’ and ‘Im’ represents the real and image part.}
  \end{figure*}
\subsection{ The Symmetry Operation Discussion}
  \begin{figure*}[]
   \includegraphics[height=3cm,width=3.6cm]{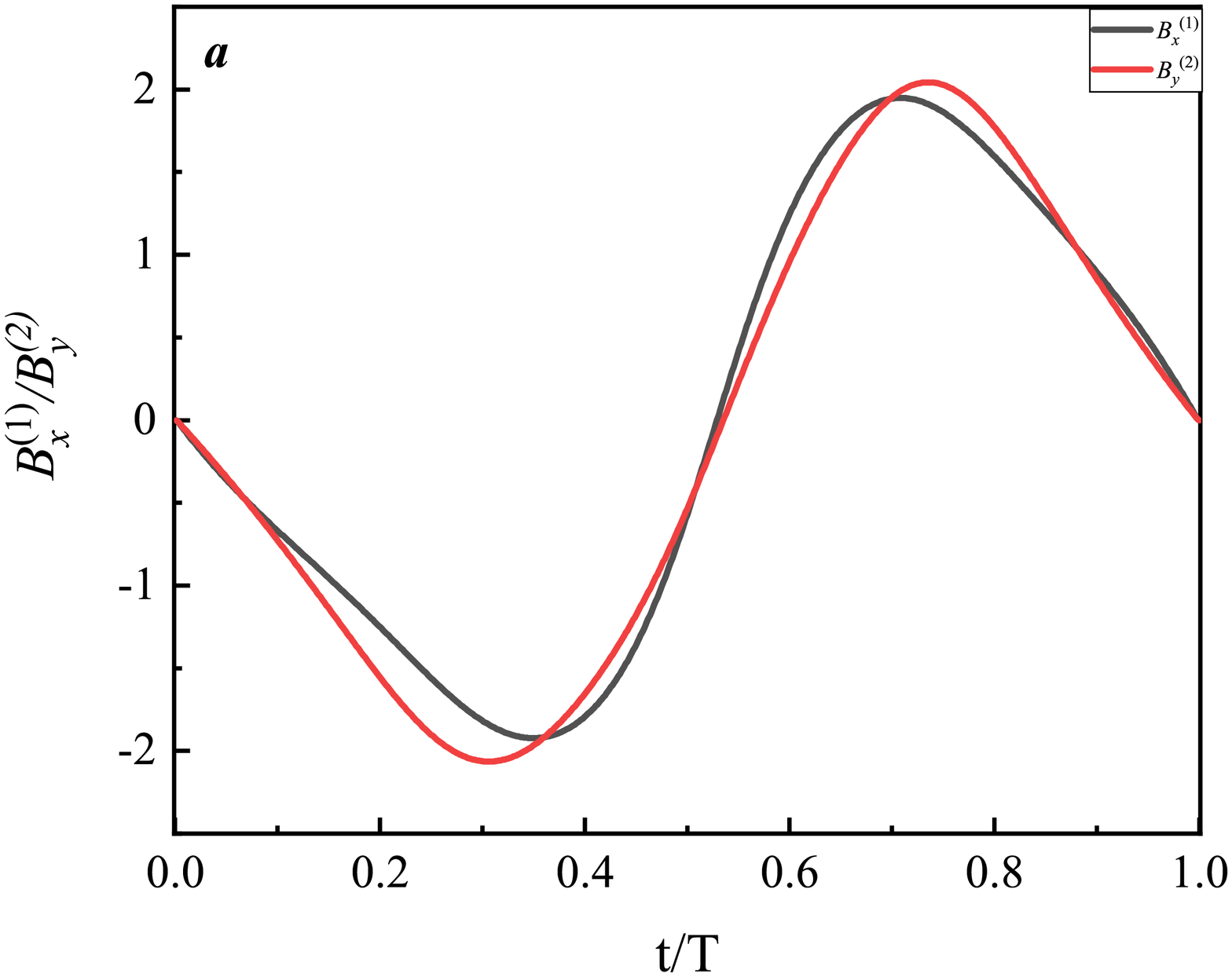}
   \includegraphics[height=3cm,width=3.6cm]{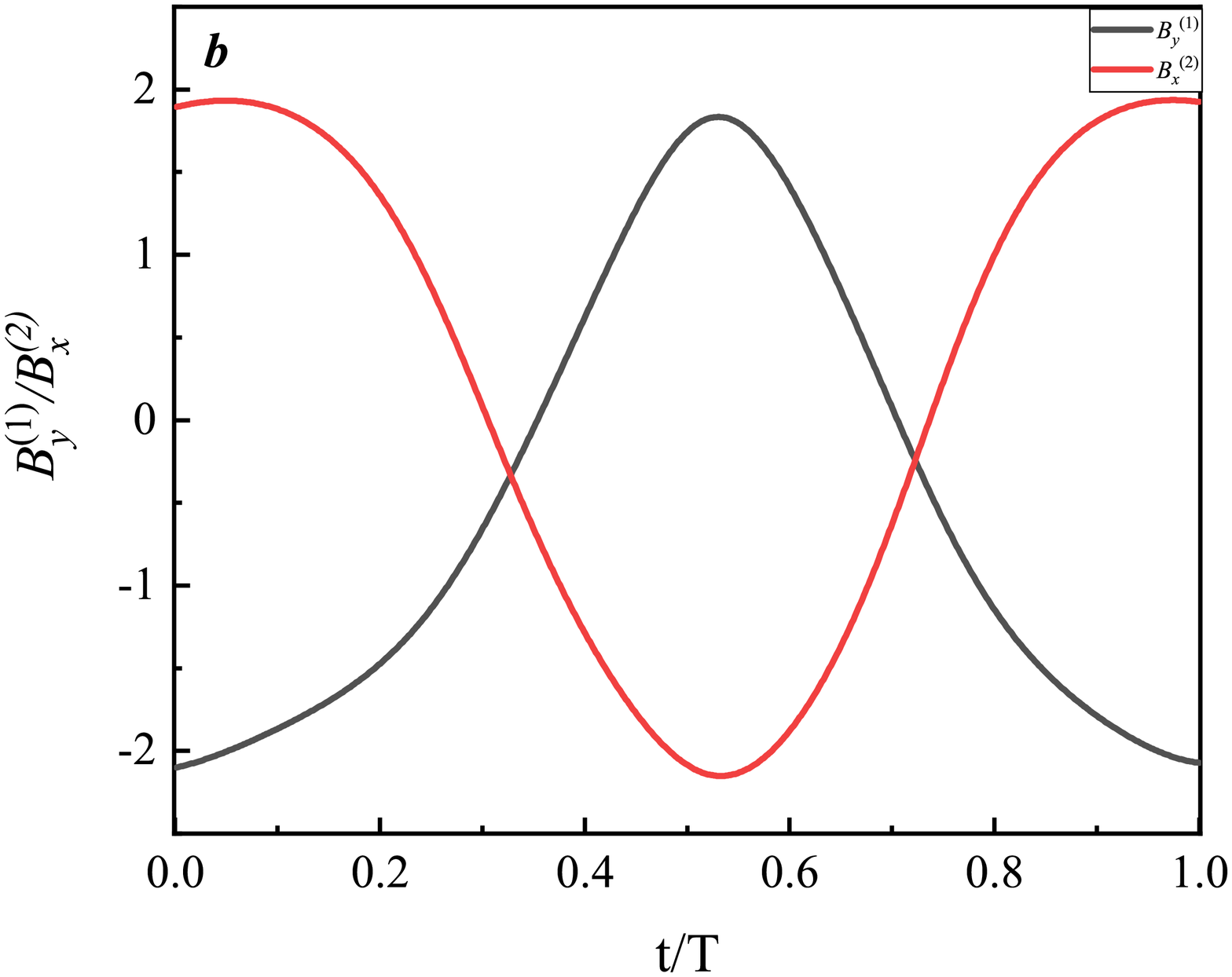}
   \includegraphics[height=3cm,width=3.6cm]{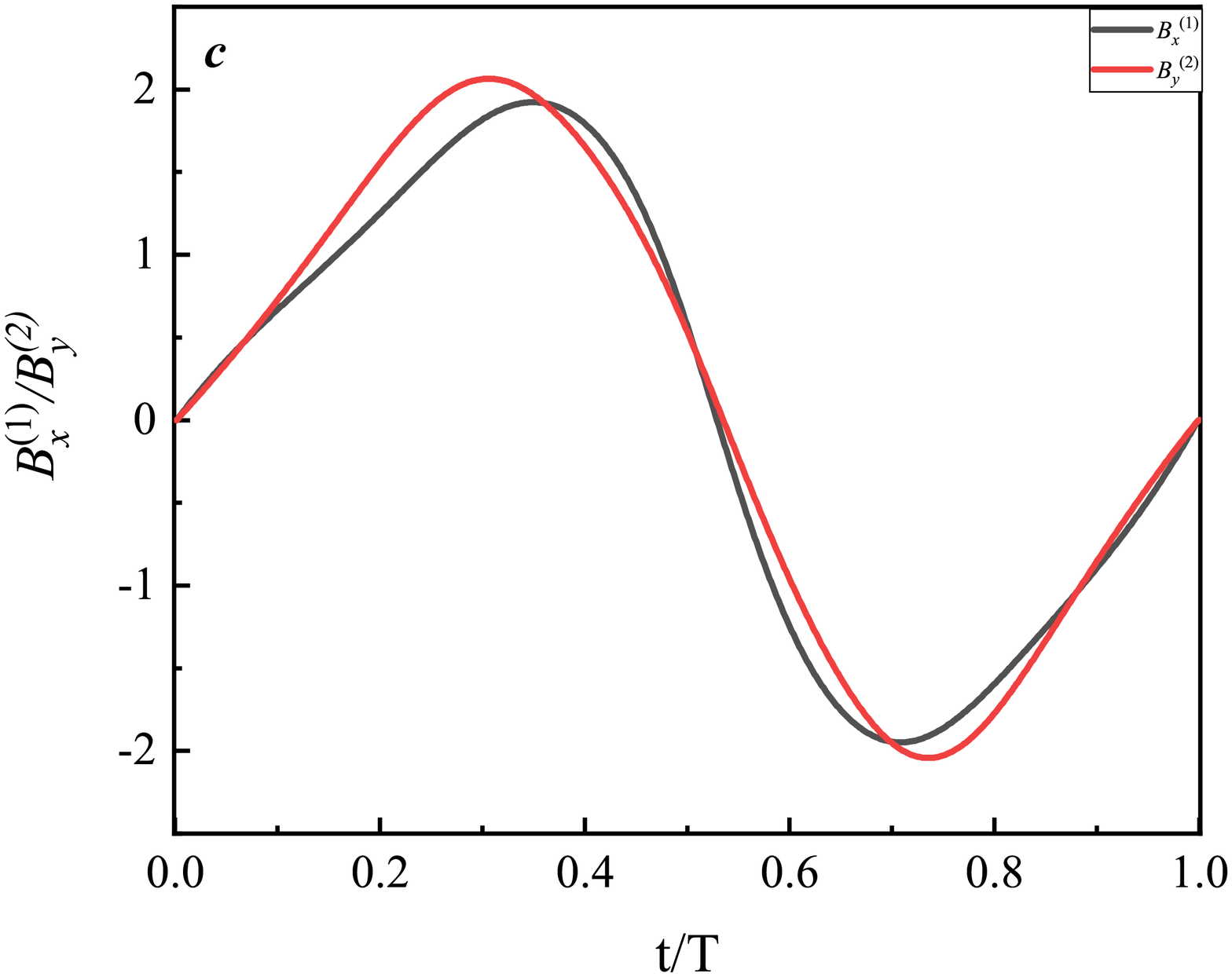}
   \includegraphics[height=3cm,width=3.6cm]{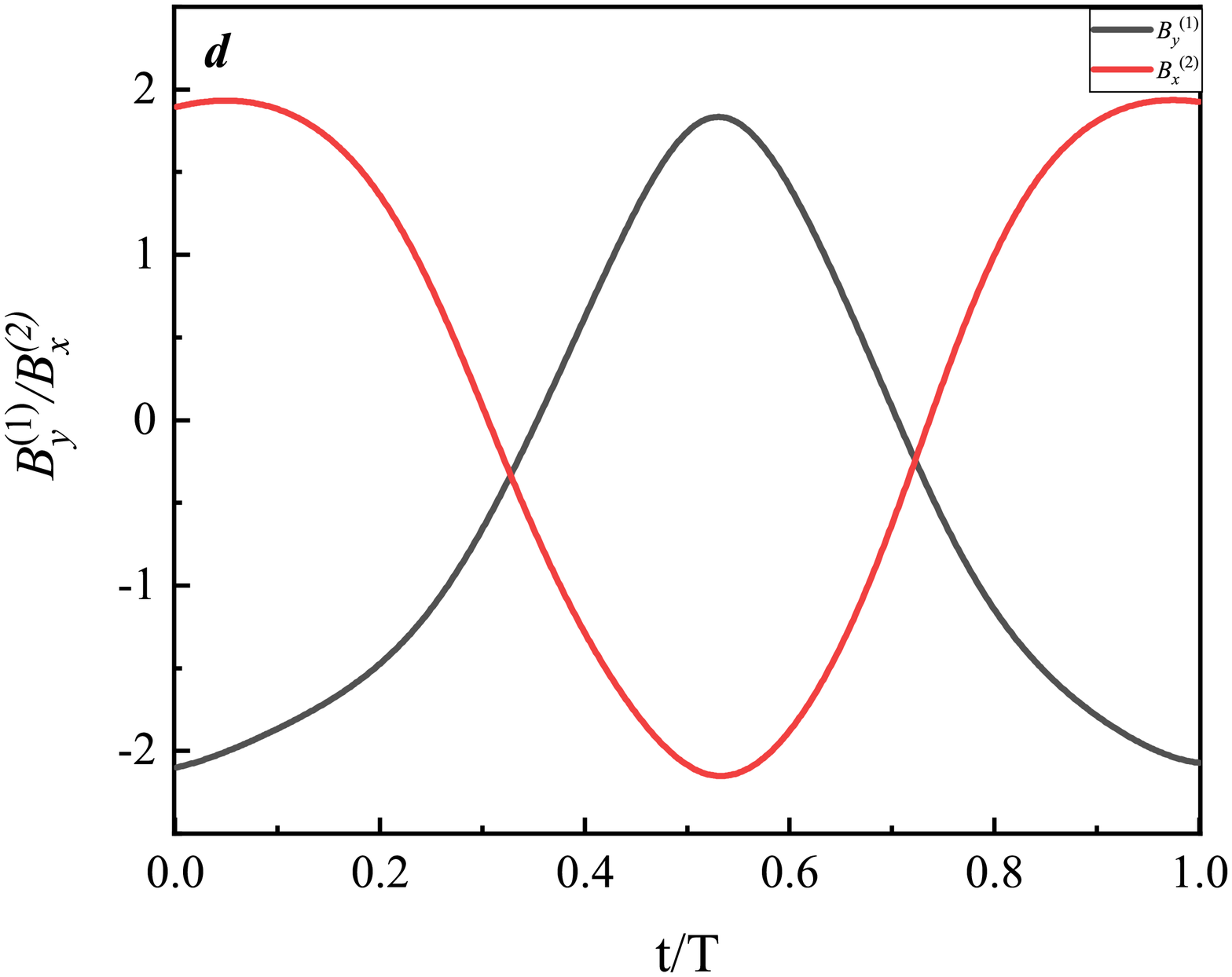}
   \includegraphics[height=3cm,width=3.6cm]{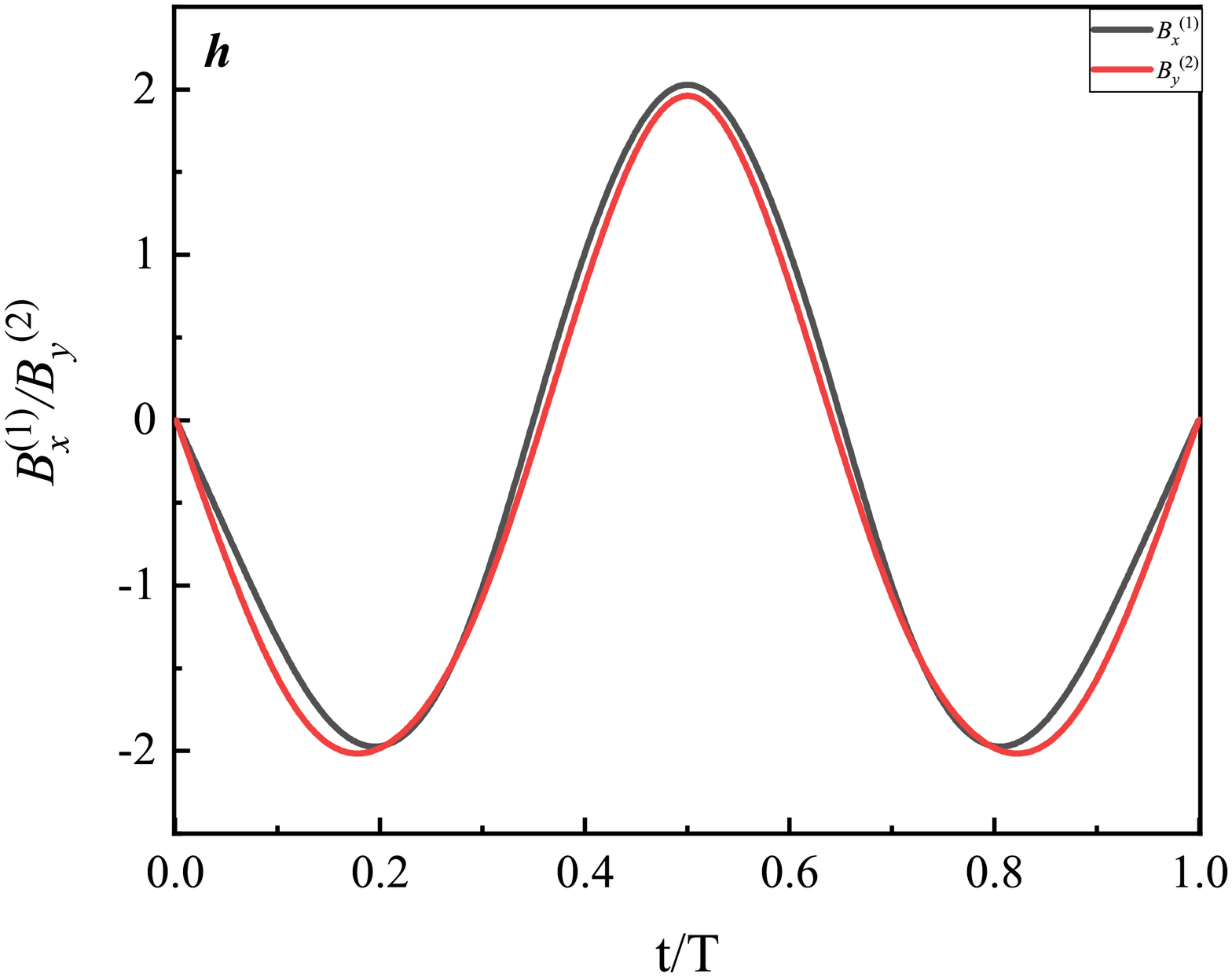}
   \includegraphics[height=3cm,width=3.6cm]{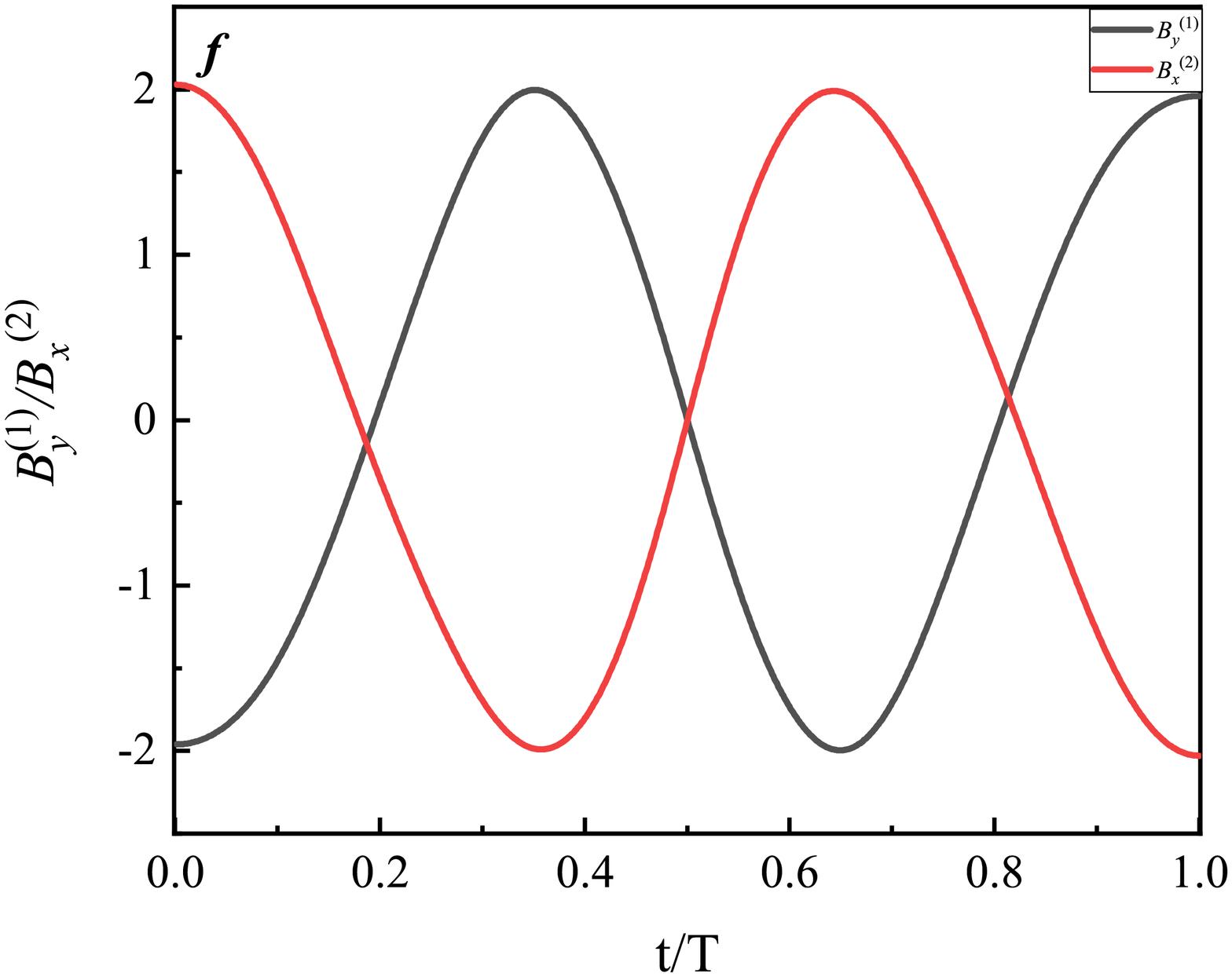}
   \includegraphics[height=3cm,width=3.6cm]{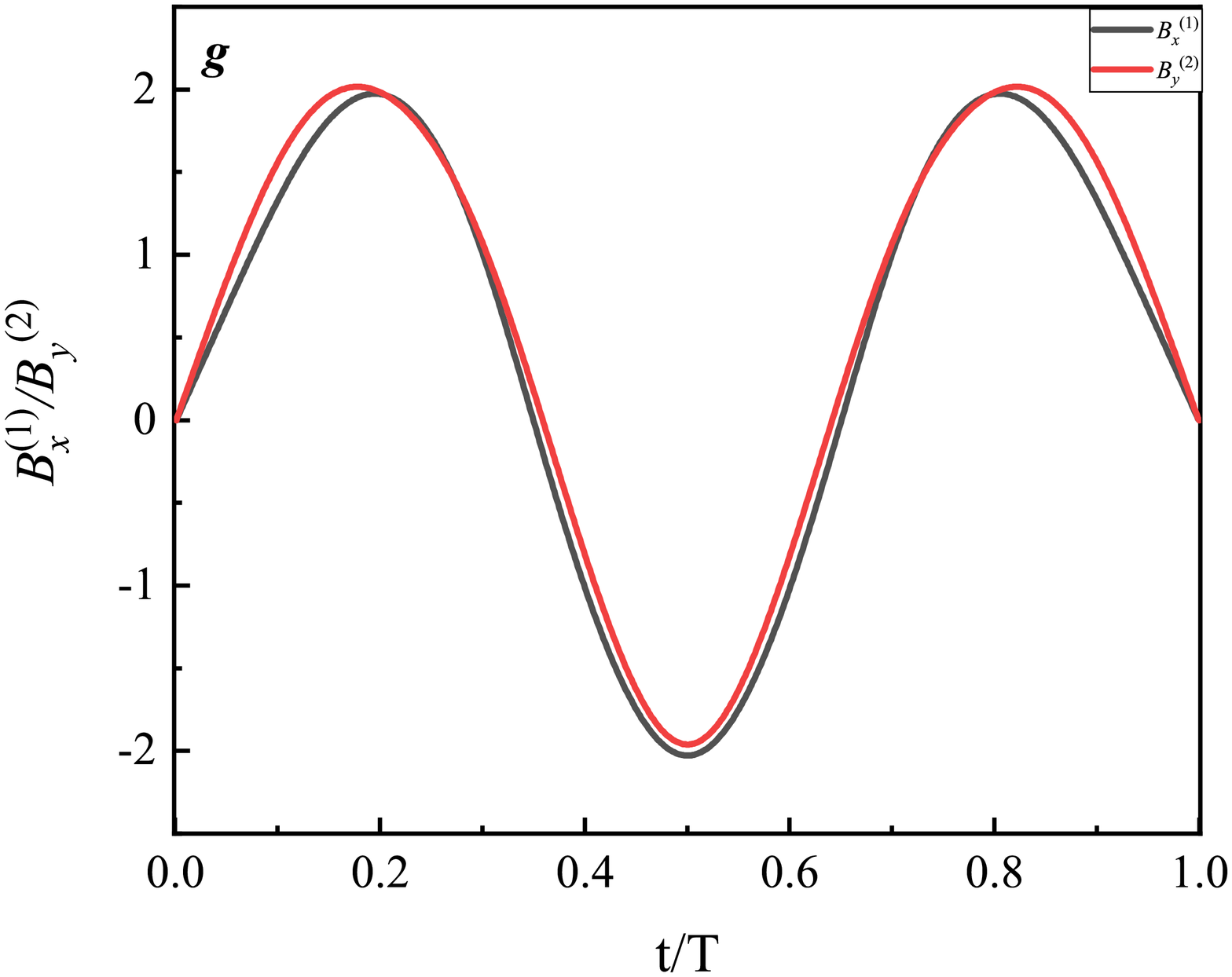}
   \includegraphics[height=3cm,width=3.6cm]{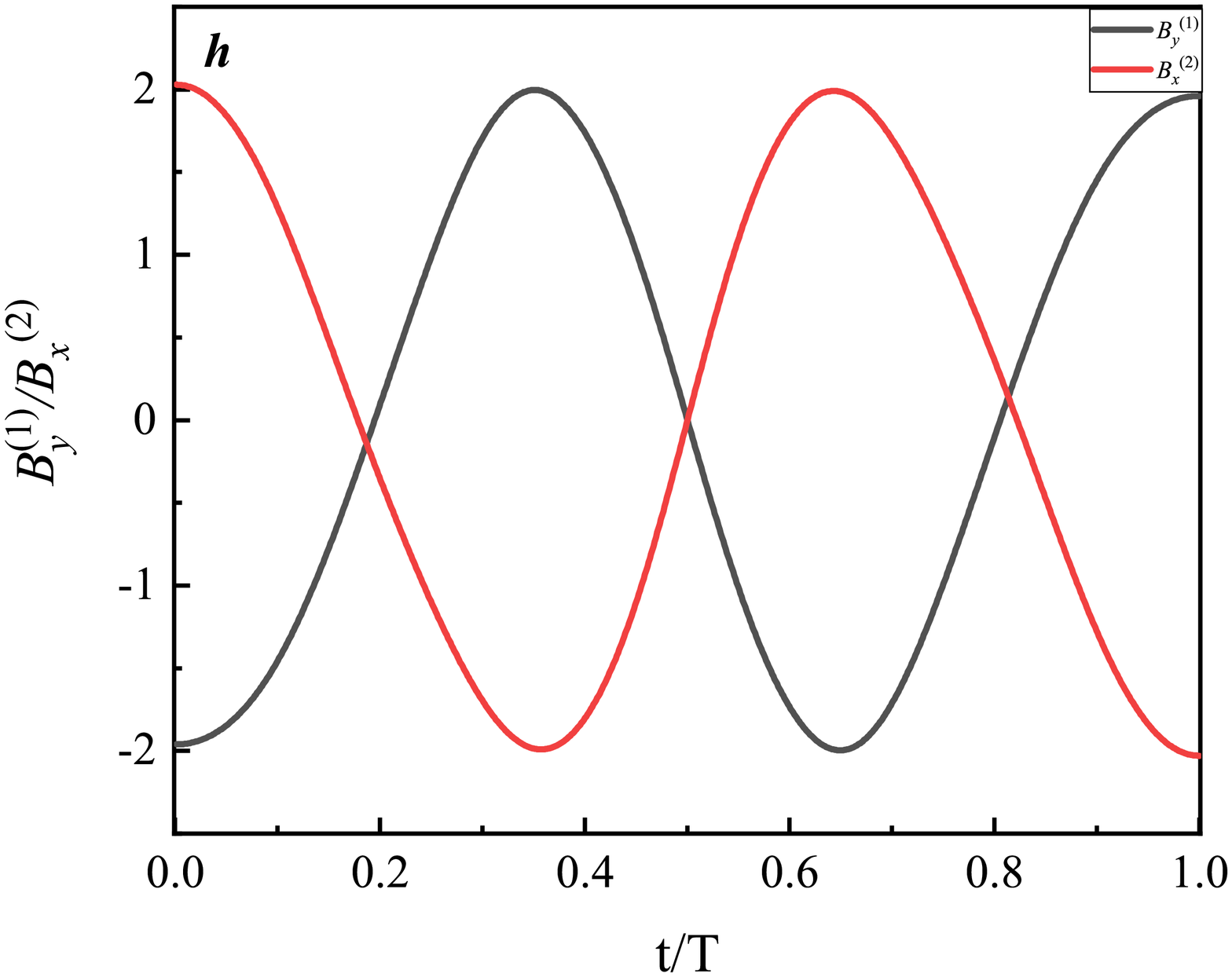}
      \caption{\label{Fig8} The function $\bm B(t)$ for L=2 ($a\sim d$) with $J_{yy}=0.05$ and L=3 ($e\sim h$)with $J_{yy}=0.2$. The black lines plot the controlled part $\bm B(t)$  for qubit one and the red lines plot the controlled part $\bm B(t)$ for the second qubit. The results of $c,~d,~g$ and $h$ were obtained for the transformed Hamiltonian and states $H'=F^T H F$, $\psi_0'=F\psi_0$, $\psi_{f}'=F\psi_{f}$.}
  \end{figure*}
\begin{figure}[]
\includegraphics[width=6cm]{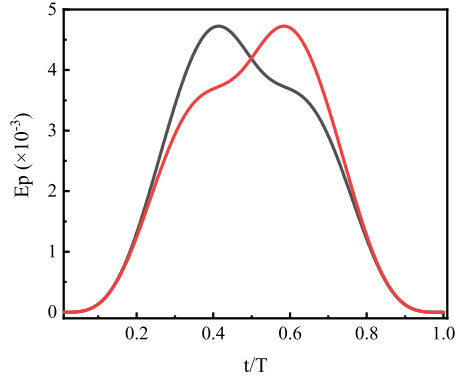}
\caption{\label{Fig7} The entanglement distribution of $\psi(t)$ (the black line) and $\psi'(t)$ (the red line) with $J_{yy}=0.05$ and L=3, $\psi'(t) $ is the symmetry counterpart of $\psi(t) $ under the action $F$.}
\end{figure}
 In this section, we check a symmetry operation of the Hamiltonian Eq.~(\ref{Hj}) and ferret out possible mechanism underlying the phenomenon discussed in Part~\ref{sec:2bitw/o} above.
  It is interesting to observe a discrete symmetry operation in Hamiltonian~(\ref{Hj}) associated with
\begin{equation}\label{ary}
  F=
\left[
 \begin{matrix}
   ~~0 &  ~~0 & ~~0 & ~~i \\
   ~~0 &  ~~0 & ~~i & ~~0 \\
   ~~0 &   -i & ~~0 & ~~0 \\
    -i &  ~~0 & ~~0 & ~~0
  \end{matrix}
  \right]
\end{equation}
which connects the initial state $\psi_0 $ and the target state $\psi_f$, {\em i.e.} $F\psi_0=-\psi_f $ in the current case. Obviously, the magnetic field $\bm B(t)$ in the Hamiltonian is artificially controlled, which should be changed accordingly (by $F^THF$) to complete the symmetry operation, as confirmed in Fig.~\ref{Fig8}.
For the case of L=1,3, especially, the middle state $\psi_{mid}$ (at t=0.5) remains invariant under this symmetry operation, namely $F\psi_{mid}=\psi_{mid}$. In fact, to search the symmetry, a clue could be directly taken from Fig.~\ref{Fig6}, where the red points represent the middle state $\psi_{mid}= (0.09483-0.08982i,~ -0.69486-0.02312i,~ -0.02309+0.69443i,~ -0.08993-0.09488i)^t$. It is easy to check that the time course of the wave function $\psi_{t}$ is also symmetric about the middle state $\psi_{mid}$. From the view of the controlled part $\bm B(t)$ of $H$, take the cases of L=2 and 3 as an example. Fig.\ref{Fig8} $a,~b~(L=2)$ and $e,~f~(L=3)$ depict the time course of $\bm B(t)$ from $\psi_0$ to $\psi_{f}$ while Fig.\ref{Fig8} $c,~d~(L=2)$ and $g,~h~(L=3)$ plot the time course of $\bm B(t)$ from $\psi_0'=F\psi_0$ to $\psi_{f}'=F\psi_{f}$ with the transformed Hamiltonian $H'= F^T H F$. Apparently, Fig.\ref{Fig8} $e$ and $f$ coincide with $g$ and $h$ respectively up to a minus sign in the relevant component. In addition, the evolution of $B^{(1)}_x$ and $B^{(2)}_y$  is axisymmteric with respect to $t=0.5$ and the evolution profiles of $B^{(1)}_y$ and $B^{(2)}_x$ have a center symmetry with respect to (0.5, 0). All the above indicates that the maximum of $E_p$ is associated with a fixed point of the operation $F$ for the case L=1, 3.

In the cases L=2, 4, the symmetry in $\psi(t)$ and $\bm B(t)$ discussed previously disappear. Taking L=2 for an example, Fig.\ref{Fig8} $a$ and $b$ do not coincide with Fig.\ref{Fig8} $c$ and $d$ and the maximum value of $E_p$ is not at t=0.5. On the other hand, as the symmetry partner $\psi'_{t}$ of the wave function $\psi_{t}$ is obtained with the symmetry operation $F$, even if the entanglement evolution of $\psi'_{t}$ (see Fig.~\ref{Fig7}, the red line) is asymmetric, those of $\psi'(t)$ and $\psi(t)$ are axially symmetric with respect to each other about t=0.5.
\begin{figure}[]
\includegraphics[width=7cm]{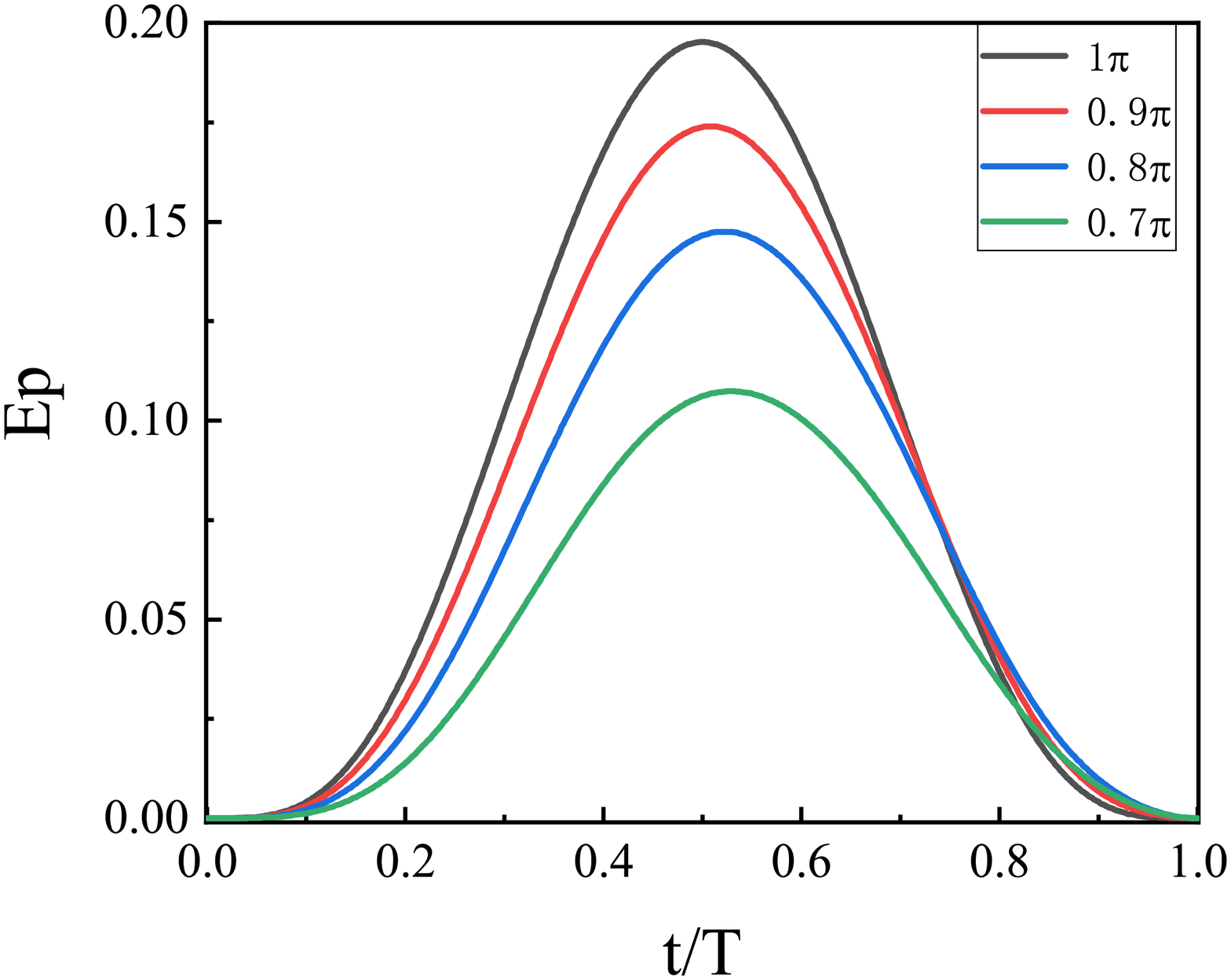}
\caption{\label{Fig10} The entanglement evolution of $\psi(t)$ with $L=1$ and $J_{yy}=0.4$ } in Eq.~(\ref{Hj}), the angle between $\psi_0$ and $\psi_f$ is from $\pi$ to $0.7\pi$.
\end{figure}

   In the case where $\psi_0$ and $\psi_f$ do not flip to each other, {\em e.g.} the angle between $\psi_0$ and $\psi_f$ is less than $\pi$, we then have $F\psi_0\neq-\psi_f$ (take the case $L=1$ as an example), which indicates that the entanglement evolution is asymmetric (shown in Fig.~\ref{Fig10}). However, the maximum entanglement still exists which is unique but not at $t=0.5$.

\section{SUMMARY AND DISCUSSION}
In this work, we derive a general set of differential equations (Eq.~(\ref{eq:evo2})) for an optimal quantum control, which is valid for single or multiple qubits with or without interaction. In the derivation and application of Eq.~(\ref{eq:evo2}), we discussed the following aspects:
First, previous numerical methods conventionally involve a large number of physically irrelevant variables derived from the commutation relation with system Hamiltonians. We thus introduce a new set of observables to eliminate the physically irrelevant part, whereby an accelerated computation becomes possible. As a result, the efficiency of computation are boosted significantly. With the help of a ‘relaxation’ idea, the drawback of the shooting method, of requiring a good initial guess, is surpassed by slowly pushing the trial solution to the correct one with a gradual restoration of the interaction. Second, we discussed symmetries of the Lagrangian formalism of the QBE. In the numerical calculation, these entail unwanted degenerate directions in the Jacobian matrix, which may cause serious trouble in the solution of the resulting boundary value problem.
Finally, groups of analytic solutions are given as rotations of qubits on Bloch spheres in the absence of interaction, which may be used as the starting point of the above relaxation scheme.

In an application of the new scheme to the case of two qubits with interaction, the solution to the QBE displays a unimodal evolution profile of the entanglement. If the initial and final state transform to each other under a symmetry operation of the Hamiltonian. The evolution of the optimal path either bears a related symmetry or has a symmetry partner. More explicitly, for the even oscillation number symmetry disappears in the profile but the symmetry operation gives another optimal path -- the symmetry partner of the original solution.

With current formalism, we are able to design optimal control strategy for multi-qubits with possibly complex interaction.
However, the evolution is not unique and many local optimal paths exist which could be indexed by the number of oscillations of the paths in simple cases. It could be interesting to classify these paths and check their bifurcation routes when system parameters change. On the other hand, this work may provide experimental researchers a practical computation tool for quantum information regulation, since an optimal control strategy of the magnetic field is readily designed based on our scheme.
\acknowledgments
This work was supported by the National Natural Science Foundation of China under Grants No. 11775035, and also by the Fundamental Research Funds for the Central Universities with Contract No.2019XD-A10.

\bibliography{Ref1}

\providecommand{\noopsort}[1]{}\providecommand{\singleletter}[1]{#1}%
\begin{thebibliography}{35}%
\makeatletter
\providecommand \@ifxundefined [1]{%
 \@ifx{#1\undefined}
}%
\providecommand \@ifnum [1]{%
 \ifnum #1\expandafter \@firstoftwo
 \else \expandafter \@secondoftwo
 \fi
}%
\providecommand \@ifx [1]{%
 \ifx #1\expandafter \@firstoftwo
 \else \expandafter \@secondoftwo
 \fi
}%
\providecommand \natexlab [1]{#1}%
\providecommand \enquote  [1]{``#1''}%
\providecommand \bibnamefont  [1]{#1}%
\providecommand \bibfnamefont [1]{#1}%
\providecommand \citenamefont [1]{#1}%
\providecommand \href@noop [0]{\@secondoftwo}%
\providecommand \href [0]{\begingroup \@sanitize@url \@href}%
\providecommand \@href[1]{\@@startlink{#1}\@@href}%
\providecommand \@@href[1]{\endgroup#1\@@endlink}%
\providecommand \@sanitize@url [0]{\catcode `\\12\catcode `\$12\catcode
  `\&12\catcode `\#12\catcode `\^12\catcode `\_12\catcode `\%12\relax}%
\providecommand \@@startlink[1]{}%
\providecommand \@@endlink[0]{}%
\providecommand \url  [0]{\begingroup\@sanitize@url \@url }%
\providecommand \@url [1]{\endgroup\@href {#1}{\urlprefix }}%
\providecommand \urlprefix  [0]{URL }%
\providecommand \Eprint [0]{\href }%
\providecommand \doibase [0]{https://doi.org/}%
\providecommand \selectlanguage [0]{\@gobble}%
\providecommand \bibinfo  [0]{\@secondoftwo}%
\providecommand \bibfield  [0]{\@secondoftwo}%
\providecommand \translation [1]{[#1]}%
\providecommand \BibitemOpen [0]{}%
\providecommand \bibitemStop [0]{}%
\providecommand \bibitemNoStop [0]{.\EOS\space}%
\providecommand \EOS [0]{\spacefactor3000\relax}%
\providecommand \BibitemShut  [1]{\csname bibitem#1\endcsname}%
\let\auto@bib@innerbib\@empty
\bibitem [{\citenamefont {Feynman}(1982)}]{Feynman1982}%
  \BibitemOpen
  \bibfield  {author} {\bibinfo {author} {\bibfnamefont {R.~P.}\ \bibnamefont
  {Feynman}},\ }\bibfield  {title} {\bibinfo {title} {Simulating physics with
  computers},\ }\href@noop {} {\bibfield  {journal} {\bibinfo  {journal} {Int.
  J. Theor. Phys.}\ }\textbf {\bibinfo {volume} {21}},\ \bibinfo {pages} {467}
  (\bibinfo {year} {1982})}\BibitemShut {NoStop}%
\bibitem [{\citenamefont {SHOR}(1994)}]{SHOR1994Algorithms}%
  \BibitemOpen
  \bibfield  {author} {\bibinfo {author} {\bibfnamefont {P.}~\bibnamefont
  {SHOR}},\ }\bibfield  {title} {\bibinfo {title} {Algorithms for quantum
  computation : Discrete logarithms and factoring},\ }in\ \href@noop {} {\emph
  {\bibinfo {booktitle} {Proceedings of 35th Annual Symposium on Foundations of
  Computer Scienece}}}\ (\bibinfo {year} {1994})\BibitemShut {NoStop}%
\bibitem [{\citenamefont {Boixo}\ \emph {et~al.}(2014)\citenamefont {Boixo},
  \citenamefont {Rønnow}, \citenamefont {Isakov}, \citenamefont {Wang},
  \citenamefont {Wecker}, \citenamefont {Lidar}, \citenamefont {Martinis},\
  and\ \citenamefont {Troyer}}]{Boix}%
  \BibitemOpen
  \bibfield  {author} {\bibinfo {author} {\bibfnamefont {S.}~\bibnamefont
  {Boixo}}, \bibinfo {author} {\bibfnamefont {T.~F.}\ \bibnamefont {Rønnow}},
  \bibinfo {author} {\bibfnamefont {S.~V.}\ \bibnamefont {Isakov}}, \bibinfo
  {author} {\bibfnamefont {Z.}~\bibnamefont {Wang}}, \bibinfo {author}
  {\bibfnamefont {D.}~\bibnamefont {Wecker}}, \bibinfo {author} {\bibfnamefont
  {D.~A.}\ \bibnamefont {Lidar}}, \bibinfo {author} {\bibfnamefont {J.~M.}\
  \bibnamefont {Martinis}},\ and\ \bibinfo {author} {\bibfnamefont
  {M.}~\bibnamefont {Troyer}},\ }\bibfield  {title} {\bibinfo {title} {Evidence
  for quantum annealing with more than one hundred qubits},\ }\href@noop {}
  {\bibfield  {journal} {\bibinfo  {journal} {Nat. Phys.}\ }\textbf {\bibinfo
  {volume} {10}},\ \bibinfo {pages} {218} (\bibinfo {year} {2014})}\BibitemShut
  {NoStop}%
\bibitem [{\citenamefont {del Campo}\ \emph {et~al.}(2013)\citenamefont {del
  Campo}, \citenamefont {Egusquiza}, \citenamefont {Plenio},\ and\
  \citenamefont {Huelga}}]{Campo2013}%
  \BibitemOpen
  \bibfield  {author} {\bibinfo {author} {\bibfnamefont {A.}~\bibnamefont {del
  Campo}}, \bibinfo {author} {\bibfnamefont {I.~L.}\ \bibnamefont {Egusquiza}},
  \bibinfo {author} {\bibfnamefont {M.~B.}\ \bibnamefont {Plenio}},\ and\
  \bibinfo {author} {\bibfnamefont {S.~F.}\ \bibnamefont {Huelga}},\ }\bibfield
   {title} {\bibinfo {title} {Quantum speed limits in open system dynamics},\
  }\href {https://doi.org/10.1103/PhysRevLett.110.050403} {\bibfield  {journal}
  {\bibinfo  {journal} {Phys. Rev. Lett.}\ }\textbf {\bibinfo {volume} {110}},\
  \bibinfo {pages} {050403} (\bibinfo {year} {2013})}\BibitemShut {NoStop}%
\bibitem [{\citenamefont {Deffner}\ and\ \citenamefont
  {Lutz}(2013)}]{Deffner2013}%
  \BibitemOpen
  \bibfield  {author} {\bibinfo {author} {\bibfnamefont {S.}~\bibnamefont
  {Deffner}}\ and\ \bibinfo {author} {\bibfnamefont {E.}~\bibnamefont {Lutz}},\
  }\bibfield  {title} {\bibinfo {title} {Quantum speed limit for non-markovian
  dynamics},\ }\href {https://doi.org/10.1103/PhysRevLett.111.010402}
  {\bibfield  {journal} {\bibinfo  {journal} {Phys. Rev. Lett.}\ }\textbf
  {\bibinfo {volume} {111}},\ \bibinfo {pages} {010402} (\bibinfo {year}
  {2013})}\BibitemShut {NoStop}%
\bibitem [{\citenamefont {Taddei}\ \emph {et~al.}(2013)\citenamefont {Taddei},
  \citenamefont {Escher}, \citenamefont {Davidovich},\ and\ \citenamefont
  {de~Matos~Filho}}]{Taddei2013}%
  \BibitemOpen
  \bibfield  {author} {\bibinfo {author} {\bibfnamefont {M.~M.}\ \bibnamefont
  {Taddei}}, \bibinfo {author} {\bibfnamefont {B.~M.}\ \bibnamefont {Escher}},
  \bibinfo {author} {\bibfnamefont {L.}~\bibnamefont {Davidovich}},\ and\
  \bibinfo {author} {\bibfnamefont {R.~L.}\ \bibnamefont {de~Matos~Filho}},\
  }\bibfield  {title} {\bibinfo {title} {Quantum speed limit for physical
  processes},\ }\href {https://doi.org/10.1103/PhysRevLett.110.050402}
  {\bibfield  {journal} {\bibinfo  {journal} {Phys. Rev. Lett.}\ }\textbf
  {\bibinfo {volume} {110}},\ \bibinfo {pages} {050402} (\bibinfo {year}
  {2013})}\BibitemShut {NoStop}%
\bibitem [{\citenamefont {Mirkin}\ \emph {et~al.}(2016)\citenamefont {Mirkin},
  \citenamefont {Toscano},\ and\ \citenamefont {Wisniacki}}]{Mirkin2016}%
  \BibitemOpen
  \bibfield  {author} {\bibinfo {author} {\bibfnamefont {N.}~\bibnamefont
  {Mirkin}}, \bibinfo {author} {\bibfnamefont {F.}~\bibnamefont {Toscano}},\
  and\ \bibinfo {author} {\bibfnamefont {D.~A.}\ \bibnamefont {Wisniacki}},\
  }\bibfield  {title} {\bibinfo {title} {Quantum-speed-limit bounds in an open
  quantum evolution},\ }\href {https://doi.org/10.1103/PhysRevA.94.052125}
  {\bibfield  {journal} {\bibinfo  {journal} {Phys. Rev. A}\ }\textbf {\bibinfo
  {volume} {94}},\ \bibinfo {pages} {052125} (\bibinfo {year}
  {2016})}\BibitemShut {NoStop}%
\bibitem [{\citenamefont {Schulte-Herbruggen}\ \emph
  {et~al.}(2005)\citenamefont {Schulte-Herbruggen}, \citenamefont {Sporl},
  \citenamefont {Khaneja},\ and\ \citenamefont {Glaser}}]{Schulte2005}%
  \BibitemOpen
  \bibfield  {author} {\bibinfo {author} {\bibfnamefont {T.}~\bibnamefont
  {Schulte-Herbruggen}}, \bibinfo {author} {\bibfnamefont {A.}~\bibnamefont
  {Sporl}}, \bibinfo {author} {\bibfnamefont {N.}~\bibnamefont {Khaneja}},\
  and\ \bibinfo {author} {\bibfnamefont {S.~J.}\ \bibnamefont {Glaser}},\
  }\bibfield  {title} {\bibinfo {title} {Optimal control-based efficient
  synthesis of building blocks of quantum algorithms: A perspective from
  network complexity towards time complexity},\ }\href
  {https://doi.org/10.1103/physreva.72.042331} {\bibfield  {journal} {\bibinfo
  {journal} {Phys. Rev. A}\ }\textbf {\bibinfo {volume} {72}},\ \bibinfo
  {pages} {042331} (\bibinfo {year} {2005})}\BibitemShut {NoStop}%
\bibitem [{\citenamefont {Nielsen}\ and\ \citenamefont
  {Chuang}(2011)}]{QuanInf}%
  \BibitemOpen
  \bibfield  {author} {\bibinfo {author} {\bibfnamefont {M.~A.}\ \bibnamefont
  {Nielsen}}\ and\ \bibinfo {author} {\bibfnamefont {I.~L.}\ \bibnamefont
  {Chuang}},\ }\bibinfo {title} {Quantum computation and quantum information:
  10th anniversary edition}\ (\bibinfo  {publisher} {Cambridge University
  Press},\ \bibinfo {year} {2011})\ \bibinfo {edition} {10th}\ ed.\BibitemShut
  {Stop}%
\bibitem [{\citenamefont {Nielsen}\ \emph
  {et~al.}(2006{\natexlab{a}})\citenamefont {Nielsen}, \citenamefont {Dowling},
  \citenamefont {Gu},\ and\ \citenamefont {Doherty}}]{Nielsen2006sci}%
  \BibitemOpen
  \bibfield  {author} {\bibinfo {author} {\bibfnamefont {M.~A.}\ \bibnamefont
  {Nielsen}}, \bibinfo {author} {\bibfnamefont {M.~R.}\ \bibnamefont
  {Dowling}}, \bibinfo {author} {\bibfnamefont {M.}~\bibnamefont {Gu}},\ and\
  \bibinfo {author} {\bibfnamefont {A.~C.}\ \bibnamefont {Doherty}},\
  }\bibfield  {title} {\bibinfo {title} {Quantum computation as geometry},\
  }\href {https://doi.org/10.1126/science.1121541} {\bibfield  {journal}
  {\bibinfo  {journal} {Science}\ }\textbf {\bibinfo {volume} {311}},\ \bibinfo
  {pages} {1133} (\bibinfo {year} {2006}{\natexlab{a}})}\BibitemShut {NoStop}%
\bibitem [{\citenamefont {Nielsen}\ \emph
  {et~al.}(2006{\natexlab{b}})\citenamefont {Nielsen}, \citenamefont {Dowling},
  \citenamefont {Gu},\ and\ \citenamefont {Doherty}}]{Nielsen2006pre}%
  \BibitemOpen
  \bibfield  {author} {\bibinfo {author} {\bibfnamefont {M.~A.}\ \bibnamefont
  {Nielsen}}, \bibinfo {author} {\bibfnamefont {M.~R.}\ \bibnamefont
  {Dowling}}, \bibinfo {author} {\bibfnamefont {M.}~\bibnamefont {Gu}},\ and\
  \bibinfo {author} {\bibfnamefont {A.~C.}\ \bibnamefont {Doherty}},\
  }\bibfield  {title} {\bibinfo {title} {Optimal control, geometry, and quantum
  computing},\ }\href {https://doi.org/10.1103/physreva.73.062323} {\bibfield
  {journal} {\bibinfo  {journal} {Phys. Rev. A}\ }\textbf {\bibinfo {volume}
  {73}},\ \bibinfo {pages} {062323} (\bibinfo {year}
  {2006}{\natexlab{b}})}\BibitemShut {NoStop}%
\bibitem [{\citenamefont {Khaneja}\ \emph {et~al.}(2005)\citenamefont
  {Khaneja}, \citenamefont {Reiss}, \citenamefont {Kehlet}, \citenamefont
  {Schulte-Herbrüggen},\ and\ \citenamefont {Glaser}}]{Khaneja2005}%
  \BibitemOpen
  \bibfield  {author} {\bibinfo {author} {\bibfnamefont {N.}~\bibnamefont
  {Khaneja}}, \bibinfo {author} {\bibfnamefont {T.}~\bibnamefont {Reiss}},
  \bibinfo {author} {\bibfnamefont {C.}~\bibnamefont {Kehlet}}, \bibinfo
  {author} {\bibfnamefont {T.}~\bibnamefont {Schulte-Herbrüggen}},\ and\
  \bibinfo {author} {\bibfnamefont {S.~J.}\ \bibnamefont {Glaser}},\ }\bibfield
   {title} {\bibinfo {title} {Optimal control of coupled spin dynamics: design
  of nmr pulse sequences by gradient ascent algorithms},\ }\href
  {https://doi.org/https://doi.org/10.1016/j.jmr.2004.11.004} {\bibfield
  {journal} {\bibinfo  {journal} {Journal of Magnetic Resonance}\ }\textbf
  {\bibinfo {volume} {172}},\ \bibinfo {pages} {296 } (\bibinfo {year}
  {2005})}\BibitemShut {NoStop}%
\bibitem [{\citenamefont {Maximov}\ \emph {et~al.}(2008)\citenamefont
  {Maximov}, \citenamefont {To$\check{s}$ner},\ and\ \citenamefont
  {Nielsen}}]{Maximov2008}%
  \BibitemOpen
  \bibfield  {author} {\bibinfo {author} {\bibfnamefont {I.~I.}\ \bibnamefont
  {Maximov}}, \bibinfo {author} {\bibfnamefont {Z.}~\bibnamefont
  {To$\check{s}$ner}},\ and\ \bibinfo {author} {\bibfnamefont {N.~C.}\
  \bibnamefont {Nielsen}},\ }\bibfield  {title} {\bibinfo {title} {Optimal
  control design of nmr and dynamic nuclear polarization experiments using
  monotonically convergent algorithms},\ }\href
  {https://doi.org/10.1063/1.2903458} {\bibfield  {journal} {\bibinfo
  {journal} {J. Chem. Phys.}\ }\textbf {\bibinfo {volume} {128}},\ \bibinfo
  {pages} {184505} (\bibinfo {year} {2008})}\BibitemShut {NoStop}%
\bibitem [{\citenamefont {Maday}\ and\ \citenamefont
  {Turinici}(2003)}]{Maday2003}%
  \BibitemOpen
  \bibfield  {author} {\bibinfo {author} {\bibfnamefont {Y.}~\bibnamefont
  {Maday}}\ and\ \bibinfo {author} {\bibfnamefont {G.}~\bibnamefont
  {Turinici}},\ }\bibfield  {title} {\bibinfo {title} {New formulations of
  monotonically convergent quantum control algorithms},\ }\href
  {https://doi.org/10.1063/1.1564043} {\bibfield  {journal} {\bibinfo
  {journal} {J. Chem. Phys.}\ }\textbf {\bibinfo {volume} {118}},\ \bibinfo
  {pages} {8191} (\bibinfo {year} {2003})}\BibitemShut {NoStop}%
\bibitem [{\citenamefont {Zhu}\ and\ \citenamefont {Rabitz}(1998)}]{Zhu1998}%
  \BibitemOpen
  \bibfield  {author} {\bibinfo {author} {\bibfnamefont {W.}~\bibnamefont
  {Zhu}}\ and\ \bibinfo {author} {\bibfnamefont {H.}~\bibnamefont {Rabitz}},\
  }\bibfield  {title} {\bibinfo {title} {A rapid monotonically convergent
  iteration algorithm for quantum optimal control over the expectation value of
  a positive definite operator},\ }\href {https://doi.org/10.1063/1.476575}
  {\bibfield  {journal} {\bibinfo  {journal} {J. Chem. Phys.}\ }\textbf
  {\bibinfo {volume} {109}},\ \bibinfo {pages} {385} (\bibinfo {year}
  {1998})}\BibitemShut {NoStop}%
\bibitem [{\citenamefont {Shi}\ and\ \citenamefont {Rabitz}(1990)}]{Shi1990}%
  \BibitemOpen
  \bibfield  {author} {\bibinfo {author} {\bibfnamefont {S.}~\bibnamefont
  {Shi}}\ and\ \bibinfo {author} {\bibfnamefont {H.}~\bibnamefont {Rabitz}},\
  }\bibfield  {title} {\bibinfo {title} {Quantum mechanical optimal control of
  physical observables in microsystems},\ }\href
  {https://doi.org/10.1063/1.458438} {\bibfield  {journal} {\bibinfo  {journal}
  {J. Chem. Phys.}\ }\textbf {\bibinfo {volume} {92}},\ \bibinfo {pages} {364}
  (\bibinfo {year} {1990})}\BibitemShut {NoStop}%
\bibitem [{\citenamefont {Khaneja}\ \emph {et~al.}(2001)\citenamefont
  {Khaneja}, \citenamefont {Brockett},\ and\ \citenamefont {Glaser}}]{KN2}%
  \BibitemOpen
  \bibfield  {author} {\bibinfo {author} {\bibfnamefont {N.}~\bibnamefont
  {Khaneja}}, \bibinfo {author} {\bibfnamefont {R.}~\bibnamefont {Brockett}},\
  and\ \bibinfo {author} {\bibfnamefont {S.~J.}\ \bibnamefont {Glaser}},\
  }\bibfield  {title} {\bibinfo {title} {Time optimal control in spin
  systems},\ }\href@noop {} {\bibfield  {journal} {\bibinfo  {journal} {Phys.
  Rev. A}\ }\textbf {\bibinfo {volume} {63}},\ \bibinfo {pages} {032308}
  (\bibinfo {year} {2001})}\BibitemShut {NoStop}%
\bibitem [{\citenamefont {Boozer}(2012)}]{Boozer2012}%
  \BibitemOpen
  \bibfield  {author} {\bibinfo {author} {\bibfnamefont {A.~D.}\ \bibnamefont
  {Boozer}},\ }\bibfield  {title} {\bibinfo {title} {Time-optimal synthesis of
  su(2) transformations for a spin-1/2 system},\ }\href
  {https://doi.org/10.1103/PhysRevA.85.012317} {\bibfield  {journal} {\bibinfo
  {journal} {Phys. Rev. A}\ }\textbf {\bibinfo {volume} {85}},\ \bibinfo
  {pages} {012317} (\bibinfo {year} {2012})}\BibitemShut {NoStop}%
\bibitem [{\citenamefont {Hegerfeldt}(2013)}]{Hegerfeldt2013}%
  \BibitemOpen
  \bibfield  {author} {\bibinfo {author} {\bibfnamefont {G.~C.}\ \bibnamefont
  {Hegerfeldt}},\ }\bibfield  {title} {\bibinfo {title} {Driving at the quantum
  speed limit: Optimal control of a two-level system},\ }\href
  {https://doi.org/10.1103/PhysRevLett.111.260501} {\bibfield  {journal}
  {\bibinfo  {journal} {Phys. Rev. Lett.}\ }\textbf {\bibinfo {volume} {111}},\
  \bibinfo {pages} {260501} (\bibinfo {year} {2013})}\BibitemShut {NoStop}%
\bibitem [{\citenamefont {Carlini}\ \emph {et~al.}(2006)\citenamefont
  {Carlini}, \citenamefont {Hosoya}, \citenamefont {Koike},\ and\ \citenamefont
  {Okudaira}}]{Carlini2006}%
  \BibitemOpen
  \bibfield  {author} {\bibinfo {author} {\bibfnamefont {A.}~\bibnamefont
  {Carlini}}, \bibinfo {author} {\bibfnamefont {A.}~\bibnamefont {Hosoya}},
  \bibinfo {author} {\bibfnamefont {T.}~\bibnamefont {Koike}},\ and\ \bibinfo
  {author} {\bibfnamefont {Y.}~\bibnamefont {Okudaira}},\ }\bibfield  {title}
  {\bibinfo {title} {Time-optimal quantum evolution},\ }\href
  {https://doi.org/10.1103/physrevlett.96.060503} {\bibfield  {journal}
  {\bibinfo  {journal} {Phys. Rev. Lett.}\ }\textbf {\bibinfo {volume} {96}},\
  \bibinfo {pages} {60503} (\bibinfo {year} {2006})}\BibitemShut {NoStop}%
\bibitem [{\citenamefont {Meiss}(1992)}]{Meiss1992}%
  \BibitemOpen
  \bibfield  {author} {\bibinfo {author} {\bibfnamefont {J.~D.}\ \bibnamefont
  {Meiss}},\ }\bibfield  {title} {\bibinfo {title} {Symplectic maps,
  variational principles, and transport},\ }\href
  {https://doi.org/10.1103/revmodphys.64.795} {\bibfield  {journal} {\bibinfo
  {journal} {Rev. Mod. Phys.}\ }\textbf {\bibinfo {volume} {64}},\ \bibinfo
  {pages} {795} (\bibinfo {year} {1992})}\BibitemShut {NoStop}%
\bibitem [{\citenamefont {Bunimovich}(1995)}]{Bunimovich1995}%
  \BibitemOpen
  \bibfield  {author} {\bibinfo {author} {\bibfnamefont {L.~A.}\ \bibnamefont
  {Bunimovich}},\ }\bibfield  {title} {\bibinfo {title} {Variational principle
  for periodic trajectories of hyperbolic billiards},\ }\href
  {https://doi.org/10.1063/1.166105} {\bibfield  {journal} {\bibinfo  {journal}
  {Chaos}\ }\textbf {\bibinfo {volume} {5}},\ \bibinfo {pages} {349} (\bibinfo
  {year} {1995})}\BibitemShut {NoStop}%
\bibitem [{\citenamefont {Lan}\ and\ \citenamefont
  {Cvitanovic}(2004)}]{Lan2004}%
  \BibitemOpen
  \bibfield  {author} {\bibinfo {author} {\bibfnamefont {Y.}~\bibnamefont
  {Lan}}\ and\ \bibinfo {author} {\bibfnamefont {P.}~\bibnamefont
  {Cvitanovic}},\ }\bibfield  {title} {\bibinfo {title} {Variational method for
  finding periodic orbits in a general flow},\ }\href
  {https://doi.org/10.1103/physreve.69.016217} {\bibfield  {journal} {\bibinfo
  {journal} {Phys. Rev. E}\ }\textbf {\bibinfo {volume} {69}},\ \bibinfo
  {pages} {016217} (\bibinfo {year} {2004})}\BibitemShut {NoStop}%
\bibitem [{\citenamefont {Ghoussoub}\ and\ \citenamefont
  {Moameni}(2007)}]{Ghoussoub2007}%
  \BibitemOpen
  \bibfield  {author} {\bibinfo {author} {\bibfnamefont {N.}~\bibnamefont
  {Ghoussoub}}\ and\ \bibinfo {author} {\bibfnamefont {A.}~\bibnamefont
  {Moameni}},\ }\bibfield  {title} {\bibinfo {title} {Selfdual variational
  principles for periodic solutions of hamiltonian and other dynamical
  systems},\ }\href {https://doi.org/10.1080/03605300600781634} {\bibfield
  {journal} {\bibinfo  {journal} {Comm. in PDE}\ }\textbf {\bibinfo {volume}
  {32}},\ \bibinfo {pages} {771} (\bibinfo {year} {2007})}\BibitemShut
  {NoStop}%
\bibitem [{\citenamefont {Dong}\ and\ \citenamefont {Lan}(2014)}]{Dong2014}%
  \BibitemOpen
  \bibfield  {author} {\bibinfo {author} {\bibfnamefont {C.}~\bibnamefont
  {Dong}}\ and\ \bibinfo {author} {\bibfnamefont {Y.}~\bibnamefont {Lan}},\
  }\bibfield  {title} {\bibinfo {title} {A variational approach to connecting
  orbits in nonlinear dynamical systems},\ }\href
  {https://doi.org/10.1016/j.physleta.2014.01.001} {\bibfield  {journal}
  {\bibinfo  {journal} {Phys. Lett. A}\ }\textbf {\bibinfo {volume} {378}},\
  \bibinfo {pages} {705} (\bibinfo {year} {2014})}\BibitemShut {NoStop}%
\bibitem [{\citenamefont {Wang}\ \emph {et~al.}(2018)\citenamefont {Wang},
  \citenamefont {Wang},\ and\ \citenamefont {Lan}}]{Wang2018}%
  \BibitemOpen
  \bibfield  {author} {\bibinfo {author} {\bibfnamefont {D.}~\bibnamefont
  {Wang}}, \bibinfo {author} {\bibfnamefont {P.}~\bibnamefont {Wang}},\ and\
  \bibinfo {author} {\bibfnamefont {Y.}~\bibnamefont {Lan}},\ }\bibfield
  {title} {\bibinfo {title} {Accelerated variational approach for searching
  cycles},\ }\href {https://doi.org/10.1103/PhysRevE.98.042204} {\bibfield
  {journal} {\bibinfo  {journal} {Phys. Rev. E}\ }\textbf {\bibinfo {volume}
  {98}},\ \bibinfo {pages} {042204} (\bibinfo {year} {2018})}\BibitemShut
  {NoStop}%
\bibitem [{\citenamefont {Carlini}\ \emph {et~al.}(2007)\citenamefont
  {Carlini}, \citenamefont {Hosoya}, \citenamefont {Koike},\ and\ \citenamefont
  {Okudaira}}]{Carlini2007}%
  \BibitemOpen
  \bibfield  {author} {\bibinfo {author} {\bibfnamefont {A.}~\bibnamefont
  {Carlini}}, \bibinfo {author} {\bibfnamefont {A.}~\bibnamefont {Hosoya}},
  \bibinfo {author} {\bibfnamefont {T.}~\bibnamefont {Koike}},\ and\ \bibinfo
  {author} {\bibfnamefont {Y.}~\bibnamefont {Okudaira}},\ }\bibfield  {title}
  {\bibinfo {title} {Time-optimal unitary operations},\ }\href
  {https://doi.org/10.1103/physreva.75.042308} {\bibfield  {journal} {\bibinfo
  {journal} {Phys. Rev. A}\ }\textbf {\bibinfo {volume} {75}},\ \bibinfo
  {pages} {042308} (\bibinfo {year} {2007})}\BibitemShut {NoStop}%
\bibitem [{\citenamefont {Carlini}\ \emph {et~al.}(2008)\citenamefont
  {Carlini}, \citenamefont {Hosoya}, \citenamefont {Koike},\ and\ \citenamefont
  {Okudaira}}]{Carlini2008}%
  \BibitemOpen
  \bibfield  {author} {\bibinfo {author} {\bibfnamefont {A.}~\bibnamefont
  {Carlini}}, \bibinfo {author} {\bibfnamefont {A.}~\bibnamefont {Hosoya}},
  \bibinfo {author} {\bibfnamefont {T.}~\bibnamefont {Koike}},\ and\ \bibinfo
  {author} {\bibfnamefont {Y.}~\bibnamefont {Okudaira}},\ }\bibfield  {title}
  {\bibinfo {title} {Time optimal quantum evolution of mixed states},\ }\href
  {https://doi.org/10.1088/1751-8113/41/4/045303} {\bibfield  {journal}
  {\bibinfo  {journal} {J. Phys. A: Math. Theor.}\ }\textbf {\bibinfo {volume}
  {41}},\ \bibinfo {pages} {045303} (\bibinfo {year} {2008})}\BibitemShut
  {NoStop}%
\bibitem [{\citenamefont {Carlini}\ \emph {et~al.}(2014)\citenamefont
  {Carlini}, \citenamefont {Mari},\ and\ \citenamefont
  {Giovannetti}}]{Carlini2014}%
  \BibitemOpen
  \bibfield  {author} {\bibinfo {author} {\bibfnamefont {A.}~\bibnamefont
  {Carlini}}, \bibinfo {author} {\bibfnamefont {A.}~\bibnamefont {Mari}},\ and\
  \bibinfo {author} {\bibfnamefont {V.}~\bibnamefont {Giovannetti}},\
  }\bibfield  {title} {\bibinfo {title} {Time-optimal thermalization of
  single-mode gaussian states},\ }\href
  {https://doi.org/10.1103/PhysRevA.90.052324} {\bibfield  {journal} {\bibinfo
  {journal} {Phys. Rev. A}\ }\textbf {\bibinfo {volume} {90}},\ \bibinfo
  {pages} {052324} (\bibinfo {year} {2014})}\BibitemShut {NoStop}%
\bibitem [{\citenamefont {Wang}\ \emph {et~al.}(2015)\citenamefont {Wang},
  \citenamefont {Allegra}, \citenamefont {Jacobs}, \citenamefont {Lloyd},
  \citenamefont {Lupo},\ and\ \citenamefont {Mohseni}}]{Xiaoting2015}%
  \BibitemOpen
  \bibfield  {author} {\bibinfo {author} {\bibfnamefont {X.}~\bibnamefont
  {Wang}}, \bibinfo {author} {\bibfnamefont {M.}~\bibnamefont {Allegra}},
  \bibinfo {author} {\bibfnamefont {K.}~\bibnamefont {Jacobs}}, \bibinfo
  {author} {\bibfnamefont {S.}~\bibnamefont {Lloyd}}, \bibinfo {author}
  {\bibfnamefont {C.}~\bibnamefont {Lupo}},\ and\ \bibinfo {author}
  {\bibfnamefont {M.}~\bibnamefont {Mohseni}},\ }\bibfield  {title} {\bibinfo
  {title} {Quantum brachistochrone curves as geodesics: Obtaining accurate
  minimum-time protocols for the control of quantum systems},\ }\href
  {https://doi.org/10.1103/physrevlett.114.170501} {\bibfield  {journal}
  {\bibinfo  {journal} {Phys. Rev. Lett.}\ }\textbf {\bibinfo {volume} {114}},\
  \bibinfo {pages} {170501} (\bibinfo {year} {2015})}\BibitemShut {NoStop}%
\bibitem [{\citenamefont {Wang}\ \emph {et~al.}(2017)\citenamefont {Wang},
  \citenamefont {Allegra}, \citenamefont {Jacobs}, \citenamefont {Lloyd},
  \citenamefont {Lupo},\ and\ \citenamefont {Mohseni}}]{Xiaoting2017}%
  \BibitemOpen
  \bibfield  {author} {\bibinfo {author} {\bibfnamefont {X.}~\bibnamefont
  {Wang}}, \bibinfo {author} {\bibfnamefont {M.}~\bibnamefont {Allegra}},
  \bibinfo {author} {\bibfnamefont {K.}~\bibnamefont {Jacobs}}, \bibinfo
  {author} {\bibfnamefont {S.}~\bibnamefont {Lloyd}}, \bibinfo {author}
  {\bibfnamefont {C.}~\bibnamefont {Lupo}},\ and\ \bibinfo {author}
  {\bibfnamefont {M.}~\bibnamefont {Mohseni}},\ }\bibfield  {title} {\bibinfo
  {title} {{Time-optimal quantum control via differential geometry}},\ }in\
  \href {https://doi.org/10.1117/12.2256267} {\emph {\bibinfo {booktitle}
  {Advances in Photonics of Quantum Computing, Memory, and Communication X}}},\
  Vol.\ \bibinfo {volume} {10118},\ \bibinfo {editor} {edited by\ \bibinfo
  {editor} {\bibfnamefont {Z.~U.}\ \bibnamefont {Hasan}}, \bibinfo {editor}
  {\bibfnamefont {P.~R.}\ \bibnamefont {Hemmer}}, \bibinfo {editor}
  {\bibfnamefont {H.}~\bibnamefont {Lee}},\ and\ \bibinfo {editor}
  {\bibfnamefont {A.~L.}\ \bibnamefont {Migdall}}},\ \bibinfo {organization}
  {International Society for Optics and Photonics}\ (\bibinfo  {publisher}
  {SPIE},\ \bibinfo {year} {2017})\ pp.\ \bibinfo {pages} {53 --
  59}\BibitemShut {NoStop}%
\bibitem [{\citenamefont {Wootters}(1998)}]{Wootters1998}%
  \BibitemOpen
  \bibfield  {author} {\bibinfo {author} {\bibfnamefont {W.~K.}\ \bibnamefont
  {Wootters}},\ }\bibfield  {title} {\bibinfo {title} {Entanglement of
  formation of an arbitrary state of two qubits},\ }\href
  {https://doi.org/10.1103/PhysRevLett.80.2245} {\bibfield  {journal} {\bibinfo
   {journal} {Phys. Rev. Lett.}\ }\textbf {\bibinfo {volume} {80}},\ \bibinfo
  {pages} {2245} (\bibinfo {year} {1998})}\BibitemShut {NoStop}%
\bibitem [{\citenamefont {Bennett}\ \emph {et~al.}(1996)\citenamefont
  {Bennett}, \citenamefont {Bernstein}, \citenamefont {Popescu},\ and\
  \citenamefont {Schumacher}}]{Charles1996}%
  \BibitemOpen
  \bibfield  {author} {\bibinfo {author} {\bibfnamefont {C.~H.}\ \bibnamefont
  {Bennett}}, \bibinfo {author} {\bibfnamefont {H.~J.}\ \bibnamefont
  {Bernstein}}, \bibinfo {author} {\bibfnamefont {S.}~\bibnamefont {Popescu}},\
  and\ \bibinfo {author} {\bibfnamefont {B.}~\bibnamefont {Schumacher}},\
  }\bibfield  {title} {\bibinfo {title} {Concentrating partial entanglement by
  local operations},\ }\href {https://doi.org/10.1103/PhysRevA.53.2046}
  {\bibfield  {journal} {\bibinfo  {journal} {Phys. Rev. A}\ }\textbf {\bibinfo
  {volume} {53}},\ \bibinfo {pages} {2046} (\bibinfo {year}
  {1996})}\BibitemShut {NoStop}%
\bibitem [{\citenamefont {Margolus}\ and\ \citenamefont
  {Levitin}(1998)}]{Margolus1998}%
  \BibitemOpen
  \bibfield  {author} {\bibinfo {author} {\bibfnamefont {N.}~\bibnamefont
  {Margolus}}\ and\ \bibinfo {author} {\bibfnamefont {L.~B.}\ \bibnamefont
  {Levitin}},\ }\bibfield  {title} {\bibinfo {title} {The maximum speed of
  dynamical evolution},\ }\href {https://doi.org/10.1016/s0167-2789(98)00054-2}
  {\bibfield  {journal} {\bibinfo  {journal} {Physica D: Nonlinear Phenomena}\
  }\textbf {\bibinfo {volume} {120}},\ \bibinfo {pages} {188} (\bibinfo {year}
  {1998})}\BibitemShut {NoStop}%
\bibitem [{\citenamefont {Levitin}\ and\ \citenamefont
  {Toffoli}(2009)}]{Levitin2009}%
  \BibitemOpen
  \bibfield  {author} {\bibinfo {author} {\bibfnamefont {L.~B.}\ \bibnamefont
  {Levitin}}\ and\ \bibinfo {author} {\bibfnamefont {T.}~\bibnamefont
  {Toffoli}},\ }\bibfield  {title} {\bibinfo {title} {Fundamental limit on the
  rate of quantum dynamics: The unified bound is tight},\ }\href
  {https://doi.org/10.1103/PhysRevLett.103.160502} {\bibfield  {journal}
  {\bibinfo  {journal} {Phys. Rev. Lett.}\ }\textbf {\bibinfo {volume} {103}},\
  \bibinfo {pages} {160502} (\bibinfo {year} {2009})}\BibitemShut {NoStop}%
\end{thebibliography}%

\appendix
\section{}
\label{sec:A}
The variation of the Hamiltonian $H$ could now be taken as multivariate change with respect to $\{\xi_j\}_{j=1,2,...m}$. Let’s respectively check different parts of the action integral Eq.~(\ref{eq:S}). First, check one term of the integrand
\begin{eqnarray}\label{a1}
I &=& \langle \phi|H|\psi\rangle+c.c. \nonumber\\
  &=& \sum_j\langle \phi|A_j|\psi\rangle \xi_j(t)+c.c. \nonumber\\
  &=& \sum_j \xi_j(t) D_j\,,
\end{eqnarray}
where $D_j=\langle \phi|A_j|\psi\rangle +c.c. $ are real numbers, thus the Hamiltonian variation of the term I is
\begin{equation}\label{a2}
  \delta I=\sum\limits_j D_j \delta \xi_j(t).
\end{equation}
Next, let’s consider the second part
\begin{equation*}
  II =g(\bm\xi,\lambda,\lambda')= \lambda(\frac{1}{2}\sum\limits_{j=1}^{m}\xi_j^2-\omega^2)+\sum\limits_{j=n+1}^m\lambda^{'}_j (\xi_j-Q_i)\,.
\end{equation*}
Let $G_j=\frac{\delta g(\bm\xi,\lambda)}{\delta\xi_j}$. We obtain the variation of the II
\begin{equation}\label{a3}
  \delta II=\sum\limits_jG_j\delta \xi_j.
\end{equation}
Third, before treating first term of the Eq.~(\ref{eq:S}), we need to do some preparation. Here, we define:
\begin{equation}\label{a4}
  F_{jk}=\langle A_jA_k+A_kA_j\rangle-2 \langle A_j\rangle\langle A_k\rangle \,,
\end{equation}
which is a symmetric tensor, and the energy variance could be written as:
\begin{eqnarray}\label{a5}
\Delta E^2 &=& \langle H^2 \rangle -\langle H \rangle^2 \nonumber\\
  &=& \sum_{jk}\xi_j \xi_k \langle A_jA_k \rangle-\xi_j\xi_k\langle A_j \rangle\langle A_k \rangle\nonumber \\
  &=& \frac{1}{2}\sum_{jk} F_{jk}\xi_j\xi_k \,,
\end{eqnarray}
leading to
\begin{equation}\label{a6}
  \delta \Delta E^2=\delta \dfrac{\bm\xi\cdot F\cdot\bm\xi}{2}=\sum\limits_{jk} F_{jk}\xi_k \delta \xi_j\,,
\end{equation}
which gives
\begin{eqnarray}\label{a7}
\delta \frac{1}{\Delta E} &=& -\frac{1}{\Delta E^2}\frac{1}{2\Delta E}\delta \Delta E^2 \nonumber \\
&=& -\frac{1}{2\Delta E^3}\sum_{jk} F_{jk}\xi_k \delta \xi_j\,.
\end{eqnarray}
So, the variation of the first term of Eq.~(\ref{eq:S}) is
\begin{eqnarray}\label{a8}
\delta\frac{\sqrt{\langle\dot{\psi}\vert(1-P)\vert\dot{\psi}\rangle}}{\Delta E}&= \sqrt{\langle\dot{\psi}\vert(1-P)\vert\dot{\psi}\rangle}\delta\dfrac{1}{\Delta E}{} \nonumber \\
&=-\dfrac{1}{2\Delta E^2}\sum\limits_{jk}F_{jk}\xi_k\delta\xi_j\,,
\end{eqnarray}
where Eq.~(\ref{Eq1}) and~(\ref{a5}) have been used. With Eqs.~(\ref{a2}), ~(\ref{a3})and~(\ref{a8}), the variation with respect to $\xi_j$ gives
\begin{equation}\label{a9}
-\dfrac{1}{2\Delta E^2}\sum\limits_k F_{jk} \xi_k +D_j+G_j=0
\end{equation}
where
\begin{equation}\label{a9}
  G_j=\left\{
  \begin{aligned}
    &\lambda \xi_j,  &(j\le n).\\
    &\lambda\xi_j-\lambda^{'}_j , &(n< j\le m).
  \end{aligned}
 \right.
\end{equation}
\section{}
\label{sec:B}
Multiplyingv $\langle \psi|A_j$ on both sides of Eq.~(\ref{Eq4}) results in
\begin{widetext}
  \begin{equation}\label{b1}
    \dfrac{i}{2\Delta E^2}(\langle A_j \dot{H} \rangle-\langle A_j \rangle\langle \dot{H} \rangle)+i(\langle A_jH\rangle-\langle A_j\rangle\langle H\rangle)\dfrac{d}{dt}\dfrac{1}{2\Delta E^2}\!-i\frac{d}{dt}\langle\psi| A_j |\phi\rangle+\langle \psi|[A_j,H]|\phi\rangle\!=\!0.
  \end{equation}
\end{widetext}
The derivative of $\dfrac{1}{2\Delta E^2}$ is
\begin{equation}\label{b2}
  \dfrac{d}{dt}\dfrac{1}{2\Delta E^2}=-\dfrac{1}{\Delta E^3}\dfrac{1}{2\Delta E}\dfrac{d}{dt}\Delta E^2.
\end{equation}
Therefore, with Eq.~(\ref{a4}) and Eq.~(\ref{a5}) and using $\bm\xi\cdot \dot{F}\cdot\bm\xi=\langle[H,H^2]\rangle=0$, we have
\begin{equation}\label{b3}
  \dfrac{d}{dt}\Delta E^2=\bm\xi\cdot F\cdot\dot{\bm\xi}=\tilde{\bm F}\cdot\dot{\bm\xi},
\end{equation}
where $\tilde{\bm F} =\bm\xi \cdot F$. And the Eq.(\ref{b2}) can be rewritten as
\begin{equation}\label{b4}
  \dfrac{d}{dt}\dfrac{1}{2\Delta E^2}=-\dfrac{1}{2\Delta E^2}\dfrac{\tilde{\bm F}\cdot\dot{\bm\xi}}{\Delta E^2}.
\end{equation}
Substituting Eq.~(\ref{b4}) into Eq.~(\ref{b1}), we arrive at
\begin{equation}\label{b5}
  \dfrac{d}{dt}D_j=\dfrac{1}{2\Delta E^2}(\sum\limits_k F_{jk}\dot{\xi}_k-\dfrac{\tilde{\bm F}\cdot\dot{\bm\xi}}{\Delta E^2}\tilde{F_j})+\sum_{k}C_{jk}D_k\,,
\end{equation}
where $C_{jk}$ is a real number and satisfies
\begin{equation}\label{b6}
  [H,A_j]=-i\sum\limits_k C_{jk} A_k.
\end{equation}
\section{ $\dfrac{d}{dt}\bm D\cdot\bm\xi=0$}
\label{sec:C}
\begin{align}
  \dfrac{d}{dt}\bm D\cdot \bm \xi
  &=\bm D\cdot\dfrac{d\bm \xi}{dt} +\dfrac{d\bm D}{dt}\cdot \bm \xi \nonumber \\
  &=\sum_j \dot{\xi}_j D_j+\dfrac{\xi_j}{2\Delta E^2}\left(\sum_k F_{jk} \dot{\xi}_k-\dfrac{\tilde{F}\cdot\dot{\xi}}{\Delta E^2}\tilde{F_j}\right)+\sum_k \xi_j C_{jk}D_k\nonumber \\
  &=\dot{\bm \xi}\cdot \bm D+\dfrac{1}{2\Delta E^2}(\tilde{\bm F}\cdot\dot{\bm \xi}-\tilde{\bm F}\cdot\dot{\bm \xi}\dfrac{\tilde{\bm F}\cdot{\bm \xi}}{\Delta E^2}) +\bm \xi\cdot \bm C\cdot \bm D \nonumber \\
  &=\dot{\bm \xi}\cdot \bm D-\dfrac{\tilde{\bm F}\cdot\dot{\bm \xi}}{2\Delta E^2} +\bm \xi\cdot \bm C\cdot \bm D\,,
\end{align}
where $\tilde{\bm F}\cdot\bm \xi=2\Delta E^2 $. As $ D_j=\dfrac{1}{2\Delta E^2}\sum_k F_{jk}\xi_k-\lambda \xi_j-\lambda_j'$, which gives
\begin{equation}
  \dfrac{d}{dt}\bm D\cdot \bm \xi=\dfrac{\tilde{\bm F}\cdot\dot{\bm \xi}}{2\Delta E^2} -\dfrac{\tilde{\bm F}\cdot\dot{\bm \xi}}{2\Delta E^2} +\bm \xi\cdot \bm C\cdot \bm D=0
\end{equation}
 with $\bm\xi\cdot\dot{\bm\xi}=0$, $\bm\lambda'\cdot\dot{\bm\xi}=\bm\lambda'\cdot\bm{\dot u}=0$ and $\bm\xi\cdot \bm C=0$ (Seeing Appendix~\ref{sec:D}).
 \section{$\dot{\bm F}=\bm C\cdot \bm F +\bm F\cdot \bm C^{T}$}\label{sec:E}
 Acorrding to the Eq.~($\ref{a4}$):
 \begin{equation}
   F_{jk}=\langle A_jA_k+A_kA_j\rangle-2 \langle A_j\rangle\langle A_k\rangle \,,
 \end{equation}
 we have
 \begin{align}\label{E1}
   \dot{F}_{jk}= & i\langle HA_jA_k+HA_kA_j-A_jA_kH-A_kA_jH\rangle-2i \langle HA_j-A_jH\rangle\langle A_k\rangle -2i \langle A_j\rangle\langle HA_k-A_kH\rangle \nonumber\\
   =&i\langle [H,A_j]A_k+A_k[H,A_j]+ A_j[H,A_k]+ [H,A_k]A_j\rangle-2i \langle [H,A_j]\rangle\langle A_k\rangle -2i \langle A_j\rangle\langle [H,A_k]\rangle\nonumber\\
   =& \sum_lC_{jl}(\langle A_lA_k+A_kA_l\rangle-2\langle A_l\rangle\langle A_k\rangle) +\sum_lC_{kl}(\langle A_jA_l+ A_lA_j\rangle -2\langle A_j\rangle\langle A_l\rangle)\nonumber\\
    =& \sum_lC_{jl}F_{lk} +\sum_lC_{kl}F_{lj}\nonumber
    \,,
 \end{align}
 then
  \begin{equation}
   \dot{\bm F}=\bm C\cdot \bm F +\bm F\cdot \bm C^{T}\,.
 \end{equation}
\section{$\bm C\cdot\bm\xi=\bm\xi\cdot\bm C=0$}
\label{sec:D}
  Here we demonstrate that $\bm C\cdot\bm\xi=\bm\xi\cdot\bm C=0$ is universally valid with respect to spin algebra. In the case of $H\in\mathcal{A}$~ i.e.~$\bm\xi=\bm u$, it yields $\bm C\cdot\bm u=\bm u\cdot \bm C=0$.
  \subsection{one qubit}
    Consider  $A_i,A_j$ in a qubit and suppose that $\{A_j\}_{j=1\,,\cdots\,,M}$ satisfy the spin algebra
    \begin{equation}
        [A_n\,,A_j]=\sum_k \dfrac{1}{\sqrt{Tr I}}~i\epsilon_{njk} A_k \,,\label{eq:aij}
    \end{equation}
    and $I$ is $2^q$ dimensional unit matrix (q is the number of qubits) and $A_k=\frac{1}{\sqrt{2}}\sigma_k$ for $q=1$. Hence
    \begin{equation*}
      \sum_k C_{jk}A_k=i[H,A_j]=\sum_n i\xi_n[A_n,A_j]=\sum_{n,k}-\dfrac{1}{\sqrt{Tr I}}\epsilon_{njk}\xi_nA_k
    \end{equation*}
    which yields $C_{jk}=\sum_n-\dfrac{1}{\sqrt{Tr I}}\epsilon_{njk}\xi_n$. Thus $\bm\xi\cdot\bm C=-\bm C\cdot\bm\xi=0$.
\subsection{many qubits with and without entanglement}
We will prove that $\bm\xi\cdot\bm C=\bm C\cdot\bm\xi=0$ remains valid. Below, we denote the unit matrix $I_0$ with $\sigma_0$ and extend our index to include 0.

\textbf{Lemma} $C_{jk}=\sum_i \Gamma_{ijk}\xi_i$, where $\Gamma$ is an antisymmetric tensor.

\underline{{Proof}} ~ First let us check some properties of Pauli matrices. With the commutation and anti-commutation relations $\{\sigma_i,\sigma_j\}=2\delta_{ij}I(i,j=1,2,3)$, it is easy to check even if $i,j=0,1,2,3$
\begin{equation}
\sigma_i^{(\alpha)}\sigma_j^{(\alpha)}=\sum_k (\theta_{ijk}^{(\alpha)}+i\epsilon_{ijk}^{(\alpha)})\sigma_k^{(\alpha)}
\,,\end{equation}
where $\sigma_0=I_0$ which is two-dimensional unit matrix and $\alpha$ is the order index of qubits,
\begin{equation}
\epsilon_{ijk}^{(\alpha)}=\left\{
\begin{aligned}
&1, ~ijk\in\{123, 231, 312\}\\
&-1, ~ijk\in\{321, 213, 132\}\\
&0, ~\mathrm{otherwise}
\end{aligned}
\right.
\,,\end{equation}
and
\begin{equation}
\theta_{ijk}^{(\alpha)}=\left\{
\begin{aligned}
&1, ~&\mathrm{one~index~is~0,~the~other~two~equal}\\
&0, ~&\mathrm{otherwise}
\end{aligned}
\right.
\,,\end{equation}
which is totally symmetric with respect to any permutation of its subscripts. Now consider the commutator of N-fold entanglement $[A_\mu,A_\nu]=[\dfrac{\sigma_{\mu_1}^{(1)}\sigma_{\mu_2}^{(2)}...\sigma_{\mu_N}^{(N)}}{\sqrt{TrI}},\dfrac{\sigma_{\nu_1}^{(1)}\sigma_{\nu_2}^{(2)}...\sigma_{\nu_N}^{(N)}}{\sqrt{TrI}}]$. Here the factor $\dfrac{1}{\sqrt{TrI}}$ is introduced to guarantee that $Tr A_j^2=1$ and $\{\mu_j, \nu_j, \gamma_j\}_{j=1, 2, ..., N}=\{0,1,2,3\}$.
Without loss of generality,
\begin{widetext}
 \begin{equation}
   TrI[A_\mu,A_\nu]\!\!=\!\!\prod_j\sigma_{\mu_j}^{(j)}\sigma_{\nu_j}^{(j)}-\prod_j\sigma_{\nu_j}^{(j)}\sigma_{\mu_j}^{(j)}\!=\!\prod_j\!\sum_{\gamma_j}(\theta_{\mu_j\nu_j\gamma_j}^{(j)}+i\epsilon_{\mu_j\nu_j\gamma_j}^{(j)})\sigma_{\gamma_j}^{(j)}-\!\prod_j\!\sum_{\gamma_j}(\theta_{\nu_j\mu_j\gamma_j}^{(j)}+i\epsilon_{\nu_j\mu_j\gamma_j}^{(j)})\sigma_{\gamma_j}^{(j)}
   \,.
 \end{equation}

Define the polynomial coefficient of the $k$-th power of $\epsilon$ as $c_k(\mu,\nu,\gamma)$
, which satisfies
\begin{equation*}
\sum_k c_k p_k(\mu,\nu,\gamma)=\prod_j(\theta_{\mu_j\nu_j\gamma_j}^{(j)}+i\epsilon_{\mu_j\nu_j\gamma_j}^{(j)})-\prod_j(\theta_{\mu_j\nu_j\gamma_j}^{(j)}-i\epsilon_{\mu_j\nu_j\gamma_j}^{(j)})
\,,\end{equation*}
\end{widetext}
\begin{equation}
[A_\mu,A_\nu]=\sum_\gamma P_{\mu\nu\gamma}A_{\gamma}
\,,\end{equation}
where $p_k(\mu,\nu,\gamma)$ is the polynomial function of $\epsilon$ and $A_{\gamma}=\dfrac{1}{\sqrt{TrI}}\prod\limits_{j}\sigma_{\gamma_{j}}^{(j)}$, $P_{\mu\nu\gamma}=\dfrac{1}{\sqrt{TrI}}\sum\limits_kc_k p_k(\mu,\nu,\gamma)$ is a pure imaginary number and holds the series of odd powers of $\epsilon$ (It is easy to check that $ c_{2k}=1-(-1)^{2k}\equiv 0, k\in\mathbb{N}$).
Here we have utilized the symmetry of $\theta$ and the anti-symmetry of $\epsilon$.

Now commutate two of the indices ${\mu,\nu,\gamma}$ of $P$. Notice that $P$ is written as series of odd powers of $\epsilon$ and each pair of $\{\mu_j,\nu_j\}$ anti-commutes. Hence P has the same anti-symmetry as $\epsilon$.
Given the definition of $\bm C$
\begin{equation}
\sum\limits_\gamma C_{\nu\gamma}A_\gamma=i[H,A_\nu]
\,,\end{equation}
we have $C_{jk}=\sum\limits_n iP_{njk}\xi_n$. \\

According to the lemma, $\bm C\cdot\bm\xi=\sum\limits_k\sum\limits_n iP_{njk}\xi_n\xi_k=-\sum\limits_k\sum\limits_n iP_{kjn}\xi_n\xi_k=-\bm C\cdot\bm\xi$. Thus $\bm C\cdot\bm\xi=0$. Furthermore, $\bm C$ is anti-symmetric in consistency with $P$. Hence ~$\bm\xi\cdot \bm C=\sum\limits_j \xi_j C_{jk}=-\sum\limits_j \xi_j C_{kj}=-\bm C\cdot\bm\xi=0$.

\section{The geometric properties of spin vector evolution}
\label{sec:F}
  \textbf{Lemma:}~$\bm D'\cdot\langle\bm\sigma\rangle=0$ for single qubit.

  \underline{{Proof:}}~From Eq.~(\ref{Eq17})in the main text, we have $ Re(\langle\phi'|\psi\rangle)=Re(\langle\phi|\psi\rangle)=0$.

 Considering the projection property of Pauli matrices
  \begin{equation}\label{F1}
    \bm \sigma\cdot\hat{\bm p} =2|\psi\rangle\langle \psi|-I\\
  \end{equation}
  where $|\psi\rangle$ is a normalized eigenvector of  $\bm\sigma$ in the direction of $\hat{\bm p} =\langle \bm \sigma\rangle$, and $I$ is the unit matrix, then we have that
  \begin{align}\label{F2}
      \bm D'\cdot\langle\bm\sigma\rangle&=2Re(\langle\phi'|\bm\sigma|\psi\rangle)\cdot\langle\bm\sigma\rangle \nonumber\\
      &=2Re(\langle\phi'|\bm\sigma\cdot\hat{\bm p}|\psi\rangle) \nonumber \\
      &=2Re(2\langle\phi'|\psi\rangle\langle \psi|\psi\rangle-\langle\phi'|\psi\rangle)\nonumber\\
      &=0\,.
  \end{align}
  Similarly, we can easily verify that
  \begin{equation}\label{F3}
      \bm D\cdot\langle\bm\sigma\rangle=2Re(\langle\phi|\bm\sigma|\psi\rangle)\cdot\langle\bm\sigma\rangle=0.
  \end{equation}

\section{}
\label{sec:G}
In this section, we're going to elaborate on the rotation matrix $ R_{\bm\Omega'}(\theta)=R^{-1}_x(\Psi)R_y(\theta)R_x(\Psi)$ in the single qubit case. According to \ref{sec:1bit_ana}, the $\bm\Omega$ lies in the $z$ direction and perpendicular to the $\bm\omega_{eff}$, and $R_{\bm\Omega'}(\theta)$ depicts the rotation of $\psi(t)$ arounds a fixed axis with  $\bm\Omega'=\bm\omega_{eff}-\bm\Omega$. Obviously, the $\bm\omega_{eff}$ lies on the $x-y$ plane, and the direction of $\bm\omega_{eff}$ is taken as the $-y$ axis for simplicity. Then the rotation $R_{\bm\Omega'}(\theta)$ can be depicted as follow, first, rotating $\psi$ around $x$-axis by $\Psi$ to make $\bm\Omega'$ and $y$-aixs coincide $(R_x(\Psi))$, then, rotating $\psi$ around $y$-axis by $\theta$ $(R_y(\theta))$, finally, rotating $\psi$ around $x$-axis back by $\Psi$ $(R^{-1}_x(\Psi))$, where $\Psi$ is the angle between $\Omega'$ and $\bm\omega_{eff}$ and tan$(\Psi)=\frac{|\bm\Omega|}{|\bm\omega_{eff}|}$.

\end{document}